\documentclass[12pt]{article}
\usepackage{amsmath,amsfonts}
\usepackage{hyperref}
\usepackage{graphicx}
\usepackage{feynmp}
\usepackage[dvips]{color}
\usepackage{amsthm}
\usepackage{amssymb}
\usepackage{epic}
\usepackage{eepic}
\usepackage[matrix,arrow]{xy}
\usepackage{epsfig}

\usepackage{psfrag}

\makeatletter\@addtoreset{equation}{section}\makeatother

\def\bC {\mathbb{C}}

\def\bR {\mathbb{R}}
\def\bZ {\mathbb{Z}}

\newcommand{\beq}{\begin{equation}}
\newcommand{\eeq}{\end{equation}}
\newcommand{\bea}{\begin{eqnarray}}
\newcommand{\eea}{\end{eqnarray}}

\newcommand{\vev}[1]{{\left< {#1} \right>}}

\newcommand{\eqn}[1]{(\ref{#1})}
\newcommand{\nn}{\nonumber}

\newcommand{\LG}{{}^L\negthinspace\hspace{.4mm} G}

\newcommand{\Tr}{{\rm Tr\,}}

\newcommand{\cC}{{\mathcal C}}

\newcommand{\cD}{{\mathcal D}}
\newcommand{\cF}{{\mathcal F}}
\newcommand{\cG}{{\mathcal G}}

\newcommand{\cL}{{\mathcal L}}

\newcommand{\cN}{{\mathcal N}}

\newcommand{\cT}{{\mathcal T}}

\newcommand{\cW}{{\mathcal W}}

\newcommand{\cZ}{{\mathcal Z}}

\newcommand{\al}{\alpha}

\newcommand{\ga}{\gamma}
\newcommand{\Ga}{\Gamma}
\newcommand{\de}{\delta}
\newcommand{\De}{\Delta}

\newcommand{\om}{\omega}

\newcommand{\si}{\sigma}
\newcommand{\up}{\Upsilon}
\newcommand{\vf}{\varphi}

\newcommand{\CA}{{\mathcal A}}

\newcommand{\CC}{{\mathcal C}}
\newcommand{\CD}{{\mathcal D}}

\newcommand{\CF}{{\mathcal F}}
\newcommand{\CG}{{\mathcal G}}
\newcommand{\CH}{{\mathcal H}}

\newcommand{\CL}{{\mathcal L}}
\newcommand{\CM}{{\mathcal M}}

\newcommand{\CO}{{\mathcal O}}

\newcommand{\CT}{{\rm T}}

\newcommand{\SC}{{\mathsf C}}

\newcommand{\SL}{{\mathsf L}}

\newcommand{\SO}{{\mathsf O}}

\newcommand{\ST}{{\mathsf T}}
\newcommand{\SU}{{\mathsf U}}
\newcommand{\SV}{{\mathsf V}}

\newcommand{\rf}[1]{(\ref{#1})}

\newcommand{\BZ}{{\mathbb Z}}

\newcommand{\ra}{\rightarrow}

\newcommand{\pa}{\partial}
\newcommand{\bz}{\bar{z}}
\newcommand{\sst}{\scriptscriptstyle}

\newcommand{\fr}[2]{{\textstyle \frac{#1}{#2} }}
\newcommand{\bra}{\langle}
\newcommand{\ket}{\rangle}
\newcommand{\sz}{{\mathsf z}}
\newcommand{\sfc}{{\mathsf c}}
\newcommand{\sq}{{\mathsf q}}
\newcommand{\spp}{{\mathsf p}}
\newcommand{\sll}{{\mathsf l}}

\renewcommand{\title}[1]{\vbox{\center\LARGE{#1}}\vspace{5mm}}
\renewcommand{\author}[1]{\vbox{\center\large#1}\vspace{5mm}}
\newcommand{\address}[1]{\vbox{\center\em#1}}

\begin{document}
\bibliographystyle{utphys}
\begin{fmffile}{graphs}

\begin{titlepage}
\begin{center}
\vspace{5mm}
\hfill {\tt HU-EP-09/40}\\
\hfill {\tt pi-strings-144}\\
\vspace{8mm}

\title{Gauge Theory Loop Operators and Liouville Theory}
\vspace{6mm}

Nadav Drukker${}^{a}$\footnote{\href{mailto:drukker@physik.hu-berlin.de}
{\tt drukker@physik.hu-berlin.de}},
Jaume Gomis${}^{b}$\footnote{\href{mailto:jgomis@perimeterinstitute.ca}
{\tt jgomis@perimeterinstitute.ca}},
Takuya Okuda${}^{b}$\footnote{\href{mailto:takuya@perimeterinstitute.ca}
{\tt takuya@perimeterinstitute.ca}},
and
J\"org Teschner${}^{c}$\footnote{\href{mailto:teschner@mail.desy.de}
{\tt teschner@mail.desy.de}}
\vskip 5mm
\address{
${}^a$Institut f\"ur Physik, Humboldt-Universit\"at zu Berlin,\\
Newtonstra\ss e 15, D-12489 Berlin, Germany
}
\address{
${}^b$Perimeter Institute for Theoretical Physics,\\
Waterloo, Ontario, N2L 2Y5, Canada}
\address{
${}^c$DESY Theory, Notkestra\ss e 85, 22603 Hamburg, Germany}

\end{center}

\vspace{5mm}
\abstract{
\noindent
We propose a correspondence between loop operators in
a family of four dimensional $\cN=2$ gauge theories on 
$S^4$ -- including Wilson, 't~Hooft and dyonic operators -- and 
Liouville theory loop operators on a Riemann surface. This 
extends the beautiful relation between the partition function 
of these $\cN=2$ gauge theories and Liouville correlators found
by Alday, Gaiotto and Tachikawa. We show that the computation
of these Liouville correlators with the insertion of a Liouville loop 
operator reproduces Pestun's formula capturing the expectation 
value of a Wilson loop operator in the corresponding 
gauge theory. We prove that our definition of Liouville 
loop operators is invariant under modular transformations,
which given our correspondence, implies the conjectured 
action of S-duality on the gauge theory loop operators.
Our computations in Liouville theory make an explicit prediction
for the exact expectation value of 't~Hooft and dyonic loop 
operators in these $\cN=2$ gauge theories. The 
Liouville loop operators are also found to admit a simple 
geometric interpretation within quantum Teichm\"uller theory 
as the quantum operators representing the length of geodesics. 
We study the algebra of Liouville loop operators and show
that it gives evidence for our proposal as well as 
providing definite predictions for the operator product expansion
of loop operators in gauge theory.}
\vfill

\end{titlepage}

\tableofcontents

\section{Introduction}
\label{sec:intro}

The concept of {\it duality} has been a major driving force in 
the study of lattice models, field theories as well as string theory. 
It is, in fact, the main idea uniting different field theories 
and string/M-theory. As an early example, the dual resonance model, 
where the search for an amplitude which can be interpreted as a sum over 
poles in the $s$-, $t$- or $u$-channels, culminated in the celebrated Veneziano 
formula. The underlying structure, {\it worldsheet duality}, is a consequence 
of the modular bootstrap of two dimensional conformal field theory, 
which associates to an amplitude multiple inequivalent representations, 
each corresponding to a different sewing of the Riemann surface from 
pairs of pants.

Ever since, duality has proliferated in supersymmetric gauge theories
and string/M-theory as a conjectured equivalence between dual
theories evaluated at   different {\it spacetime} couplings. These dualities
have greatly advanced our nonperturbative understanding of gauge
theories and string/M-theory. An elegant example is the S-duality
of four dimensional ${\cal N}=4$ super Yang-Mills
\cite{Montonen:1977sn, Witten:1978ma,Osborn:1979tq}. The theory
with gauge group $G$ conjecturally admits the action of a duality
group $\Gamma_G\subset SL(2,\bR)$, which contains elements
that  exchange  the theory with gauge group $G$ with the theory
with dual gauge group $\LG$ \cite{Goddard:1976qe}.

Dualities are also expected to act on four dimensional ${\cal N}=2$
supersymmetric conformal field theories. Unlike gauge theories with
${\cal N}=4$ supersymmetry, which are unique given a choice of
gauge group $G$, the space of four dimensional ${\cal N}=2$
supersymmetric conformal field theories is very rich. The first
duality proposal was made for ${\cal N}=2$ $SU(2)$ super
Yang-Mills with $4$ fundamental hypermultiplets, conjectured by
Seiberg and Witten to admit the action of an $SL(2,\bZ)$ duality
group \cite{Seiberg:1994aj}. Other proposals for various
${\cal N}=2$ gauge theories have been made since then
(see e.g. \cite{Argyres:1997cg,Witten:1997sc}), strengthening the 
hypothesis of duality for
${\cal N}=2$ supersymmetric conformal field theories.

Gaiotto \cite{Gaiotto-N=2} has recently constructed a class of
${\cal N}=2$ theories, denoted by ${\cal T}_{g,n}$, from
the data associated with a genus $g$ Riemann surface with
$n$-punctures, denoted by $C_{g,n}$.\footnote{Important 
ingredients for this construction are the results presented in 
\cite{Witten:1997sc,Gaiotto:2009hg}.}
For this class of
${\cal N}=2$ gauge theories, which are based on $SU(2)$ gauge
groups,\footnote{This construction generalizes to higher rank gauge
groups. This requires enriching the types of punctures on the
Riemann surface. See \cite{Gaiotto-N=2} for more details.}
Gaiotto \cite{Gaiotto-N=2} has conjectured that the {\it spacetime}
duality of the gauge theory ${\cal T}_{g,n}$ corresponds to
{\it worldsheet} duality on the associated Riemann surface $C_{g,n}$:
\begin{center}
S-duality $\quad\longleftrightarrow\quad$ worldsheet duality\,.
\end{center}
The construction identifies the parameter space of ${\cal T}_{g,n}$
with the moduli space of complex structures of $C_{g,n}$,
as well as the duality group $\Gamma({\cal T}_{g,n})$ of the four
dimensional gauge theory ${\cal T}_{g,n}$ with the mapping class group of
$C_{g,n}$
\beq
\Gamma({\cal T}_{g,n}) \simeq {\rm MCG}(C_{g,n})\,.
\eeq

Given a theory ${\cal T}_{g,n}$ encoded by a Riemann surface
$C_{g,n}$, Gaiotto associates a generalized quiver gauge theory
to a pants decomposition of $C_{g,n}$, describing the sewing
of the Riemann surface from pairs of pants. A choice of pants 
decomposition corresponds to a choice of duality frame for 
${\cal T}_{g,n}$. At certain corners of
the moduli space of $C_{g,n}$ the surface degenerates to
a collection of $3g-3+n$ thin tubes connected to each other by
pairs of pants. In such singular limits the surface reduces to a
graph with tri-valent vertices, where the theory ${\cal T}_{g,n}$
has a weakly coupled Lagrangian description as a generalized
quiver gauge theory.%
\footnote{For the case of the four punctured sphere, denoted by
$C_{0,4}$, there are three boundaries
corresponding to sewing the surface along the $s$-,$t$- and
$u$-channels.}
Hence, the ${\cal T}_{g,n}$ theory admits multiple descriptions in
terms of generalized quivers, each valid in different corners of
parameter space and involving different degrees of freedom.

The partition function of ${\cal N}=2$ gauge theories
on $S^4$ was derived by Pestun in \cite{Pestun:2007rz} using localization
with respect to the $Osp(2|2)$ supergroup.
He showed that the partition function reduces to a matrix integral
over the zero modes of the scalar fields in the ${\cal N}=2$ vector
multiplets of the product of the Nekrasov partition function
$Z_{\text{Nekrasov}}$ \cite{Nekrasov:2002qd,Nekrasov:2003rj}
with its complex conjugate.
The partition function has an elegant interpretation in terms of the
product of a holomorphic contribution corresponding to the North
pole and an antiholomorphic contribution corresponding to the
South pole \cite{Pestun:2007rz}, akin to the holomorphically
factorized representation of two dimensional conformal field theory correlators.

In a paper by Alday, Gaiotto and Tachikawa \cite{Alday:2009aq},
the partition function of a family of
${\cal T}_{g,n}$ theories on $S^4$ was identified with a Liouville
conformal field theory correlation function on the associated Riemann
surface $C_{g,n}$
\beq
{\cal Z}_{{\cal T}_{g,n}}=\Big<\prod_{a=1}^n V_{m_a} \Big>_{C_{g,n}}\,,
\label{correl-parti}
\eeq
where $V_{m_a}$ are vertex operators in Liouville theory associated
with the punctures on the Riemann surface.\footnote{As we review in 
section \ref{sec:review}, $m$ is related to the mass of a hypermultiplet.} 
This observation points
towards an intriguing connection between four dimensional
${\cal N}=2$ gauge theories and the rich subject of two dimensional
conformal field theory.
The identification was made in a particular choice of duality frame
of ${\cal T}_{g,n}$, corresponding to a particular sewing of the
Riemann surface from pairs of pants. The modular bootstrap of Liouville
conformal field theory correlators
\cite{Teschner:2001rv,Teschner:2003en,Teschner:2008qh}, which
associates to a correlator a different representation for each choice
of sewing of the Riemann surface, can then be used to demonstrate
that the partition function ${\cal Z}_{{\cal T}_{g,n}}$ is invariant under the
corresponding duality group $\Gamma({\cal T}_{g,n})$.
This provides strong evidence for the action of S-duality for this
class of ${\cal N}=2$ gauge theories.

Since the supersymmetries imposed in the localization procedure are
preserved by a supersymmetric circular Wilson loop $W_R$ in a 
representation $R$ of the ${\cal N}=2$ gauge theory, 
Pestun's computation also yields the
exact result for the expectation value of a Wilson loop \cite{Pestun:2007rz}
\beq
\vev{W_R}=\int[da]\, \hbox{Tr}_R\,e^{2\pi i a}\,
\overline{Z}_\text{Nekrasov}\, Z_\text{Nekrasov}\,,
\label{pestun}
\eeq
where $a$ denotes the scalar components of the ${\cal N}=2$ vector
multiplets in the gauge theory. 
By inserting the trivial Wilson loop, one may recover the
expression of the partition function $\cZ_{{\cal T}_{g,n}}$. 

Loop operators in gauge theories play a pivotal r\^ole in understanding
dualities. Since they are characterized by a set of electric and magnetic
charges, the action of S-duality -- being a non-abelian extension of
electric-magnetic duality -- naturally acts on them. The study of Wilson
\cite{Maldacena-wl,Rey-Yee,Erickson,Dru-Gross,Pestun:2007rz} and
't~Hooft loop operators
\cite{Kapustin:2005py,Kapustin:2006pk,Gomis:2009ir,Gomis:2009xg}
in ${\cal N}=4$ super Yang-Mills has yielded quantitative evidence in
favour of the S-duality conjecture, demonstrating that the correlation
function of Wilson loops and 't~Hooft loops are exchanged under
S-duality \cite{Gomis:2009ir,Gomis:2009xg}.\footnote{The explicit
comparison was performed to next to leading order in the coupling
constant expansion.}
Characterizing loop operators in ${\cal N}=2$ gauge theories
and computing their expectation values provides a theoretical
framework in which to study S-duality for this class of theories.

In \cite{Drukker:2009tz} a complete classification of loop operators
in the ${\cal T}_{g,n}$ theories was performed. It was shown that
loop operators in ${\cal T}_{g,n}$ are geometrically in one-to-one
correspondence with homotopy classes of non-self-intersecting
curves on the Riemann surface $C_{g,n}$. The charges of a loop
operator, which can be labeled by a vector $d$ in a given duality
frame, were identified with the  Dehn-Thurston data $d$ of a closed curve
$\gamma_{d}$ on the Riemann surface $C_{g,n}$ in the corresponding
choice of pants decomposition.

In this paper we identify loop operators $L_{d}$ in the ${\cal N}=2$
gauge theory ${\cal T}_{g,n}$ with loop operators $\CL(\gamma_{d})$
we construct in the Liouville conformal field theory on $C_{g,n}$.
Our proposal gives the following Liouville conformal field theory
realization for the expectation value of a gauge theory loop operator
\beq
\vev{L_{d}}_{{\cal T}_{g,n}}
=\Big<\prod_{a=1}^n V_{m_a}
\cdot \CL(\gamma_{d})\Big> _{C_{g,n}}\,.
\label{proposal}
\eeq
The identification of loop operators in ${\cal N}=2$ gauge theories with
two dimensional conformal field theory correlation functions, introduces
a theoretical framework in which to compute {\it exactly} the expectation value
of loop operators in certain ${\cal N}=2$ gauge theories.

The modular properties of the Liouville loop operators combined with
modular invariance of Liouville correlation functions
\cite{Teschner:2001rv,Teschner:2003en,Teschner:2008qh}
implies that the correlator on the right hand side of (\ref{proposal})
is invariant under a change of pants decomposition of the Riemann
surface $C_{g,n}$. In contrast, the precise identification of the
loop operator $L_{d}$ on the left hand side of \eqn{proposal} --
whether it is a Wilson, 't~Hooft or dyonic
operator -- depends very sensitively on the choice of duality frame
of ${\cal T}_{g,n}$. We will show that, given a curve $\gamma_d$ in
$C_{g,n}$, there always exists a choice of pants decomposition $\sigma_W$ 
such that the result of the Liouville correlator exactly reproduces
Pestun's computation for the Wilson loop expectation value (\ref{pestun})
\begin{equation}
{\vev{W_R}}_{{\cal T}_{g,n}}
=\Big<\prod_{a=1}^n V_{m_a}
\cdot \CL(\gamma_{d})\Big> _{C_{g,n}}\,.
\end{equation}
This fits rather nicely with the field theory result stating that any
dyonic operator in the ${\cal N}=2$ theories ${\cal T}_{g,n}$ can
always be mapped to a Wilson loop by the action of S-duality
\cite{Kapustin:2006hi,Drukker:2009tz}. In other duality frames,
the Liouville correlator (\ref{proposal}) computes the expectation
value of a dyonic operator.

The modular invariance of Liouville correlation functions 
in the presence of loop operators, together with
our proposal (\ref{proposal}) implies for the gauge theory observables that
\begin{equation}
\vev{L_{d,\sigma}}_{{\cal T}_{g,n}}=\vev{L_{d',\sigma'}}_{{\cal T}_{g,n}}\,,
\end{equation}
where $d$ and $d'$ are the charges of the loop operator in the
${\cal T}_{g,n}$ duality frames $\sigma$ and $\sigma'$ respectively.
Therefore, our proposal automatically incorporates the conjectured
action of the S-duality group $\Gamma({\cal T}_{g,n})$ on gauge
theory loop operators.

The quantization of the Teichm\"uller space of the Riemann surface $C_{g,n}$,
referred to as quantum Teichm\"uller theory, provides a dual
description of Liouville theory leading to a geometric
description of loop operators in ${\cal T}_{g,n}$. In recent years, a
precise correspondence between Liouville conformal field theory
and quantum Teichm\"uller theory has been established
\cite{Teschner:2003em,Teschner:2003at,Teschner:2005bz}\footnote{The existence of such a correspondence had been predicted in \cite{Verlinde:1989ua}}.
We will show that the loop operators ${\cal L}(\ga)$ are in one-to-one 
correspondence with the operators $\SL_\ga$ representing the geodesic 
length functions in the quantum theory of the Teichm\"uller spaces, and 
that the expectation values \rf{proposal}
are equal to the expectation values 
\begin{equation}
\langle\,q\,|\,\SL_\ga\,|\,q\,\rangle\,,
\label{qLq}
\end{equation}
with $|\,q\,\rangle$ being the coherent state associated to the complex structure of the
surface $C_{g,n}$ in quantum Teichm\"uller theory.

It follows that the algebra of length operators,
which has been well-studied in the literature, 
should be identified with the algebra of
loop operators in gauge theory.
We show as a particular consequence that
the gauge theory loop operators
captured by Liouville theory
satisfy the 't~Hooft commutation relations,
which are known to be obeyed by
gauge theory loop operators that are Hopf-linked in a constant time slice.
We also demonstrate that 
the charges of the operators that appear in the OPE,
as determined by the algebra of length operators,
precisely matches the prediction of S-duality,
thus providing a non-trivial consistency check
of our proposal.

The plan of the rest of the paper is as follows. In Section~\ref{sec:review}
we briefly introduce the necessary ingredients entering the construction
of the ${\cal N}=2$ gauge theories ${\cal T}_{g,n}$ given a Riemann surface
$C_{g,n}$ \cite{Gaiotto-N=2} and the relation to Liouville theory found in
\cite{Alday:2009aq}. We then recall the classification of loop operators in
${\cal T}_{g,n}$ and identification of loop operators with non-self-intersecting
closed curves in $\gamma$. In Section~\ref{sec:pert} we introduce the
notion of loop operators in Liouville field theory, construct them in terms
of the Lax connection and relate the holonomy of the Lax connection
around a closed curve with the monodromy of a degenerate
operator in the Liouville conformal field theory as it goes around the curve.

In Section~\ref{sec:bootstrap} we make the explicit identification
relating loop operators in gauge theory with loop operators in Liouville
field theory. We put forth an exact prescription for calculating
the monodromies of degenerate operators and compare the result
to the known gauge theory expression for the Wilson loop.
In Section~\ref{sec:calculations} we implement this general prescription,
first by providing the necessary tools and then study specific examples
of Wilson loops and 't~Hooft loops in $\cN=2^*$ theory and ${\cal N}=2$
$SU(2)$ gauge theory with four fundamental flavours

A more abstract point of view on the problem is presented in
Section~\ref{liou-teich} which reviews quantum-Teichm\"uller theory and
the realization of loop operators as geodesic length operators.
Section~\ref{sec:algebra} uses the algebra of geodesic length operators
to study the algebra of gauge theory loop operators, and 
compares it to the explicit expressions we derived for the loop operators
in Section~\ref{sec:calculations}.

We conclude with a discussion of our results in Section~\ref{sec:discuss}. Some technical calculations and relevant formulae are relegated to the appendices.

\medskip\medskip

\noindent
{\it Note:}
We have coordinated the submission to the arXiv of our work
and the 
related work by Fernando Alday, Davide Gaiotto, Sergei Gukov, 
Yuji Tachikawa and Herman Verlinde 
 \cite{Alday:2009fs}. 
We would like to thank 
them for  arranging the joint release of the two papers.

\section{Gauge Theories, Liouville Theory and Curves on Riemann Surfaces}
\label{sec:review}

In this section we review the key elements in Gaiotto's construction of the
$\cN=2$ theories ${\cal T}_{g,n}$ and of the
connection with Liouville theory \cite{Alday:2009aq}.
We also recount the classification of loop operators in ${\cal T}_{g,n}$
in terms of curves on the Riemann surface $C_{g,n}$ \cite{Drukker:2009tz}.
Once equipped with the key ingredients, we will proceed with the
Liouville conformal field theory realization of the gauge theory loop
operators in the following sections.

\subsection{Gauge Theories and Liouville Theory}
\label{sec:AGT}

In \cite{Gaiotto-N=2}, Gaiotto has put forward an algorithm which
associates a four dimensional ${\cal N}=2$ theory to a Riemann
surface. Furthermore, he has interpreted the theory as the low
energy description of coincident M5-branes wrapping the Riemann
surface. In this paper we consider
a special class of these theories, namely,
those whose gauge group is the product of several
$SU(2)$ groups.
They arise from the $A_1$ $(2,0)$ six dimensional
conformal field theory on a genus $g$ Riemann surface with $n$
punctures $C_{g,n}$ in the presence of $n$ codimension-two defect
operators in the $(2,0)$ theory. The theory associated with the
Riemann surface $C_{g,n}$ is denoted by ${\cal T}_{g,n}$. The Riemann
surface $C_{g,n}$ plays a key r\^ole in the construction. There
is a canonical double cover of $C_{g,n}$ corresponding to the 
Seiberg-Witten curve of ${\cal T}_{g,n}$ which determines the
prepotential of the gauge theory. The curve $C_{g,n}$ is therefore of
direct relevance for the physics of the four-dimensional gauge theory ${\cal T}_{g,n}$.

A weakly coupled description of the ${\cal T}_{g,n}$ theory is
encoded in a trivalent graph, also
referred as a (generalized) quiver diagram.
Given a choice of sewing $\sigma$ of $C_{g,n}$ from $3g-3+n$ pairs of pants
one may associate a trivalent graph $\Gamma_\sigma$, whose thickening
spans $C_{g,n}$ in the corresponding region in Teichm\"uller space.%
\footnote{More precisely, in order to fully specify
the duality frame in which the weakly coupled description
is valid, we need to specify both a trivalent graph $\Gamma_\sigma$
and a pants decomposition. The pair of these data is called
a ``marking'' (see {\it e.g.}, \cite{Teschner:2005bz}).
In this paper $\sigma$ really denotes a marking, but we will often
refer to it by the more familiar term ``pants decomposition''.
\label{comment-pants}
}
The trivalent graph has $3g-3+n$ internal edges and $n$ external ones.
The field content of this description of the ${\cal N}=2$ ${\cal T}_{g,n}$
theory can be read off from the trivalent graph by
associating to each internal edge an $SU(2)$ gauge group and to each
cubic vertex eight half-hypermultiplets in the trifundamental
fundamental representation of the $SU(2)^3$ group associated to the
three incoming edges. When the edge is external the $SU(2)$
symmetry corresponds to a global symmetry. Different choices of
pants decomposition of $C_{g,n}$ give rise to different weakly
coupled descriptions of $ {\cal T}_{g,n}$.

\begin{figure}[ht]
\begin{center}
\begin{tabular}{cccccc}
\raisebox{0mm}{\parbox{4cm}{
\hspace{5mm}
\includegraphics[scale=.5]{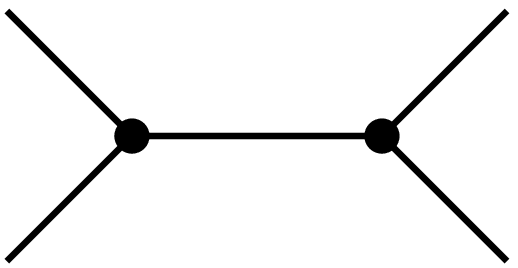}
}}
&&
\raisebox{0mm}{\parbox{4cm}{
\hspace{5mm}
\includegraphics[scale=.5]{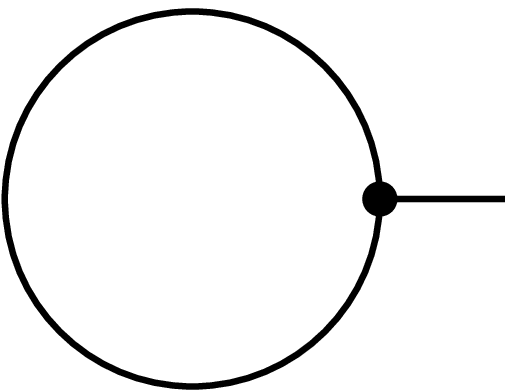}
}}
&&
\hspace{0mm}
\raisebox{0mm}{\parbox{4cm}{
\includegraphics[scale=.5]{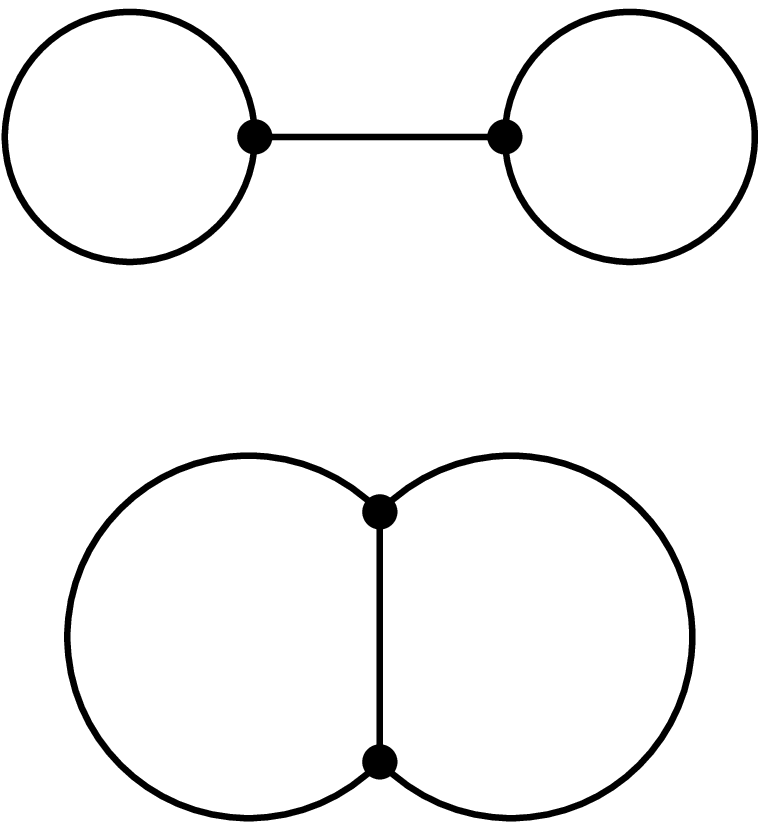}
}}
\\
\\
(a)&&(b)&&(c)
\end{tabular}
\parbox{5in}{
\caption{Examples of trivalent graphs (quiver diagrams)
corresponding to Riemann surfaces:
(a) a 4-puncture sphere, (b) a once-punctured torus,
(c) a genus two surface with no puncture.
\label{quivers}}}
\end{center}
\end{figure}

The two simplest theories of this class are $\cT_{1,1}$ and $\cT_{0,4}$.
The first case, $\cT_{1,1}$ corresponds to $\cN=2^*$ super Yang-Mills, the mass
deformation of $\cN=4$ super Yang-Mills. The relevant Riemann surface $C_{1,1}$ is
the one-punctured torus whose pants decomposition involves only a single
pair of pants with two legs glued to each other. A quiver diagram for
this theory is shown in Figure~\ref{quivers}b.
The $\cT_{0,4}$ theory also has a single gauge group and it has four
fundamental hypermultiplets. The corresponding surface is $C_{0,4}$ -- the
four-punctured sphere -- which can be decomposed into two pairs
of pants glued at a single leg. This leads to the quiver diagram
in Figure~\ref{quivers}a.

The Riemann surface $C_{g,n}$ encodes more information
than the field content of the gauge
theory ${\cal T}_{g,n}$. The space of
couplings of the gauge theory is
identified with the Teichm\"uller space of $C_{g,n}$,
{\it i.e.}, the space of deformations
of the complex structure of $C_{g,n}$
through the formula
\beq
q_i=\exp \left(2\pi i \tau_i\right)
\qquad i=1,\ldots, 3g-3+n
\eeq
where
\beq
\tau_i\equiv \frac{\theta_i}{2\pi}
+\frac{4\pi i}{g^2_i}
\eeq
are the gauge coupling constants
and $q_i$ are the gluing parameters
associated with the pants decomposition $\sigma$ of $C_{g,n}$.
Thus the physical parameter space of the gauge theory
is precisely the moduli space of $C_{g,n}$,
obtained by modding the Teichm\"uller space by
${\rm MCG}(C_{g,n})=\Gamma(\cT_{g,n})$. Furthermore, the duality group 
$\Gamma({\cal T}_{g,n})$ acting on ${\cal T}_{g,n}$ has an elegant geometrical
interpretation as the mapping class group ${\rm MCG}(C_{g,n})$ of $C_{g,n}$.

In \cite{Alday:2009aq} the correspondence between a four dimensional
${\cal N}=2$ gauge theory and a Riemann surface was extended to a
correspondence between the four dimensional gauge theory and the
Liouville conformal field theory living on the Riemann surface.

The action describing Liouville field theory on an arbitrary genus $g$
Riemann surface with $n$ punctures -- henceforth denoted by $C_{g,n}$ --
is given by
\beq
S=\frac{1}{ 4\pi}\int d^2z\left(g^{ab}\partial_a\phi\partial_b\phi
+QR\phi+4\pi \mu e^{2b\phi}\right)\,,
\label{liouacti}
\eeq
where $(z,{\overline z})$ is a local coordinate system on the Riemann
surface, $Q=b+1/b$ and the Liouville central charge is $c=1+6Q^2$
(See reference \cite{Teschner:2001rv} for a 
review of Liouville field theory). The operators creating delta function
normalizable states are labeled by a quantum number
$\alpha\in Q/2+ i\bR^+$, and give rise to a continuous spectrum.
These operators, which are denoted by $V_\alpha$, describe
the quantization of the semiclassical expression
\beq
V_\alpha(z,\overline{z})\simeq e^{2\alpha \phi(z,\overline{z})}\,
\eeq
and have conformal dimension $\Delta(\alpha)=\alpha(Q-\alpha)$.

Within the conformal bootstrap approach
\cite{Belavin:1984vu}, any correlation function can be constructed by
"sewing" three point functions according to the pattern
given by a pants decomposition of the Riemann surface $C_{g,n}$, 
where ``sewing" is defined by summing over the vertex operator
insertions representing a complete set of intermediate states.
The summation  
splits into the integration over the different Virasoro primary fields,
and the summation over their descendants. The result of the summation over the
descendants factorizes into a holomorphic and an anti-holomorphic part,
the so-called conformal blocks.
For the correlation function (\ref{correl-parti}), we denote the
conformal block
\cite{Belavin:1984vu} in the sewing of $C_{g,n}$ given by $\sigma$ by
\beq
{\cal F}_{\alpha,E}^{(\sigma)}\,,
\label{BPZblock}
\eeq
where $\alpha\equiv(\alpha_1,\ldots,\alpha_{3g-3+n})$ are the labels
of the representations associated to the internal edges of $\Gamma_\sigma$ while
$E\equiv(m_1,\ldots,m_{n})$ label the external vertex operator insertions.

In this formulation the correlation function of Liouville theory is written as
the integral over the holomorphic and anti-holomorphic conformal blocks
\beq
\Big<\prod_{a=1}^n V_{m_a}\Big>_{C_{g,n}}=\int d\nu(\alpha)\,
{\overline \cF^{(\sigma)}_{\alpha,E}}\,\cF^{(\sigma)}_{\alpha,E}\,,
\label{correla-F}
\eeq
where the measure $\nu(\alpha)$ includes for each pair of pants the
three point functions of the primary fields, for which the explicit expression was
proposed in \cite{Dorn:1994xn,Zamolodchikov:1995aa}, and 
derived in \cite{Teschner:2003en}.

The conformal blocks ${\cal F}^{(\sigma)}_{\alpha,E}$
depend on the choice of pants decomposition $\sigma$, but the complete
correlator does not. This is a consequence of the modular properties in two
dimensional conformal field theory. The conformal blocks corresponding
to two different trivalent graphs $\Gamma_\sigma$ and $\Gamma_\sigma'$
are related by a change of basis, as proved in \cite{Teschner:2003en} for $g=0$.

A key ingredient in the identification between ${\cal T}_{g,n}$ and
Liouville theory is played by the Nekrasov partition function
$Z_\text{instanton}$ \cite{Nekrasov:2002qd,Nekrasov:2003rj}.
This partition function, which is defined by localizing the ${\cal N}=2$
gauge theory with respect to a $U(1)\times U(1)$ subgroup of the
$SO(4)$ rotation group with rotation parameters
$(\epsilon_1,\epsilon_2)$, is the product of the classical contribution
$Z_\text{classical}$, the perturbative one loop contribution
$Z_\text{1-loop}$ and the instanton contribution $Z_\text{instanton}$
\beq
Z_\text{Nekrasov}= Z_\text{classical} Z_\text{1-loop} Z_\text{inst}\,.
\eeq

Given
a choice of pants decomposition $\sigma$ of the Riemann
surface $C_{g,n}$,
the equivariant instanton partition function \cite{Nekrasov:2002qd,Nekrasov:2003rj}
of the gauge theory ${\cal T}_{g,n}$ associated
with the generalized quiver diagram $\Gamma_\sigma$
equals the BPZ conformal block \cite{Belavin:1984vu} on $C_{g,n}$
in the choice of pants decomposition $\sigma$ \cite{Alday:2009aq}%
\footnote{The definition of a conformal block also requires
specifying how the graph $\Gamma_\sigma$
is drawn on the decomposed Riemann surface. See footnote
\ref{comment-pants}.
}
\beq
Z_\text{instanton}^{(\sigma)}={\cal F}_{\alpha,E}^{(\sigma)}\,.
\label{nekra-liou}
\eeq
The partition function $Z_\text{instanton}$ depends on the zero mode
of the $3g-3+n$ scalar fields in the ${\cal N}=2$ vector multiplets%
\footnote{In the gauge theory
the parameters $a_i$ are traceless anti-hermitean matrices, and
$\hat m_a$ imaginary numbers, of the same magnitude as the physical
mass. Up to $SU(2)$ transformations $a_i$ is given by one of the two
eigenvalues and we identify the matrix and its eigenvalue.
In \eqn{alphaa} $a_i$ is the eigenvalue and in \eqn{pestun}
and \eqn{pestuna} it is a matrix with the appropriate Haar measure.}
$\vec{a}$, on $n$ mass parameters $\vec{\hat m}$ and on the equivariant
localization parameters $\epsilon_1$ and $\epsilon_2$. The BPZ
conformal block \cite{Belavin:1984vu}
depends on the conformal dimension of the
vertex operators decorating the $3g-3+n$ internal edges and $n$
external edges of $\Gamma_\sigma$, which carry Liouville momenta
$\vec{\alpha}$ and $\vec{m}$ respectively,
as well as on the Liouville central charge $c$.
With the proposed dictionary
\begin{equation}
\begin{aligned}
\label{alphaa}
\alpha_i &= Q/2+a_i&\qquad i&=1,\ldots, 3g-3+n\\
m_a &= Q/2+\hat m_a&\qquad a&=1,\ldots, n\\
b&=\epsilon_1\\
1/b&=\epsilon_2\,,
\end{aligned}
\end{equation}
the relation (\ref{nekra-liou}) was demonstrated up to several orders in
a Taylor series expansion in $q_i=\exp(2\pi i \tau_i)$ 
\cite{Alday:2009aq}.\footnote{The 
correspondence between the instanton partition in gauge theory and 
Liouville theory has been extended to a class of
asymptotically free theories in \cite{Gaiotto:2009ma}.}

The complete Liouville theory correlator on $C_{g,n}$ also
admits an elegant gauge theory interpretation.
It was found in \cite{Alday:2009aq} that
the Liouville correlator on $C_{g,n}$ (\ref{correl-parti}) for $b=1$
(or equivalently $c=25$) corresponds to the partition function of
${\cal T}_{g,n}$ on $S^4$ computed in the duality frame $\sigma$
\cite{Pestun:2007rz}\footnote{The precise agreement holds after a overall
normalization constant for the vertex operators is removed.}
\beq
\cZ^{(\sigma)}_{{\cal T}_{g,n}}=\int[da]\, \overline{Z}^{(\sigma)}_\text{Nekrasov}
Z^{(\sigma)}_\text{Nekrasov}
=\Big<\prod_{a=1}^n V_{m_a}\Big>_{C_{g,n}}\,.
\label{pestuna}
\eeq
Pestun's partition function resembles the holomorphically factorized
representation of two dimensional conformal field theory correlators.
Formula (\ref{pestuna}) makes this observation precise. In making the
identification, the authors of \cite{Alday:2009aq}
noted that the product of three point functions of primary operators in
Liouville \cite{Dorn:1994xn,Zamolodchikov:1995aa,Teschner:2003en} 
precisely assemble themselves to realize the measure in Pestun's 
partition function (\ref{pestuna}).

It is important to note that the Liouville correlator
$\vev{\prod_{a=1}^n V_{m_a}}$
does not depend on the choice of pants decomposition of
$C_{g,n}$. Correlators in Liouville theory
are invariant with respect to the action of ${\rm MCG}(C_{g,n})$,
which relate different sewings of the surface
\cite{Teschner:2001rv,Teschner:2003en,Teschner:2008qh}. This implies that the
partition function of ${\cal T}_{g,n}$ on $S^4$ is invariant under
the S-duality group $\Gamma({\cal T}_{g,n})$
\beq
\cZ^{(\sigma)}_{{\cal T}_{g,n}}=\cZ^{(\sigma')}_{{\cal T}_{g,n}}\,,
\label{pestunb}
\eeq
where $\sigma$ and $\sigma'$ denotes a choice of pants
decomposition of $C_{g,n}$.

\subsection{Loop Operators in Gauge Theory and Curves on 
Riemann Surfaces}
\label{sec:geodesics}

The $(2,0)$ theory in six dimensions has surface observables
(strings), which arise
in M-theory from M2-branes ending on the M5-branes whose low energy
dynamics
are governed by this conformal field theory.
Upon dimensional reduction, these observables become local operators,
loop operators or surface operators, depending on how many
of the dimensions are along the compact manifold and how many in the four
dimensional space. In this paper we focus on the case where one of
the directions is on the Riemann surface and the other is in the four
dimensional
space.

As far as the four dimensional theory is concerned, these are 
operators supported on one dimensional curves, like 
Wilson loops. In fact in the present
framework we can study Wilson loops,
't~Hooft loops and dyonic loop operators in a unified fashion. We
focus on loop operators supported on a circle,
which have the potential of preserving the
most supersymmetries, and study the possible shapes they can take on the
internal manifold.

One would therefore expect that the classification of maximally
supersymmetric loop operators in the $\cN=2$ gauge theory
$\cT_{g,n}$ is related to the classification of loop
operators in Liouville theory
on the two dimensional surface $C_{g,n}$. Moreover, as
mentioned above, the S-duality symmetries of the
gauge theory are the modular symmetries of the Riemann surface.
Understanding the duality map for loop operators in Liouville theory
on $C_{g,n}$ will teach us the precise working of
S-duality for gauge theory loop operators,
generalizing the exchange of a fundamental Wilson loop and an
't~Hooft loop in $\cN=4$ super Yang-Mills.

The classification of loop operators in these gauge theories and the
mapping to curves on Riemann surfaces,
as well as their transformation rules under S-duality
were presented in \cite{Drukker:2009tz}, as we review now.

Loop operators in the gauge theory measure the response of the
system to an external particle which may be charged both electrically
and
magnetically under any of the gauge groups. This gives the set of
integers $(p_i,q_i)$, where $i=1,\cdots,3g-3+n$,
with $p_i$ being the magnetic charge%
\footnote{In \cite{Drukker:2009tz} a generalization was presented which
allowed for magnetic charges on the flavor groups, but we will suppress
this possibility in the present paper.}
in the $i^{\rm th}$ group and $q_i$ the electric charge.%
\footnote{It should be clear from the context whether
$q_i$ refers to the electric charge or the gluing parameter/modulus
$\exp(2\pi i\tau_i)$.
}
Some extra conditions are imposed on the set of magnetic charges
$p_i$, due to the presence of a Dirac string in a monopole background.
In order for the loop operator to be well-defined, all of the fields of
the theory, in particular the hypermultiplets in fundamental
representations of different $SU(2)$ factors,
should be single-valued even around the Dirac string.
This implies that for each trivalent vertex
in the quiver diagram, the sum of the three magnetic charges
associated with the edges connected to it is even:
\beq
p_i+p_j+p_k \in 2 \bZ\quad\hbox{if $i,j,k$ are three boundaries of a pair of pants.}
\label{dirac-cond}
\eeq
Beyond that, in each gauge group
there is an unbroken $\bZ_2$ Weyl symmetry which we have to mod out by,
and therefore we may assume that $p_i\geq0$,
and if $p_i=0$ that $q_i\geq0$.

This set of data should match the classification of
loop operators on
the
surface $C_{g,n}$, or more precisely loop operators for the Liouville theory
on this Riemann surface. At the purely topological level the observation
of \cite{Drukker:2009tz} is that there is a one-to-one map between the
possible sets of charges of loop operators in $\cT_{g,n}$ and
homotopy classes of curves, not necessarily connected,%
\footnote{In this section, we mean by a ``curve'' a one-dimensional
submanifold that may have more than one connected component.
In the rest of the paper, a ``curve'' is assumed to be
connected.
\label{foot:curve}
}
without
self-intersections.

There is a beautiful classification of all such curves due to Dehn
\cite{dehn-breslau} and independently to Thurston \cite{MR956596},
in terms of precisely the same data
\beq
d\equiv 
(p_i,q_i)_{i=1,\ldots,3g-3+n}
\eeq
called {\it Dehn-Thurston parameters},
as those  which  label the charges of gauge theory loop operators.
The classification is performed for a given pants decomposition
of the surface and therefore is perfectly suited to our purposes
of matching with the gauge theory data in a given duality frame.

For a given pants decomposition one can choose open segments
on the three boundaries of each pair of pants. By a homotopy it
is possible to arrange that all the components of
the curve going through the pair of
pants cross the boundaries through these segments. Then each
component will follow one of the six possible paths illustrated in
Figure~\ref{fig:arcs} (the open segments are the upper
halves of each of the circles).
How many times the curve
crosses each boundary of the pair of pants
uniquely specifies the combination of paths
that the curve contains.
See \cite{Drukker:2009tz} for more detailed
explanation.

Clearly the total number of times the curve passes the
boundaries of the pair of pants is even,
matching the Dirac condition on the magnetic charges $p_i$,
which are also all non-negative.
In gluing two pairs of pants there is a
canonical identification of the curves on the two sides due to the
open segments, but one may choose to do a cyclic permutation
(in other words a twist around the boundary)
on the lines before the identification, which can then be associated
to $q_i\in \bZ$.
Lastly when $p_i=0$, there are no lines crossing the
boundary between two pairs of pants, and it is possible to introduce
$q_i\geq 0$ loops around the boundary as components of the curve
such that there is no intersection among components.
Thus we have a natural map between free homotopy classes%
\footnote{A free homotopy class is the set of closed curves that are
related by a homotopy, {\it without} the condition that
the homotopy fixes a base point.
The set of free homotopy classes coincides with the
set of conjugacy classes of the fundamental group.}
of non-intersecting closed curves on $C_{g,n}$
and the charges of loop operators in the gauge theory $\cT_{g,n}$.

In particular the simple connected curve
corresponding to the $i^{\rm th}$ edge of the trivalent graph represents
the purely electric loop in the $i^\text{th}$
group---a Wilson loop.
It is also important for us to note that for any non-self-intersecting
curve, there is a choice of pants decomposition
such that it is homotopic to the boundary of a pair of pants.
Thus any loop operator in the gauge theory
corresponding to such a curve
is a Wilson loop in some duality frame.

The discussion above completely classifies the allowed set of charges carried by the
loop operators. Let us comment, though, that it does not specify fully what the operator is,
as a Wilson loop may be in the irreducible representation of that dimension
or a product of lower irreducible representations and similar 
refinements should apply also to 't~Hooft and dyonic loops. 
Indeed as we shall see below, in Liouville theory it is also possible to 
associate different loop operators to the same set of Dehn-Thurston data. 
The distinction between them is simply different choices of bases for 
loop operators. For Wilson loop operators, where the classification of 
loop operators in the gauge theory is complete, we have an exact 
map between different bases in the gauge theory and in Liouville.%
\footnote{There does not exist to date a complete 
classification of 't~Hooft and dyonic loop operators in these gauge theories 
which addresses these distinctions, so we can at best  
identify them by their charges.}

For the most part we will concentrate below on the basis that 
mirrors the discussion above and which is presented in the
next section. We will comment about  other
bases and their possible advantages in Section~\ref{sec:ell}.

\section{Holonomy and Monodromy in Liouville Theory}
\label{sec:pert}

In this section we introduce loop operators in classical Liouville theory
and relate them to the monodromy of a special class of operators in
Liouville theory, the so called degenerate operators. In the next section
we will explain how to calculate the action of the Liouville loop operators
on conformal blocks in quantum Liouville theory, and use these operators
to calculate the expectation value of loop operators in ${\cal N}=2$ gauge
theories.

The classical equations of motion of the Liouville field theory 
can be recast as the zero curvature representation of a connection%
\footnote{The Liouville equation of
motion on an arbitrary background metric $g$ determines the Weyl factor
$e^{2b\phi}$ which renders the curvature of the metric
$\hat{g}_{ab}=e^{2b\phi}g_{ab}$ constant and negative.}
\beq
{\cal F}=d{\cal A}+{\cal A}\wedge {\cal A}=0
\quad\Longrightarrow\quad
\text{Equations of motion}\,.
\eeq
The flatness of the connection on the Riemann surface arises from
consistency of the following system of linear equations, known as the Lax pair
\begin{equation}
\begin{aligned}
\label{linear}
\frac{\partial \Psi}{ \partial z}&+{\cal A}_z\Psi=0\,,\\
\frac{\partial \Psi}{\partial {\overline z}}&+{\cal A}_{\overline z}\Psi=0\,.
\end{aligned}
\end{equation}
This system of equations implies that $\Psi$ is covariantly constant
under parallel transport with respect to the Lax connection ${\cal A}$
on the Riemann surface $C_{g,n}$
\begin{equation}
\label{covconstant}
(d+{\cal A}) \Psi=0\,.
\end{equation}

The chiral quantum fields associated to the classical auxiliary fields
$\Psi$ represent building blocks in the construction 
of Liouville primary fields, and will play a key r\^ole in our
Liouville realization of loop operators in ${\cal N}=2$ gauge
theories with $SU(2)$ gauge groups.

For the Liouville field theory, the Lax connection ${\cal A}$ takes values
in the Lie algebra $SL(2,\bR)$. The Lax connection can be recast in
terms of the Liouville field $\phi$ through
\beq
{\cal A}= e T^{+}+ \overline{e} T^{-} +w T^{3}\,,
\eeq
where $e$ and $w$ are the vielbein and spin connection of the two
dimensional metric on the Riemann surface $C_{g,n}$ given by
$\hat{g}_{ab}=\exp(2b\phi)g_{ab}$, which has constant negative
curvature. $\{T^3,T^+,T^-\}$ are $SL(2,\bR)$ Lie algebra generators
in the $j=1/2$ representation.

The existence of an $SL(2,\bR)$ Lax connection in Liouville field theory
allows us to 
parallel transport the fields $\Psi$ between any two points
$x$ and $y$ on the Riemann surface $C_{g,n}$. Since the Lax
connection is flat, the holonomy depends only on the homotopy
class of the path connecting $x$ and $y$, and allows us to relate
fields at different points on $C_{g,n}$
\beq
\Psi(x)=U(x,y) \Psi(y)\,,
\qquad
U(x,y)=\hbox{P}\exp\int_x^y{\cal A}\,.
\eeq
Of particular importance is the holonomy obtained by integrating the
Lax connection around an arbitrary non-self-intersecting closed loop
$\gamma$ on the Riemann surface $C_{g,n}$
\beq\label{holodef}
U(\gamma)=\hbox{P}\exp\oint_\gamma {\cal A}\,.
\eeq
In particular, the holonomy around a closed curve encircling a tube in
the pants decomposition of $C_{g,n}$ is obtained by integrating
\beq
{\cal A}_x=\left(\begin{matrix}
b\hskip+0.5pt\partial_t\phi & \mu e^{b\phi} \\
\mu e^{b\phi} &-b\hskip 0.5pt \partial_t\phi
\end{matrix}\right)\,
\label{holoa}
\eeq
around the circle parametrized by $x$, where $(t,x)$ are
coordinates on the tube. Heuristically, the holonomy around the
tube measures the momentum of the state propagating across it.
The connection (\ref{holoa}) is gauge equivalent to 
\begin{equation}
{\cal A}_x\,=\,\left(\begin{matrix} 0 & -b^2T\\ 1 & 0\end{matrix}\right)\,,
\label{laxgauge}
\end{equation}
where $T$ is the Liouville stress tensor.
This form will be useful when relating the auxiliary fields $\Psi$ with degenerate fields.

The discussion has thus far been restricted to classical Liouville
theory, where $b\ll 1$ (or equivalently $b\gg 1$).
We are interested in performing the computation of the insertion
of a Wilson loop in the Liouville conformal field theory correlator
in the quantum regime. In particular we are interested in the case
$b=1$, when the Liouville correlator corresponds to the partition
function of the dual gauge theory on $S^4$, which is very quantum in Liouville.

It is not easy, but possible to define  
the Wilson loop in the quantum theory by means of a lattice
regularization \cite{Bytsko:2009mg}. An alternative approach is to
first quantize the chiral fields $\Psi$, and to build the Liouville field
out of these building blocks later \cite{Teschner:2001rv,Teschner:2003em}.
The loop operator associated to the loop $\gamma$ can then be defined
in terms of the monodromy of chiral fields $\Psi$. We introduce 
it here in the quantum
theory (and define rigorously in the following section) and explain why it
reduces to the holonomy in the classical limit.

We recall that the fields $V_\alpha$ in Liouville giving rise to delta function
normalizable states are labeled by a quantum number
$\alpha\in Q/2+ i\bR$. Degenerate operators/fields, on the other hand, are
labeled by a pair of integers $(r,s)\in \bZ_{>0}\times \bZ_{>0}$. Their corresponding quantum numbers are given by
\beq
2\alpha_{r,s}=Q-\frac{r}{ b} -s b\,.
\label{momentadegenerate}
\eeq
In this paper, we denote the {\it chiral degenerate fields} 
by $V_{r,s}$. The  complete, conformal primary degenerate 
field is obtained by combining $V_{r,s}$ with its anti-holomorphic 
counterpart, but in this paper
only the {\it chiral degenerate field}   $V_{r,s}$ is relevant. 
The defining property of a degenerate field is that it contains a
Virasoro descendant that is a null state (a state of zero norm).
Decoupling of null states imposes strong constraints on the
correlation functions of the theory. A chiral degenerate field $V_{r,s}$ contains a null
state at Virasoro level $rs$, and decoupling of such a state is
captured by the action of a differential operator of order $rs$ on
the correlation functions of the theory containing the chiral 
degenerate field $V_{r,s}$. 
The differential operators constraining the correlation
functions are of Fuchsian type and, therefore, their solutions 
are not single valued, rather they acquire monodromies 
when encircling the singularities in
the differential equation.

The monodromies of the chiral degenerate fields 
correspond precisely to the 
holonomy acquired by transporting the Lax connection around
a closed curve $\gamma$, 
\beq\label{holo}
\Psi(x+2\pi)=U(\gamma)\cdot \Psi(x)\,,
\eeq
where $x\in C_{g,n}$ is a  base point of the 
curve $\gamma$ in $C_{g,n}$. This can be verified 
by noting that the 
null decoupling differential equation
\beq
\left(\partial^2+b^2T(z)\right)V_{1,2}=0\,,
\label{null}
\eeq
is precisely the equation for parallel
transport with respect to the Lax connection in the spin $1/2$
representation (\ref{laxgauge}).

The relation between the
differential equation associated to the  $V_{1,2j+1}$ chiral degenerate
field and the parallel transport equation extends to arbitrary
spin $j$.\footnote{As another example of this correspondence, the
degenerate field $V_{1,3}$ satisfies the semiclassical null
decoupling equation $(\partial^3+4 b^2 T+2b^2\partial T)V_{1,3}=0$,
which translates into flatness of the Lax connection in the $j=1$
representation.}
The Lax connection has a counterpart in the spin $j$ representation of
$SL(2,\bR)$, and the corresponding system
of $2j+1$ first order linear differential equations (\ref{linear})
for $2j+1$ fields $\Psi_{j,m}$ can be
recast in terms of a single differential equation of order $2j+1$. This
differential equation precisely coincides with the null decoupling
equation for the $V_{1,2j+1}$ chiral degenerate field in the semiclassical
approximation. This identifies the  chiral  degenerate field $V_{1,2j+1}$ 
as the quantization of the auxiliary fields $\Psi_{j,m}$.%
\footnote{If, in particular, we are considering the complete, conformal primary 
 degenerate field with label $(1,2j+1)$, it turns out to be possible to reconstruct 
this  degenerate field  from the $2j+1$ linearly independent solutions
$\Psi_{j,m}(z)$, $m=-j,\dots,j$ of the null vector decoupling equations and 
their anti-holomorphic counterparts $\overline{\Psi}_{j,m}(\bar z)$.}
 This establishes the correspondence
\beq
\hbox{Tr}_j U(\gamma)
\quad\Longleftrightarrow\quad
\text{Monodromy of  $V_{1,2j+1}$ degenerate field}\,,
\eeq
where $\hbox{Tr}_j U(\gamma)$ is a Wilson loop in the spin $j$
representation of $SL(2,\bR)$.

While we perform our calculations for arbitrary values of $b$, we eventually focus
our analysis to $b=1$, for which a gauge theory interpretation of the Liouville
correlators is available.
Inspection of (\ref{momentadegenerate}) shows
that there is a large redundancy in the labeling of degenerate fields
for $b=1$, as they depend actually only on the sum $r+s$. Without loss of
generality, we have chosen a basis of chiral degenerate fields spanned 
by $V_{1,2j+1}$.

There are different types of Wilson loops that can be inserted into
the Liouville conformal field theory correlator, depending upon
whether the Wilson loop along the curve $\gamma$ is in the spin
$j$ representation, multiply wound or multi-traced. The
corresponding statement in terms of degenerate fields is
\begin{equation}
\begin{aligned}
\label{map}
\Tr_j U(\gamma)
&\quad\Longleftrightarrow\quad&&
\text{Monodromy of $V_{1,2j+1}$ along $\gamma$}\\
{\rm Tr}[ U(\gamma)^k]
&\quad\Longleftrightarrow\quad&&
\text{Monodromy of $V_{1,2}$ along $k\cdot \gamma$}\\
[\Tr U(\gamma)]^k
&\quad\Longleftrightarrow\quad&&
\text{$k$ times the Monodromy of $V_{1,2}$ along $\gamma$}
\end{aligned}
\end{equation}

One point that is worth mentioning is that based on the analysis of
\cite{Drukker:2009tz} we focus our discussion on
non-intersecting curves,%
\footnote{Note though that when resolved, the multi-wound
loop $\Tr U^k$ has intersections while $(\Tr U)^k$ does not.}
even though there is no impediment to considering the classical 
holonomy or a monodromy
along curves with intersections. There are two justifications for that. First
based on the comparison with the gauge theory in \cite{Drukker:2009tz}, we
expect non-intersecting loops to provide a complete basis for loop operators.
Moreover intersecting curves lead to more severe divergences than
smooth curves and require more care to regularize in the quantum theory.

We expect the resolution of these singularities to invoke mixing between
different degenerate fields, which we comment on in Section~\ref{sec:ell}.
We also address the issue of intersecting curves in Section~\ref{sec:algebra},
where we indeed show that they can be represented
as linear combinations of non-intersecting ones.

In the next section 
we make the discussion of the monodromy more precise
and provide the explicit connection to the gauge theory loop operators.

\section{Loop Operators in Gauge Theories from Liouville Theory}
\label{sec:bootstrap}

In this section we give an exact prescription for calculating the
monodromy associated to transporting a degenerate field  along a
simple connected curve $\gamma$ on
the Riemann surface by introducing the notion of a quantum loop
operator $\CL(\gamma)$ associated with the closed curve.
In a duality frame where $\gamma$ represents a Wilson loop,
the definition we present exactly reproduces the result of
Pestun \eqn{pestun} and for all other loops it provides a
prediction for the gauge theory observable. In the next section
we follow this procedure in some simple examples.

\subsection{Loop Operators in Liouville Theory}
\label{sec:basic}

Our goal is to study the effect of inserting the monodromy of a degenerate field
into the Liouville conformal field theory correlator
$\vev{\prod_{a=1}^n V_{m_a}}_{C_{g,n}}$. We perform the calculations
using the conformal bootstrap approach as reviewed in Section~\ref{sec:review}.
We first define the calculational procedure somewhat abstractly and then
spell it out in explicit detail.

In order to calculate the monodromy of a degenerate field, we must first think
about enriching a conformal block with extra punctures corresponding to
the degenerate fields in question.
Consider the simplest situation where two degenerate fields of type
$V_{1,2}$ are used%
\footnote{It is clear how the construction generalizes. One can add
pairs of higher degenerate fields $V_{r,s}$.}
(the reason why two fields rather than one will become apparent shortly).
We denote the space of conformal blocks on $C_{g,n}$ by ${\cal C}(C_{g,n})$
and those on $C_{g,n+2}$ when two extra punctures are of type $V_{1,2}$
by $\hat{\cal C}(C_{g,n+2})$,

An important property of degenerate fields is that their OPE with a Liouville
primary operator truncates
\beq
[V_{1,2}]\cdot [V_{\alpha}]= [V_{\alpha+b/2}]+[V_{\alpha-b/2}]\,.
\label{opedegen}
\eeq
This equation implies that
$\hat{\cal C}(C_{g,n+2})$ is isomorphic to ${\cal C}(C_{g,n})\otimes \bC^4$
as a vector space. The key observation to be made is that
there is an embedding
\begin{equation}\label{embed}
\imath_{g,n}:\CC(C_{g,n})\hookrightarrow \hat\CC(C_{g,n+2})\,,
\end{equation}
coming from the fact that the fusion of the two degenerate fields
$V_{1,2}$ contains the vacuum representation.

This can be seen more explicitly as follows. Consider a pants
decomposition $\hat\sigma$ of the Riemann surface $C_{g,n+2}$ in
which both representations $V_{1,2}$ are in the same pair of pants.
The OPE of these two degenerate fields contains only the identity
state and the $V_{1,3}$ state. The conformal block
$\hat\cF^{(\hat\sigma)}_{\alpha,E}$ associated to this pants decomposition
vanishes unless the edge connecting this pair of pants to the
rest of the surface carries the representation $V_0$ or $V_{1,3}$.
The subspace spanned by the elements with the identity $V_0$ as the
intermediate state is isomorphic to $\hat\CC(C_{g,n+1})$, with the $(n+1)^{\rm st}$
boundary component being assigned the vacuum representation
$V_0$. This space is canonically isomorphic to $\CC(C_{g,n})$.

It follows from the existence of the embedding \rf{embed}
that the mapping class group action
on $\hat\CC(C_{g,n+2})$ can be projected onto $\CC(C_{g,n})$.
The mapping class group ${\rm MCG}(C_{g,n+2})$ contains in particular
the monodromies generated by moving the insertion point of one of the
representations $V_{1,2}$
along a connected simple closed curve $\ga$ on $C_{g,n}$. The projection
of the action of these elements on $\hat\CC(C_{g,n+2})$ down to
$\CC(C_{g,n})$ defines operators on
$\CC(C_{g,n})$. These are the {\it Liouville loop operators} $\CL(\gamma)$
we are interested in.

A key property of the Liouville loop operators, which is an immediate
consequence of this definition, is their modular invariance. The
description above does not depend on a pants decomposition of the
Riemann surface $C_{g,n}$, it only requires to separate the two extra
degenerate fields on a pair of pants to get the canonical identification
between $\hat\CC(C_{g,n+2})$ and $\CC(C_{g,n})$. Since Liouville theory
is modular invariant \cite{Teschner:2001rv,Teschner:2003en,Teschner:2008qh},
so are the loop operators. In the explicit
calculations below we will employ specific pants decompositions
for the entire surface and the details of the calculation will depend
on it, but by this argument, the final result is modular invariant.

We turn now to presenting a more concrete form of the loop operators 
in terms of their action on the conformal blocks. Since we chose for the 
construction the chiral degenerate operators, the monodromy acts 
non-trivially only on the holomorphic conformal block 
\eqn{BPZblock} while it acts trivially on the antiholomorphic one%
\footnote{One could equally well choose to act on the antiholomorphic 
blocks. They get acted on by the adjoint of the operators 
$\cD_\alpha^{(\sigma)}$ defined in \eqn{CLeigenval}. The distinction 
completely disappears when the discussing length operators in 
quantum Teichm\"uller theory \eqn{qLq}, see Section~\ref{liou-teich}.}
\begin{equation}
\begin{aligned}
{\cal F}^{(\sigma)}_{\alpha,E}&\longrightarrow 
[{\cal L(\gamma)}\cdot {\cal F}^{(\sigma)}]_{\alpha,E}\,,\\
{\overline {\cal F}^{(\sigma)}_{\alpha,E}}&\longrightarrow
{\overline {\cal F}^{(\sigma)}_{\alpha,E}}\,.
\label{actionmon}
\end{aligned}
\end{equation}

Once we have an explicit expression for $\cL(\gamma)$, the
Liouville correlation function \eqn{correla-F} 
gets modified by the insertion of Liouville loop operator to
\begin{equation}
\label{liouloop}
\big\langle\CL(\gamma)\big\rangle_{C_{g,n}}^{}
\,=\,\int d\nu(\alpha)\,\overline{{\cal F}}^{(\sigma)}_{\alpha,E}\,
\left(\CL(\gamma)\cdot {\cal F}^{(\sigma)}_{\alpha,E}\right)\,.
\end{equation}

As we discussed in Section~\ref{sec:pert}, it is also possible
to consider the monodromy of a higher degenerate field
$V_{1,2j+1}$.
Clearly it is possible to defined a more general Liouville loop operator
by replacing the lowest degenerate field $V_{1,2}$ by $V_{1,2j+1}$
in the above definition of $\cL(\gamma)$.
We will denote the resulting loop operator by $\cL_{j}(\gamma)$.
In the special case $j=1/2$, we obtain $\cL(\gamma)\equiv \cL_{1/2}(\gamma)$
back. 
We will comment on different bases for loop operators 
which utilize $V_{1,2}$ or more general $V_{1,2j+1}$ fields in 
Section~\ref{sec:calculations} and in Section~\ref{sec:algebra}. 
For the most part, though, we will focus on the basis mirroring 
the discussion in Section~\ref{sec:geodesics}, transporting $V_{1,2}$ 
along non-intersecting curves on the surface. This also corresponds 
to the last line in \eqn{map},

\subsection{Calculational Scheme}
\label{sec:scheme}

In order to explicitly calculate the action of the Liouville loop operator
${\cal L(\gamma)}$ on the conformal blocks ${\cal F}^{(\sigma)}_{\alpha,E}$
we follow the following algorithm
that was originally used to derive the Verlinde formula
\cite{Verlinde:1988sn}:

\begin{enumerate}
\item
Start by inserting the identity operator at a point in the trivalent graph which
is spanned by the curve $\gamma$. 
The corresponding conformal blocks transform under
changes of the marking exactly as ${\cal {F}}^{(\sigma)}_{\alpha,E}$. 

\item
We represent the identity operator as the projection to the identity of the
fusion of two degenerate fields. This operation adds two external edges
to the original trivalent graph $\Gamma_\sigma$, associated
to these degenerate fields, and yields a new
trivalent graph $\Gamma_{\hat \sigma}$. We denote the conformal
block corresponding to this enriched trivalent graph by
$\hat{\cal {F}}^{(\hat\sigma)}_{\alpha,E}$.

\item
We transport one of the degenerate fields around the trivalent graph
$\Gamma_{\sigma}$ to span the curve $\gamma$, and return
it to its original position. Each intermediate step involves a different
trivalent graph $\Gamma_{\hat\sigma'}$, where the degenerate field
is connected to different edges in $\Gamma_\sigma$.

\item
We fuse again the two degenerate fields to produce the identity operator.
\end{enumerate}

Following the strategy we have described provides a concrete computational
framework in which to calculate
$ \big\langle\CL({\ga_{d}})\big\rangle_{C_{g,n}}^{}$. In order to evaluate
the monodromy (specifically step 3) we must find how conformal blocks
based on inequivalent trivalent graphs are related to each other.
Trivalent graphs associated with a given Riemann surface ${C_{g,n}}$
can be transformed into each other by the action of the so-called
Moore-Seiberg groupoid \cite{Moore:1988qv} (see \cite{Moore:1989vd} for a review). The Moore-Seiberg groupoid is generated
by three basic moves, which act on the space of trivalent graphs. They are

\begin{itemize}
\item
Fusion move
\item
Braiding move
\item
S-move
\end{itemize}

It is an important result in Liouville theory that the
generators of the Moore-Seiberg groupoid act on the space of conformal
blocks \cite{Teschner:2001rv,Teschner:2003en,Teschner:2008qh}. 
Once the action of the generators on the conformal blocks is
known, the monodromy action on the conformal block
$[{\cal L(\gamma)}\cdot {\cal F}^{(\sigma)}]_{\alpha,E}$
can be obtained by concatenating the moves that are required to create
the pair of degenerate fields, move them around the trivalent graph and
annihilate them again.

In Section~\ref{sec:calculations} we write down the expressions for the fusion
and braiding moves on the Liouville conformal blocks and use them to
calculate the expectation value of various loop operators.

The general fusion matrices are quite complicated, but
they simplify significantly when one of the legs is a degenerate
operator. Due to the OPE property of the $V_{1,2}$ degenerate field
\eqn{opedegen}, the basis change in each step in the calculation
involves a sum over two terms, of conformal blocks with representations
shifted by $\pm b/2$.

At the end of the calculation, after the degenerate fields are projected out
we are back at the original basis for $\CC(C_{g,n})$ being generated by the
conformal blocks ${\cal F}^{(\sigma)}_{\alpha,E}$, but the representations
$\alpha$ carried by
the edges in the graph $\Gamma_\sigma$ have been shifted by integer
multiples of $-b/2$. We conclude that the operator $\cL(\gamma)$
can be represented as a difference operator
$\CD_\alpha^{(\si)}(\gamma)$ which acts by shifting the indices $\alpha$
on ${\cal F}^{(\sigma)}_{\alpha,E}$ and by multiplying the conformal blocks
with $\alpha$ and $E$-dependent factors:
\begin{equation}
\label{CLeigenval}
[{\cal L(\gamma)}\cdot {\cal F}^{({\sigma})}]_{\alpha,E}
=\,\CD_\alpha^{(\si)}(\gamma){\cal F}^{({\sigma})}_{\alpha,E}\,.
\end{equation}

Within this framework one may demonstrate the
modular invariance of the Liouville loop operators
more explicitly as follows.
The operators $\cD_{\alpha}^{(\sigma)}(\gamma)$
were defined as projections of
an operator $M_{\hat{\sigma}}(\gamma)$ on $\hat{\cal C}(C_{g,n+2})$,
which represents the
element in ${\rm MCG}(C_{g,n+2})$ corresponding to the
monodromy associated to transporting a degenerate field along $\gamma$.
We are using the notation $\hat\sigma$ for a pants decomposition
(more precisely a marking, see footnote 8) on $C_{g,n+2}$
which becomes a pants decomposition on $C_{g,n}$ by splitting off the pair of
pants containing the insertions of the two degenerate fields.
Choosing another pants decomposition $\sigma'$ would yield an operator
$D_{\alpha}^{(\sigma')}(\gamma)$. The conformal blocks in
$\hat{\cal C}(C_{g,n+2})$
associated to $\hat\sigma$ and $\hat\sigma'$ are related by an operator
$M_{\hat\sigma'\hat\sigma}$. The fact that the operators
$M_{\hat\sigma'\hat\sigma}$
represent the modular groupoid implies in particular the relation
\begin{equation}
M_{\hat\sigma'\hat\sigma}\cdot M_{\hat\sigma}(\gamma)\,=\,
M_{\hat\sigma'}(\gamma)\cdot M_{\hat\sigma'\hat\sigma}\,.
\end{equation}
Projecting this relation to ${\cal C}(C_{g,n})$ yields
\begin{equation}
\label{Msigsig}
M_{\sigma'\sigma}\cdot \CD_{\alpha}^{(\sigma)}(\gamma)\,=\,
\CD_{\alpha}^{(\sigma')}(\gamma)\cdot M_{\sigma'\sigma}\,,
\end{equation}
where now $M_{\sigma'\sigma}$ relates the conformal blocks in
${\cal C}(C_{g,n})$ associated to
$\sigma $ and $\sigma'$. This shows that the operators
$\CD_{\alpha}^{(\sigma)}(\gamma)$
are mapped to each other by the change of representation
induced by change of pants decomposition. They therefore represent
an operator ${\cal L}(\gamma)$ that is independent of the choice of
pants decomposition on the Riemann surface.

As mentioned before, it is only the concrete realization of the operator
as a difference operator acting on $\alpha$ that may change when
changing the pants decomposition $\si$. The modular invariance of
Liouville correlation functions
\cite{Teschner:2001rv,Teschner:2003en,Teschner:2008qh} therefore
implies the modular invariance of the Liouville loop operator expectation
values. The fact that the expectation values
$\langle\CL({\ga})\rangle_{C_{g,n}}^{}$ defined in \rf{liouloop}
are modular invariant allows us to regard the $\CL(\ga)$ as natural
observables in Liouville theory.

\subsection{Relation to the Gauge Theory}
\label{sec:gauge-liouville}

As reviewed in Section~\ref{sec:review}, loop operators in
${\cal T}_{g,n}$ are in one-to-one correspondence with
non-self-intersecting geodesics in $C_{g,n}$. Our proposal
is that the expectation value of a loop operator $L_d$ in the
${\cal N}=2$ theory ${\cal T}_{g,n}$ is captured by inserting
into the Liouville correlator the loop operator $\CL(\gamma_d)$
\begin{equation}\label{main}
\vev{L_{d,\sigma}}_{{\cal T}_{g,n}}\,=\,
\big\langle\CL({\ga_{d,\si}})\big\rangle_{C_{g,n}}^{}
\qquad
\end{equation}

We note that for all the theories ${\cal T}_{g,n}$, any loop
operator $L_d$ can always be transformed into a duality frame
$\sigma$ where the operator is purely electric, and
corresponds to a Wilson loop in the four dimensional gauge
theory. In this case the corresponding
conformal block ${\cal F}^{({\sigma})}_{\alpha,E}$ is an eigenstate
of the loop operator ${\cal L(\gamma)}$
\begin{equation}
[{\cal L(\gamma)}\cdot {\cal F}^{({\sigma})}]_{\alpha,E}
=\lambda(\alpha) {\cal F}^{({\sigma})}_{\alpha,E}\,.
\label{}
\end{equation}
By explicit calculation, when the monodromy is that of
a $V_{1,2}$ degenerate field around the curve $\gamma$ the action is \eqn{w01}
\begin{equation}
[{\cal L(\gamma)}\cdot {\cal F}^{({\sigma})}]_{\alpha,E}=
\frac{\cos\left(\pi b(2\alpha-Q)\right)}{\cos\left(\pi b Q\right)}
{\cal F}^{({\sigma})}_{\alpha,E}\,.
\label{wilsonj=1/2}
\end{equation}

Using our proposal (\ref{main}) we find that the Liouville correlator
$\big\langle\CL({\ga_{d,\si}})\big\rangle_{C_{g,n}}^{}$ \eqn{liouloop}
in the limit $b=1$ exactly reproduces Pestun's formula (\ref{pestun}) for
the expectation value of a Wilson  loop  in the corresponding ${\cal N}=2$
theory.

Gauge theory operators carrying magnetic charge in a duality
frame $\sigma$ are captured by the monodromy acquired by the conformal 
blocks when a degenerate field is transported around a curve $\gamma$,
homotopic to a path in $\Gamma_\sigma$.
This implies that the quantum numbers of the edges traversed by the
degenerate field are shifted due to their OPE with the degenerate
field which is being transported. Therefore, the conformal blocks
are not eigenstates of Liouville loop operators exploring the
trivalent graph. Nevertheless, their action on the conformal blocks is
rather simple, and involves elementary shifts of the quantum numbers
labeling the conformal block. These shifted conformal blocks capture
't~Hooft and dyonic operators in the gauge theory.

We also note that our proposal, together with the modular
invariance of the Liouville loop operator correlators, implies the
conjectured action of the S-duality duality group $\Gamma({\cal T}_{g,n})$
on gauge theory loop operators, which states that
\begin{equation}
\vev{L_{d,\sigma}}_{{\cal T}_{g,n}}=\vev{L_{d',\sigma'}}_{{\cal T}_{g,n}}\,,
\end{equation}
where $d$ and $d'$ are the charges of the loop operator in the
${\cal T}_{g,n}$ duality frames $\sigma$ and $\sigma'$ respectively.

\section{Computation of Gauge Theory Loop Operators}
\label{sec:calculations}

In this section we explicitly use the identification of loop operators 
in Liouville theory on $C_{g,n}$ with loop operators in ${\cal T}_{g,n}$ 
to calculate the {\it exact} expectation value of Wilson loop operators 
in ${\cal T}_{g,n}$ and the expectation value of 't~Hooft operators in 
two interesting examples: the ${\cal N}=2$ $SU(2)$ theory with $N_F=4$ and 
the ${\cal N}=2^*$ theory. We then proceed to define operators on a pair of 
pants, which by composing them, computes the expectation value 
of complicated loop operator in any of the ${\cal T}_{g,n}$ theories.

\subsection{Preliminaries}
\label{sec:prelim}

As explained in the previous section, the calculation of the 
monodromy of a degenerate field as it moves 
around a closed curve in $C_{g,n}$ can be broken into a set 
of moves on the trivalent graphs on which the conformal 
blocks are defined. It is a non-trivial fact of two dimensional 
conformal field theory that any allowed move can be obtained 
by composing the basic moves: fusion, braiding and S-move. 
In this subsection we introduce the key ingredients and formulas 
behind the basic moves we need. 

The moves are local on the trivalent graph. They involve at most 
five consecutive edges. On such a graph we can define the 
$s$-channel and $t$-channel 4-pt BPZ conformal blocks (\ref{BPZblock}) 
\beq
\CF_{\al}^{(s)}\big[\begin{smallmatrix} \al_3 & \al_2\\ \al_4 & \al_1\end{smallmatrix}\big]
=\raisebox{0mm}[10mm][12mm]{\parbox{26mm}{\begin{center}
\begin{fmfgraph*}(20,15)
\fmfbottom{bl,br}
\fmftop{tl,tr}
\fmf{plain,width=2}{bl,b1,b2,b3,br}
\fmf{phantom}{tl,t1,t2,t3,tr}
\fmffreeze
\fmf{plain,width=2}{t1,b1}
\fmf{plain,width=2}{t3,b3}
\fmfv{label=$\alpha_4$,label.angle=-90,label.dist=8}{bl}
\fmfv{label=$\alpha_3$,label.angle=180}{t1}
\fmfv{label=$\alpha_2$,label.angle=0}{t3}
\fmfv{label=$\alpha$,label.angle=-90,label.dist=8}{b2}
\fmfv{label=$\alpha_1$,label.angle=-90,label.dist=8}{br}
\end{fmfgraph*}\end{center}}}
\eeq

\beq
\CF_{\al}^{(t)}\big[\begin{smallmatrix} \al_3 & \al_2\\ \al_4 & \al_1\end{smallmatrix}\big]
=
\raisebox{0mm}[10mm][12mm]{\parbox{24mm}{\begin{center}
\begin{fmfgraph*}(15,15)
\fmfbottom{bl,br}
\fmftop{tl,tr}
\fmf{plain,width=2}{bl,b1,b2,b3,br}
\fmf{phantom}{tl,t1,tt1,t2,tt3,t3,tr}
\fmffreeze
\fmf{plain,width=2}{t1,c2,t3}
\fmf{plain,width=2,tension=2}{c2,b2}
\fmfv{label=$\alpha_4$,label.angle=-90,label.dist=8}{bl}
\fmfv{label=$\alpha_1$,label.angle=-90,label.dist=8}{br}
\fmfv{label=$\alpha$,label.angle=-30}{c2}
\fmfv{label=$\alpha_3$,label.angle=180}{t1}
\fmfv{label=$\alpha_2$,label.angle=0}{t3}
\end{fmfgraph*}\end{center}}}
\eeq

In the explicit calculations in this section we employ for calculational 
convenience a different set of conformal blocks $\cG^{(\sigma)}_{\alpha,E}$ 
which absorb most of the Liouville three-point function 
\cite{Dorn:1994xn,Zamolodchikov:1995aa, Teschner:2003en} 
into the blocks themselves.%
\footnote{These are also slightly different from the conventions of
\cite{Alday:2009aq}.}

In terms of these conformal blocks, the Liouville correlation function 
\eqn{correla-F} is written as
\beq
\Big<\prod_{a=1}^n V_{m_a}\Big>_{C_{g,n}}
=\int d\mu(\alpha)\,
{\overline {\cal G}^{(\sigma)}_{\alpha,E}}\,{\cal G}^{(\sigma)}_{\alpha,E}\,,
\label{correla}
\eeq
where the measure is
\beq
d\mu (\alpha) =\prod_{i=1}^{3g-3+n} d\alpha_i
\left( 4 \sin\left(\pi b\left(2\alpha_i-Q\right)\right)
\sin\left(\frac{\pi}{ b}\left(Q-2\alpha_i\right)\right)\right)
\label{measure}
\eeq
The precise relation between the two involves multiplying the $\cF$ blocks 
by one factor of $N(\al_3,\al_2,\al_1)$ for each vertex in the graphs 
$\Gamma_\sigma$ where \cite{Ponsot:1999uf}
\begin{equation}
\label{Nexp}
N(\al_1 ,\al_2,\al_3)= \frac{\Ga_b(2Q-\al_1-\al_2-\al_3)
\Ga_b(\al_1+\al_2-\al_3)\Ga_b(\al_1+\al_3-\al_2)\Ga_b(\al_2+\al_3-\al_1)}
{\Ga_b(2Q-2\al_1)\Ga_b(2\al_2)\Ga_b(2\al_3)}\,,
\end{equation}
and $\Ga_b(x)$ is the Barnes double Gamma function (see 
Appendix~\ref{spefunct} for the definition and a collection of 
properties of this function  and the closely related 
$S_b(x)$ and $\Upsilon(s)$). The function $N(\al_1,\al_2,\al_3)$ 
is related to the Liouville three-point function $C(\al_1,\al_2,\al_3)$ 
\cite{Dorn:1994xn,Zamolodchikov:1995aa} through \cite{Ponsot:1999uf}
\beq
C(\al_1,\al_2,\al_3)=
\left(\pi\mu\frac{\Gamma(b^2)}{\Gamma(1-b^2)}b^{2-2b^2}\right)
^{\frac{1}{b}(Q-\alpha_1-\alpha_2-\alpha_3)}
\Upsilon'(0)|N(\al_1,\al_2,\al_3)|^2
\prod_{i=1}^3\frac{\Ga_b(2Q-2\al_i)}{ \Ga_b(Q-2\al_i)}\,.
\label{DOZZ}
\eeq
In the product of all the three-point functions for the different vertices, 
the only terms which depend on the internal labels $\alpha$ are the 
$N(\alpha_1,\alpha_2,\alpha_3)$ factors and some of the ratios of 
Barnes functions, which give the measure \eqn{measure}. 
All the rest of the terms can be absorbed in the normalizations 
of the external legs. 
Note also that for normalizable states, when $\bar\alpha_1=Q-\alpha_1$, 
then $|N(\alpha_1,\alpha_2,\alpha_3)|^2=|N(Q-\alpha_1,\alpha_2,\alpha_3)|^2$.

In the case of the four-point s-channel and t-channel conformal blocks above 
the explicit mapping is
\beq
\begin{aligned}
\label{def-cG}
\cG_\alpha^{(s)}\big[\begin{smallmatrix} \al_3 & \al_2\\ \al_4 & \al_1\end{smallmatrix}\big]
&=N(\al_4,\al_3,\al)N(\al,\al_2,\al_1)
\cF_\alpha^{(s)}\big[\begin{smallmatrix} \al_3 & \al_2\\ \al_4 & \al_1\end{smallmatrix}\big]\\
\cG_\alpha^{(t)}\big[\begin{smallmatrix} \al_3 & \al_2\\ \al_4 & \al_1\end{smallmatrix}\big]
&=N(\al_4,\alpha,\al_1)N(\al,\al_3,\al_2)
\cF_\alpha^{(t)}\big[\begin{smallmatrix} \al_3 & \al_2\\ \al_4 & \al_1\end{smallmatrix}\big]\,,
\end{aligned}
\eeq

A {\it fusion} move is a transformation that relates the $s$-channel 
conformal blocks to the $t$-channel conformal blocks. They are 
related by a change of basis. The moves on the BPZ conformal 
blocks for Liouville theory are
\beq
\label{genfus}
\raisebox{0mm}[10mm][12mm]{\parbox{25mm}{\begin{center}
\begin{fmfgraph*}(20,15)
\fmfbottom{bl,br}
\fmftop{tl,tr}
\fmf{plain,width=2}{bl,b1,b2,b3,br}
\fmf{phantom}{tl,t1,t2,t3,tr}
\fmffreeze
\fmf{plain,width=2}{t1,b1}
\fmf{plain,width=2}{t3,b3}
\fmfv{label=$\alpha_4$,label.angle=-90,label.dist=8}{bl}
\fmfv{label=$\alpha_1$,label.angle=-90,label.dist=8}{br}
\fmfv{label=$\alpha$,label.angle=-90,label.dist=8}{b2}
\fmfv{label=$\alpha_3$,label.angle=180}{t1}
\fmfv{label=$\alpha_2$,label.angle=0}{t3}
\end{fmfgraph*}\end{center}}}
\to
\raisebox{0mm}[10mm][12mm]{\parbox{24mm}{\begin{center}
\begin{fmfgraph*}(15,15)
\fmfbottom{bl,br}
\fmftop{tl,tr}
\fmf{plain,width=2}{bl,b1,b2,b3,br}
\fmf{phantom}{tl,t1,tt1,t2,tt3,t3,tr}
\fmffreeze
\fmf{plain,width=2}{t1,c2,t3}
\fmf{plain,width=2,tension=2}{c2,b2}
\fmfv{label=$\alpha_4$,label.angle=-90,label.dist=8}{bl}
\fmfv{label=$\alpha_1$,label.angle=-90,label.dist=8}{br}
\fmfv{label=$\alpha'$,label.angle=-30}{c2}
\fmfv{label=$\alpha_3$,label.angle=180}{t1}
\fmfv{label=$\alpha_2$,label.angle=0}{t3}
\end{fmfgraph*}\end{center}}}
:=\quad
\CF_{\al}^{(s)}\big[\begin{smallmatrix} \al_3 & \al_2\\ \al_4 & \al_1\end{smallmatrix}\big]
=\int d\al'\;
F_{\al\al'}^{}\big[\begin{smallmatrix} \al_3 & \al_2\\ \al_4 & \al_1\end{smallmatrix}\big]
\CF_{\al'}^{(t)}\big[\begin{smallmatrix} \al_3 & \al_2\\ \al_4 & \al_1\end{smallmatrix}\big]\,,
\eeq
\beq
\raisebox{0mm}[10mm][12mm]{\parbox{24mm}{\begin{center}
\begin{fmfgraph*}(15,15)
\fmfbottom{bl,br}
\fmftop{tl,tr}
\fmf{plain,width=2}{bl,b1,b2,b3,br}
\fmf{phantom}{tl,t1,tt1,t2,tt3,t3,tr}
\fmffreeze
\fmf{plain,width=2}{t1,c2,t3}
\fmf{plain,width=2,tension=2}{c2,b2}
\fmfv{label=$\alpha_4$,label.angle=-90,label.dist=8}{bl}
\fmfv{label=$\alpha_1$,label.angle=-90,label.dist=8}{br}
\fmfv{label=$\alpha$,label.angle=-30}{c2}
\fmfv{label=$\alpha_3$,label.angle=180}{t1}
\fmfv{label=$\alpha_2$,label.angle=0}{t3}
\end{fmfgraph*}\end{center}}}
\to
\raisebox{0mm}[10mm][12mm]{\parbox{25mm}{\begin{center}
\begin{fmfgraph*}(20,15)
\fmfbottom{bl,br}
\fmftop{tl,tr}
\fmf{plain,width=2}{bl,b1,b2,b3,br}
\fmf{phantom}{tl,t1,t2,t3,tr}
\fmffreeze
\fmf{plain,width=2}{t1,b1}
\fmf{plain,width=2}{t3,b3}
\fmfv{label=$\alpha_4$,label.angle=-90,label.dist=8}{bl}
\fmfv{label=$\alpha_1$,label.angle=-90,label.dist=8}{br}
\fmfv{label=$\alpha'$,label.angle=-90,label.dist=5}{b2}
\fmfv{label=$\alpha_3$,label.angle=180}{t1}
\fmfv{label=$\alpha_2$,label.angle=0}{t3}
\end{fmfgraph*}\end{center}}}
:=\quad
\CF_{\al}^{(t)}\big[\begin{smallmatrix} \al_3 & \al_2\\ \al_4 & \al_1\end{smallmatrix}\big]
=\int d\al'\;
F^{-1}_{\al\al'}\big[\begin{smallmatrix} \al_3 & \al_2\\ \al_4 & \al_1\end{smallmatrix}\big]
\CF_{\al'}^{(s)}\big[\begin{smallmatrix} \al_3 & \al_2\\ \al_4 & \al_1\end{smallmatrix}\big]\,,
\eeq
where 
$F_{\al\al'}\big[\begin{smallmatrix} \al_1 & \al_2\\ \al_4 & \al_3\end{smallmatrix}\big]$ 
are called the fusion matrices.
The forward and inverse transformation are simply related by
\beq
F^{-1}_{\al\al'}\big[\begin{smallmatrix} \al_3 & \al_2\\ \al_4 & \al_1\end{smallmatrix}\big]
=F_{\al\al'}\big[\begin{smallmatrix} \al_1 & \al_2\\ \al_4 & \al_3\end{smallmatrix}\big]\,.
\eeq

The fusion matrices for the $\cG$ blocks (\ref{def-cG}) are therefore
\begin{equation}\label{G-F}
G_{\al\alpha'}^{}\big[\begin{smallmatrix} \al_3 & \al_2\\ \al_4 & \al_1\end{smallmatrix}\big]=
\frac{N(\al_4,\al_3,\al)N(\al,\al_2,\al_1)}
{N(\al_4,\al',\al_1)N(\al',\al_3,\al_2)}
F_{\al\alpha'}^{}\big[\begin{smallmatrix} \al_3 & \al_2\\ \al_4 & \al_1\end{smallmatrix}\big]\,
\end{equation}
and satisfy the same symmetry properties as the fusion matrices for the $\cF$ blocks 
\beq
G_{\al\al'}\big[\begin{smallmatrix} \al_3 & \al_2\\ \al_4 & \al_1\end{smallmatrix}\big]
=G_{\al\al'}\big[\begin{smallmatrix} \al_2 & \al_3\\ \al_1 & \al_4\end{smallmatrix}\big]
=G_{\al\al'}\big[\begin{smallmatrix} \al_4 & \al_1\\\al_3 & \al_2\end{smallmatrix}\big]
=G_{\al\al'}\big[\begin{smallmatrix} \al_1 & \al_4\\\al_2 & \al_3\end{smallmatrix}\big]\,.
\eeq

The other operation we will employ is the {\it braiding} move. This 
move exchanges two consecutive edges in a planar graph. In what 
follows, we will usually perform braiding twice, so we return to a diagram 
which looks like the original one, only that we do have to remember that 
a ``Dehn-twist'' was performed along a tube on the Riemann surface. 
Pictorically, the braiding move is given by
\beq
\raisebox{0mm}[10mm][8mm]{\parbox{24mm}{\begin{center}
\begin{fmfgraph*}(15,15)
\fmfbottom{bl,br}
\fmftop{tl,tr}
\fmf{phantom}{bl,b1,b2,b3,br}
\fmf{phantom}{tl,t1,tt1,t2,tt3,t3,tr}
\fmffreeze
\fmf{plain,width=2}{t1,c2,t3}
\fmf{plain,width=2,tension=2}{c2,b2}
\fmfv{label=$\alpha_1$,label.angle=30}{b2}
\fmfv{label=$\alpha_2$,label.angle=180}{t1}
\fmfv{label=$\alpha_3$,label.angle=0}{t3}
\end{fmfgraph*}\end{center}}}
\to
\raisebox{0mm}[10mm][8mm]{\parbox{24mm}{\begin{center}
\begin{fmfgraph*}(15,15)
\fmfbottom{bl,br}
\fmftop{tl,tr}
\fmf{phantom}{bl,b1,b2,b3,br}
\fmf{phantom}{tl,t1,tt1,t2,tt3,t3,tr}
\fmffreeze
\fmf{plain,width=2}{t1,c2,t3}
\fmf{plain,width=2,tension=2}{c2,b2}
\fmfv{label=$\alpha_1$,label.angle=30}{b2}
\fmfv{label=$\alpha_3$,label.angle=180}{t1}
\fmfv{label=$\alpha_2$,label.angle=0}{t3}
\end{fmfgraph*}\end{center}}}
\eeq
Braiding introduces the following phase into the conformal blocks
\beq
B_{\alpha_1}^{\alpha_2\alpha_3}
=e^{i\pi(\Delta(\alpha_1)-\Delta(\alpha_2)-\Delta(\alpha_3))}\,,
\label{braid}
\eeq
where $\Delta(\alpha)=\alpha(Q-\alpha)$ is the conformal dimension 
of the operator $V_\alpha$. Equipped with the action of fusion and 
braiding on the conformal blocks we now proceed to calculate the 
expectation value of Liouville loop operators.

\subsection{Wilson Loops from Liouville Theory}
\label{sec:wilson}

In \cite{Drukker:2009tz} 
the Wilson loop operators in ${\cal T}_{g,n}$ were identified with 
closed curves that wind around any of the $3g-3+n$ cycles along 
which the surface $C_{g,n}$ is sewn from pairs of pants. Let 
us label a curve along one such cycle by $\gamma_{0,1}$.

We first start by considering the simplest Wilson loop, which 
transforms in the $j=1/2$ representation with respect
to the $SU(2)$ gauge group corresponding to the tube that 
the curve $\gamma_{0,1}$ encircles. The relevant gauge 
group is identified with an internal 
edge in the generalized quiver $\Gamma_\sigma$.

We need to calculate the monodromy of a $V_{1,2}$ 
degenerate field as it moves around the curve $\gamma_{0,1}$. 
As the operation is local on $\Gamma_\sigma$, we ignore the 
rest of the trivalent graph and suppress all other indices in the 
conformal blocks, except for the quantum number labeling the 
internal edge in the generalized quiver $\Gamma_\sigma$ that 
corresponds to the desired $SU(2)$ gauge group, which we
label by the quantum number $\alpha$.

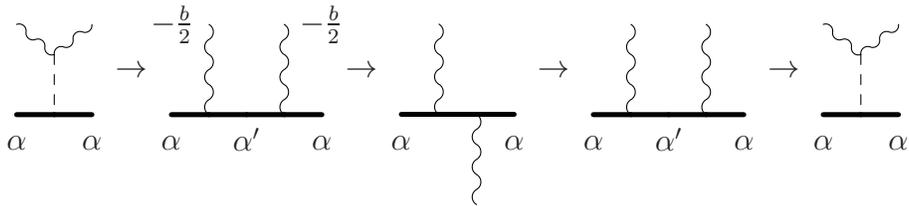
\begin{figure}[t]
\begin{center}
$$
\raisebox{0mm}[10mm][9mm]{\parbox{14mm}{\begin{center}
\begin{fmfgraph*}(10,15)
\fmfbottom{bl,br}
\fmftop{tl,tr}
\fmf{plain,width=2}{bl,b1,br}
\fmf{phantom}{tl,t1,tr}
\fmffreeze
\fmf{boson,width=.5}{tl,c1,tr}
\fmf{dashes,width=.5}{c1,b1}
\fmfv{label=$\alpha$,label.angle=-90,label.dist=8}{bl}
\fmfv{label=$\alpha$,label.angle=-90,label.dist=8}{br}
\end{fmfgraph*}\end{center}}}
\to
\raisebox{0mm}[10mm][9mm]{\parbox{24mm}{\begin{center}
\begin{fmfgraph*}(20,15)
\fmfbottom{bl,br}
\fmftop{tl,tr}
\fmf{plain,width=2}{bl,b1,b2,b3,br}
\fmf{phantom}{tl,t1,t2,t3,tr}
\fmffreeze
\fmf{boson,width=.5}{t1,b1}
\fmf{boson,width=.5}{t3,b3}
\fmfv{label=$\alpha$,label.angle=-90,label.dist=8}{bl}
\fmfv{label=$\alpha$,label.angle=-90,label.dist=8}{br}
\fmfv{label=$\alpha'$,label.angle=-90,label.dist=5}{b2}
\fmfv{label=$-\frac{b}{2}$,label.angle=180}{t1}
\fmfv{label=$-\frac{b}{2}$,label.angle=0}{t3}
\end{fmfgraph*}\end{center}}}
\to
\raisebox{-6mm}[10mm][12mm]{\parbox{19mm}{\begin{center}
\begin{fmfgraph*}(15,30)
\fmfbottom{bl,br}
\fmftop{tl,tr}
\fmf{phantom}{bl,cl,tl}
\fmf{phantom}{br,cr,tr}
\fmffreeze
\fmf{plain,width=2}{cl,c1,c3,cr}
\fmf{phantom}{tl,t1,t3,tr}
\fmf{phantom}{bl,b1,b3,br}
\fmffreeze
\fmf{boson,width=.5}{t1,c1}
\fmf{boson,width=.5}{b3,c3}
\fmfv{label=$\alpha$,label.angle=-90,label.dist=8}{cl}
\fmfv{label=$\alpha$,label.angle=-90,label.dist=8}{cr}
\end{fmfgraph*}\end{center}}}
\to
\raisebox{0mm}[10mm][9mm]{\parbox{24mm}{\begin{center}
\begin{fmfgraph*}(20,15)
\fmfbottom{bl,br}
\fmftop{tl,tr}
\fmf{plain,width=2}{bl,b1,b2,b3,br}
\fmf{phantom}{tl,t1,t2,t3,tr}
\fmffreeze
\fmf{boson,width=.5}{t1,b1}
\fmf{boson,width=.5}{t3,b3}
\fmfv{label=$\alpha$,label.angle=-90,label.dist=8}{bl}
\fmfv{label=$\alpha$,label.angle=-90,label.dist=8}{br}
\fmfv{label=$\alpha'$,label.angle=-90,label.dist=5}{b2}
\end{fmfgraph*}\end{center}}}
\to
\raisebox{0mm}[10mm][9mm]{\parbox{14mm}{\begin{center}
\begin{fmfgraph*}(10,15)
\fmfbottom{bl,br}
\fmftop{tl,tr}
\fmf{plain,width=2}{bl,b1,br}
\fmf{phantom}{tl,t1,tr}
\fmffreeze
\fmf{boson,width=.5}{tl,c1,tr}
\fmf{dashes,width=.5}{c1,b1}
\fmfv{label=$\alpha$,label.angle=-90,label.dist=8}{bl}
\fmfv{label=$\alpha$,label.angle=-90,label.dist=8}{br}
\end{fmfgraph*}\end{center}}}
$$
\parbox{5in}{
\caption{Calculating the Wilson loop in the fundamental representation. 
The solid lines carry arbitrary states, the wiggly ones are degenerate 
fields and the dashed line carries the identity state.
\label{fig:wilson}}}
\end{center}
\end{figure}
In order to probe the edge labeled by $\alpha$ we need to 
introduce a degenerate operator $V_{1,2}$ and braid it around. 
Introducing a single operator will necessarily change the edge in 
a way that if it carries on one side an intermediate state 
$\alpha$, then on the other it will carry $\alpha\pm b/2$, which 
follows from the OPE of the $V_{1,2}$ degenerate field 
(\ref{opedegen}). To avoid this, as explained in Section~\ref{sec:bootstrap},
we introduce a pair of $V_{1,2}$ degenerate operators with their 
OPE projected to the identity. This configuration is represented by 
the left-most diagram in Figure~\ref{fig:wilson}.

By a fusion move we can then express the same configuration 
in the $s$-channel, where we sum over the intermediate states 
$\alpha'=\alpha-sb/2$ between the pair of degenerate fields, 
where $s=\pm1$. Now we perform a non-trivial 
operation of moving one of the degenerate fields around, 
corresponding to a braiding move, fusing it back together 
with the other degenerate field and projecting onto the identity.

Given this sequence of moves, we can calculate the action 
of the loop operator ${\cal L}(\gamma_{0,1})$ on the conformal 
block by using the explicit form of the fusion matrices (see 
Appendix~\ref{app:fusion} for the explicit expression of all the 
fusion matrices required in our calculations). Since the only 
relevant label in the conformal block is the one associated to 
the quantum number $\alpha$, we do not write explicitly all the 
other quantum numbers in the conformal block. The action of 
the loop operator ${\cal L}(\gamma_{0,1})$ on the conformal 
block is given by the following sequence of fusion and 
braiding steps 
\begin{align}
\label{w01}
[\cL(\gamma_{0,1})\cdot\cG]_\alpha
&=
\sum_{s=\pm} 
G^{-1}_{0,\alpha'}
\left[\begin{smallmatrix}
-\frac{b}{2} & -\frac{b}{2}\\
\alpha &\alpha
\end{smallmatrix}\right]
\left(B_{\alpha'}^{-\frac{b}{2},\alpha}\right)^2
G_{\alpha',0}
\left[\begin{smallmatrix}
-\frac{b}{2} & -\frac{b}{2}\\
\alpha & \alpha
\end{smallmatrix}\right]\cG_\alpha
\nonumber\\
&
=\sum_{s=\pm} 
G_{-,s}
\left[\begin{smallmatrix}
\alpha & -\frac{b}{2}\\
\alpha &-\frac{b}{2}
\end{smallmatrix}\right]
e^{\pi b i(Q+s(2\alpha-Q))}
G_{s,-}
\left[\begin{smallmatrix}
-\frac{b}{2} & -\frac{b}{2}\\
\alpha & \alpha
\end{smallmatrix}\right]\cG_\alpha
\\\nonumber
&\hskip-1cm
=\sum_{s=\pm}
\frac{\sin\left(\pi b^2\right)}
{\sin\left(\pi b(2\alpha-Q)\right)}
e^{\pi b i(Q+s(2\alpha-Q))}
\frac{\sin\left(\pi b(2\alpha'-Q)\right)}
{\sin\left(2\pi b^2\right)}\,\cG_\alpha
=\frac{\cos\left(\pi b(2\alpha-Q)\right)}
{\cos\left(\pi bQ\right)}\,\cG_\alpha\,.
\end{align}
Taking the $b\to1$ limit, which corresponds to the partition 
function of ${\cal T}_{g,n}$ on $S^4$, gives
\beq
[\cL(\gamma_{0,1})\cdot\cG]_\alpha
=\cos (2\pi a)\,\cG_\alpha\,,
\qquad \alpha=\frac{Q}{2}+a\,.
\eeq

An interesting point about this formula is the normalization. It exactly matches 
the Wilson loop in the spin $1/2$ representation when the trace is normalized 
by dividing by the dimension of the representation, which is two for $j=1/2$. More 
generally, away from $b=1$, if we multiply the numerator by $2$ then the denominator is 
$2\cos(\pi bQ)=e^{i\pi bQ}+e^{-i\pi bQ}$, which is the quantum dimension 
of the spin $1/2$ representation of q-deformed $SL(2,\bR)$.

The same expression would apply also to the conformal blocks with 
any of the other normalizations, since the operator $\cL(\gamma_{0,1})$ 
diagonalizes the conformal blocks, and acts by a multiplicative factor.

Using the expression for the correlator of a Liouville loop operator 
(\ref{liouloop}), we find that the result exactly matches Pestun's 
formula \cite{Pestun:2007rz} for the expectation value of a Wilson 
loop in the $j=1/2$ representation for one of the gauge groups%
\footnote{We stress that the insertion of $\cos(2\pi a)$ is for only one 
of the gauge fields.}
\beq
\big\langle\CL(\gamma_{0,1})\big\rangle_{C_{g,n}}^{}
\,= 
\vev{W_{j=1/2}}_{{\cal T}_{g,n}}
=\int[da]\, \cos(2\pi a) \, \overline{Z}_\text{Nekrasov}
Z_\text{Nekrasov}\,,
\label{eq:pestun}
\eeq
where we have used the identification between the Liouville 
correlator and the partition function of the gauge theory 
without a loop operator found in \cite{Alday:2009aq}.

We now proceed to calculate the expectation value of a Wilson 
loop in the spin $j=q/2$ representation of one of the $SU(2)$ 
groups. As discussed in Section~\ref{sec:pert}, the degenerate 
operator that must be moved around the Riemann surface is 
$V_{1,q+1}$. Since the product of two identical degenerate 
fields $V_{r,s}$ always contains the identity operator, we can 
carry out the same operation we did for the $V_{1,2}$ degenerate field. 
A pair of them can also be created in the same way as before, 
as discussed in detail in Appendix~\ref{app:higher}. The setup 
is precisely the same as the one in Figure~\ref{fig:wilson}, but now 
the degenerate fields have quantum numbers $-qb/2$.
In this case, between the 
two degenerate fields $V_{1,q+1}$ the state $\alpha$ is shifted to 
$\alpha-kb/2$ with $k=-q,-q+2,\cdots,q$, due to the OPE expansion 
of a $V_{1,q+1}$ degenerate field and a primary field. The action of the 
loop operator $\CL(\gamma_{0,1})$ on the conformal blocks is therefore%
\footnote{This expression was checked up to $q=5$, 
see also equation \eqn{qfs}.}
\begin{align}
\label{w0q}
[\cL_{q/2}(\gamma_{0,1})\cdot\cG]_\alpha
&=
\sum_k
G^{-1}_{0,\alpha-\frac{kb}{2}}
\left[\begin{smallmatrix}
-\frac{qb}{2} & -\frac{qb}{2}\\
\alpha &\alpha
\end{smallmatrix}\right]
\left(B_{\alpha-k\frac{b}{2}}^{-q\frac{b}{2},\alpha}\right)^2
G_{\alpha-\frac{kb}{2},0}
\left[\begin{smallmatrix}
-\frac{qb}{2} & -\frac{qb}{2}\\
\alpha & \alpha
\end{smallmatrix}\right]\cG_\alpha
\nonumber
\\&=
\frac{\sum_k \exp[2\pi i kb\,a]}
{\sum_k \exp[\pi ikbQ]}\,
\cG_\alpha\,.
\end{align}
As in the case of 
$j=1/2$, the denominator is the quantum dimension 
of the spin $q/2$ representation of the q-deformed $SL(2,\bR)$ 
(or the classical dimension for $b=1$).

In the limit $b\to1$, where the Liouville correlator has an interpretation 
in term of ${\cal N}=2$ gauge theories on $S^4$ we get
\beq
[\cL_{q/2}(\gamma_{0,1})\cdot\cG]_\alpha
=\frac{1}{q+1}\sum_k \exp[2\pi ik\,a]\,\cG_\alpha\,.
\eeq
This result exactly matches Pestun's formula \cite{Pestun:2007rz} for the expectation 
value of the Wilson loop in the spin $j=q/2$ representation for one of the gauge groups
\beq
\big\langle\CL_{q/2}(\gamma_{0,1})\big\rangle_{C_{g,n}}^{}
\,= 
\vev{W_{q/2}}_{{\cal T}_{g,n}}\,.
\eeq

As discussed in Section~\ref{sec:pert}, there are two other 
natural bases  of Wilson loop observables with quantum number $q$, 
corresponding to the multi-wrapped loops and the multi-trace 
loops. They also have a simple realization in terms of Liouville 
loop operators.

We can use the same degenerate field, and note that wrapping a 
single $V_{1,2}$ field $q$ times around the curve gives extra braiding 
factors in \eqn{w01}. The result is
\beq
\begin{aligned}[]
[\cL(q\cdot \gamma_{0,1})\cdot \cG]_\alpha
&=\sum_{s=\pm}
e^{q\pi b i(Q+s(2\alpha-Q))}
\frac{\sin\left(\pi b^2\right)}
{\sin\left(2\pi b^2\right)}
\frac{\sin\left(\pi b(2\alpha'-Q)\right)}
{\sin\left(\pi b(2\alpha-Q)\right)}\,\cG_\alpha
\\&\mathop{\longrightarrow}_{b\to1}\quad
\cos\left(2\pi q a\right)\,\cG_\alpha\,.
\end{aligned}
\eeq
This yields the expectation value of the multi-wrapped Wilson loop 
in the ${\cal N}=2$ gauge theory%
\footnote{On the right-hand side we are using 
notations from Pestun's matrix model calculation to represent the 
corresponding field theory observable.}
\beq
\big\langle\CL(q\cdot \gamma_{0,1})\big\rangle_{C_{g,n}}^{}
\,= 
\vev{\frac{1}{2}\Tr e^{2\pi i qa}}_{{\cal T}_{g,n}}\,.
\label{}
\eeq

Instead, if we consider $q$ pairs of independent degenerate fields we just get 
the $q^\text{th}$ power of the previous result
\beq
[\cL(\gamma_{0,1})^q\cdot \cG]_\alpha
=\left[-\frac{\cos\left(\pi b(2\alpha-Q)\right)}{\cos\left(\pi b^2\right)}\right]^q\,
\cG_\alpha
\quad \mathop{\longrightarrow}_{b\to1}\quad
\cos^q\left(2\pi a\right)\,\cG_\alpha\,.
\eeq
This yields the expectation value of the multi-trace Wilson loop 
in the ${\cal N}=2$ gauge theory
\beq
\big\langle\cL(\gamma_{0,1})^q)\big\rangle_{C_{g,n}}^{}
\,= 
\vev{\big[W_{1/2}\big]^q}_{{\cal T}_{g,n}}\,.
\label{}
\eeq

We thus have found a Liouville description for each of the three natural 
Wilson loops with quantum number $q$:
\begin{enumerate}
\item
The Wilson loop in the irreducible $j=q/2$ representation is given by 
generating a pair of $V_{1,q+1}$ degenerate fields, braiding one of 
them around and fusing them back into the identity state.
\item
The Wilson loop wrapped $q$ times around the curve is given by 
generating a pair of the basic $V_{1,2}$ degenerate fields, braiding one of 
them around $q$ times and fusing them back into the identity state. 
This can be thought of as a $q$-times wound curve.
\item
The $q^\text{th}$ product of the fundamental Wilson loop 
is achieved by generating $q$ pairs of the basic $V_{1,2}$ degenerate 
fields, braiding one of each pair and fusing them back into the identity states. 
This can be thought of as $q$ parallel curves on the surface.
\end{enumerate}
Combinations of any of these can be used to create all other Wilson loop 
operators.

\subsection{'t~Hooft Loop in $\cN=2^*$ Theory}
\label{sec:torus}

\begin{figure}[t]
\begin{center}
$$
\raisebox{0mm}[10mm][12mm]{\parbox{40mm}{
\begin{fmfgraph*}(40,40)
\fmfsurroundn{o}{8}
\fmf{phantom,tension=.5}{o1,v1}
\fmf{phantom,tension=.5}{o2,v2}
\fmf{phantom,tension=.5}{o3,v3}
\fmf{phantom,tension=.5}{o4,v4}
\fmf{phantom,tension=.5}{o5,v5}
\fmf{phantom,tension=.5}{o6,v6}
\fmf{phantom,tension=.5}{o7,v7}
\fmf{phantom,tension=.5}{o8,v8}
\fmfcyclen{plain,width=2,right=0.20}{v}{8} 
\fmffreeze
\fmf{phantom}{v5,c}
\fmf{plain,width=2}{c,v1}
\fmf{phantom}{o4,o45,o5,o55,o6}
\fmffreeze
\fmf{boson,width=.5}{o45,oo,o55}
\fmf{dashes,width=.5}{oo,v5}
\fmfv{label=$\alpha$,label.angle=-90,label.dist=6}{v7}
\fmfv{label=$m$,label.angle=-90,label.dist=6}{c}
\end{fmfgraph*}}}
\hspace{-8mm}
\to
\raisebox{0mm}[10mm][12mm]{\parbox{40mm}{
\begin{fmfgraph*}(40,40)
\fmfsurroundn{o}{8}
\fmf{phantom,tension=.5}{o1,v1}
\fmf{phantom,tension=.5}{o2,v2}
\fmf{phantom,tension=.5}{o3,v3}
\fmf{phantom,tension=.5}{o4,v4}
\fmf{phantom,tension=.5}{o5,v5}
\fmf{phantom,tension=.5}{o6,v6}
\fmf{phantom,tension=.5}{o7,v7}
\fmf{phantom,tension=.5}{o8,v8}
\fmfcyclen{plain,width=2,right=0.20}{v}{8} 
\fmffreeze
\fmf{phantom}{v5,c}
\fmf{plain,width=2}{c,v1}
\fmf{phantom}{o4,o45,o5,o55,o6}
\fmffreeze
\fmf{boson,width=.5}{o45,v4}
\fmf{boson,width=.5}{o55,v6}
\fmfv{label=$\alpha$,label.angle=-90,label.dist=6}{v7}
\fmfv{label=$\alpha'$,label.angle=180,label.dist=6}{v5}
\fmfv{label=$m$,label.angle=-90,label.dist=6}{c}
\end{fmfgraph*}}}
\hspace{-8mm}
\to
\raisebox{0mm}[10mm][12mm]{\parbox{40mm}{
\begin{fmfgraph*}(40,40)
\fmfsurroundn{o}{8}
\fmf{phantom,tension=.5}{o1,v1}
\fmf{phantom,tension=.5}{o2,v2}
\fmf{phantom,tension=.5}{o3,v3}
\fmf{phantom,tension=.5}{o4,v4}
\fmf{phantom,tension=.5}{o5,v5}
\fmf{phantom,tension=.5}{o6,v6}
\fmf{phantom,tension=.5}{o7,v7}
\fmf{phantom,tension=.5}{o8,v8}
\fmfcyclen{plain,width=2,right=0.20}{v}{8} 
\fmffreeze
\fmf{phantom}{v5,c}
\fmf{plain,width=2}{c,v1}
\fmf{phantom}{o4,o45,o5,o55,o6}
\fmf{phantom}{o8,o85,o1}
\fmffreeze
\fmf{boson,width=.5}{o55,v6}
\fmf{boson,width=.5}{o85,v8}
\fmfv{label=$\alpha$,label.angle=-90,label.dist=6}{v7}
\fmfv{label=$\alpha'$,label.angle=180,label.dist=6}{v5}
\fmfv{label=$\alpha''$,label.angle=-60,label.dist=6}{v1}
\fmfv{label=$m$,label.angle=-90,label.dist=6}{c}
\end{fmfgraph*}}}
\hspace{-4mm}
\to
\raisebox{0mm}[10mm][12mm]{\parbox{40mm}{
\begin{fmfgraph*}(40,40)
\fmfsurroundn{o}{8}
\fmf{phantom,tension=.5}{o1,v1}
\fmf{phantom,tension=.5}{o2,v2}
\fmf{phantom,tension=.5}{o3,v3}
\fmf{phantom,tension=.5}{o4,v4}
\fmf{phantom,tension=.5}{o5,v5}
\fmf{phantom,tension=.5}{o6,v6}
\fmf{phantom,tension=.5}{o7,v7}
\fmf{phantom,tension=.5}{o8,v8}
\fmfcyclen{plain,width=2,right=0.20}{v}{8} 
\fmffreeze
\fmf{phantom}{v5,c}
\fmf{plain,width=2}{c,v1}
\fmf{phantom}{o4,o45,o5,o55,o6}
\fmffreeze
\fmf{boson,width=.5}{o45,oo,o55}
\fmf{dashes,width=.5}{oo,v5}
\fmfv{label=$\alpha'$,label.angle=-90,label.dist=6}{v7}
\fmfv{label=$m$,label.angle=-90,label.dist=6}{c}
\end{fmfgraph*}}}
$$
\parbox{5in}{
\caption{Calculating the 't~Hooft loop on the one-punctured torus.
\label{fig:torus}}}
\end{center}
\end{figure}
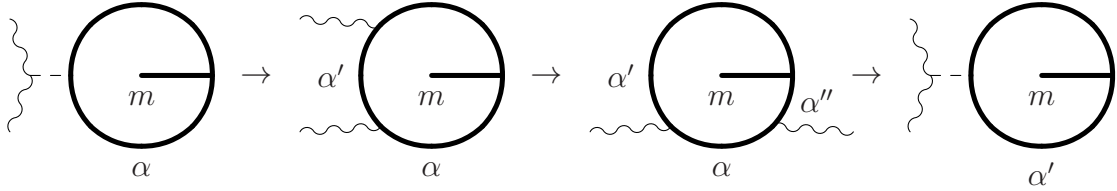

As discussed earlier, from the Liouville point of view 't~Hooft 
loop operators are also described by the monodromy of the conformal 
block associated to a degenerate field moving around a closed curve. 
Although these are essentially the same calculations we performed 
for the Wilson loop, once we choose a different pants-decomposition 
of the Riemann surface the details change drastically.

Given a choice of pants decomposition $\sigma$ of the Riemann 
surface, the curve that describes a loop operator carrying magnetic 
charges must move nontrivially around the corresponding trivalent 
graph $\Gamma_\sigma$. For the case of the Wilson loop just 
considered, the associated curve simply corresponds to a rotation 
around one the edges of $\Gamma_\sigma$. Since the degenerate 
field does not move in $\Gamma_\sigma$, one would expect that 
the loop operator does not change the quantum number 
propagating along the edge, which is what we have found.
For an operator carrying magnetic charge the curve must explore 
the trivalent graph $\Gamma_\sigma$, which suggests that the 
quantum numbers associated to the edges traversed by the curve 
are now affected by the action of the loop operator. As we show, 
this intuitive picture is materialized in the actual computation of the 
monodromy.

We now consider an 't~Hooft operator in ${\cal N}=2^*$, which 
corresponds to the once-punctured torus. As usual we create a 
pair of degenerate fields which project to the identity operator 
(see the first picture in Figure \ref{fig:torus}). If in the second step, 
after fusion, we move one of the degenerate fields around the 
graph without changing the location of the degenerate field on 
the graph, then this corresponds to the Wilson loop calculation 
we have already carried out. 

Transporting the degenerate field around the entire graph 
corresponds instead to the 't~Hooft operator. The procedure is 
illustrated in Figure~\ref{fig:torus}. A pair of degenerate 
fields, which combine to the identity are rewritten in the dual channel with 
an intermediate state $\alpha'=\alpha-sb/2$, where $s=\pm 1$. 
Then one of the degenerate fields is fused across the puncture 
(with label $m$). In principle one should allow for a new state 
$\alpha''$ there, but since we then combine again 
the two degenerate fields and project onto the identity, this 
identifies the two quantum numbers $\alpha''=\alpha'$.

There is a subtle point about this calculation, which is the exact 
identification of the initial and final state. Great care has to 
be taken to guarantee that we follow exactly the same 
path as in the Wilson loop calculation (in the dual channel), 
and do not permute the two degenerate fields. Otherwise 
one ends up with extra phases \eqn{braid} due to braiding 
between these two fields. Note that the ambiguity is independent 
of $\alpha$ and $m$, so we can  examine the case of 
$m=0$ (torus without a puncture), where the operator 
$M_{\si'\si}$ in \eqn{Msigsig} is just Fourier-transformation. 
In that case  the transformation of the difference operator 
$\cD_\alpha^{(\sigma)}$ is obvious, and gives the
formula below. This fixes the possible phase.%
\footnote{Also, since we know that the loop operators are Hermitian 
and positive definite (see Section~\ref{liou-teich}) this, in principle 
fixes the phase.}

This sequence of moves induces a monodromy on the conformal 
block of the once punctured torus, which we denote by $\cG_{\alpha,m}$. 
The fusion and braiding moves yield 
\begin{align}
[\cL(\gamma_{1,0})\cdot&\cG]_{\alpha,m}
=\sum_{s=\pm} 
G_{0,\alpha-\frac{sb}{2}}
\left[\begin{smallmatrix} 
\alpha & -\frac{b}{2}\\
\alpha & -\frac{b}{2}
\end{smallmatrix}\right]
G^{-1}_{\alpha,\alpha-\frac{sb}{2}}
\left[\begin{smallmatrix} 
\alpha- s\frac{b}{2} & -\frac{b}{2}\\
m & \alpha
\end{smallmatrix}\right]
G_{\alpha,0}
\left[\begin{smallmatrix} 
\alpha-s\frac{b}{2} & \alpha- s\frac{b}{2}\\
-\frac{b}{2} & -\frac{b}{2}
\end{smallmatrix}\right]
\cG_{\alpha-s b/2,m}
\label{thooft-G}
\nonumber\\&=
\sum_{s=\pm} 
G_{-,s}
\left[\begin{smallmatrix} 
\alpha & -\frac{b}{2}\\
\alpha & -\frac{b}{2}
\end{smallmatrix}\right]
G_{-s,s}
\left[\begin{smallmatrix} 
\alpha & -\frac{b}{2}\\
m & \alpha- s\frac{b}{2}
\end{smallmatrix}\right]
G_{-s,-}
\left[\begin{smallmatrix} 
-\frac{b}{2} & -\frac{b}{2}\\
\alpha-s\frac{b}{2} & \alpha- s\frac{b}{2}
\end{smallmatrix}\right]
\cG_{\alpha-s b/2,m}
\\\nonumber&=
\frac{\sin(\pi b^2)}{\sin(2\pi b^2)}
\left(\frac{\sin(\pi b(2\alpha-Q-m))}{\sin(\pi b(2\alpha-Q))}\cG_{\alpha-b/2,m}
+\frac{\sin(\pi b(2\alpha-Q+m))}{\sin(\pi b(2\alpha-Q))}\cG_{\alpha+b/2,m}
\right)
\end{align}
We note that, as expected, the loop operator shifts the value of 
the quantum number associated with the edge 
along which the degenerate field is transported.
Hence, unlike the calculation of the Wilson loop, 
the functional prefactors are then 
dependent on the normalization conventions for the conformal 
blocks, whether $\cG$, $\cF$ or $Z_\text{Nekrasov}$.%
\footnote{To be explicit, for the one-punctured torus 
$\cG_{\alpha,m}=N(\alpha,m,\alpha)\cF_{\alpha,m}$ \eqn{Nexp}.}

In the limit $b\to1$, where the Liouville correlator has an interpretation in term of ${\cal N}=2^*$ on $S^4$, this yields
\beq
\big\langle\CL(\gamma_{1,0})\big\rangle_{C_{g,n}}^{}
\,= 2
\int d \alpha \sin(2\pi \alpha)\,\overline{\cG}_{\alpha,m}
\left[ \sin(\pi (2\alpha-m))\cG_{\alpha-1/2,m}
+\sin(\pi (2\alpha+m))\cG_{\alpha+1/2,m}
\right].
\eeq
Our proposal (\ref{proposal}) identifies the result of this 
calculation as the expectation value of an 't~Hooft 
operator $T\equiv L_{1,0}$ in the ${\cal N}=2^*$ theory
\beq
\vev{T}_{{\cal T}_{g,n}}= \big\langle\CL(\gamma_{1,0})\big\rangle_{C_{g,n}}\,.
\eeq
The 't~Hooft loop is given by a sum of two terms, one where the 
argument of the holomorphic contribution is shifted by $1/2$ and the 
other by $-1/2$. In section \ref{sec:discuss}, we provide a possible 
gauge theory interpretation of this observation.

It is straightforward to generalize the calculation to the dyonic 
loop operator $\cL(\gamma_{1,q})$ by including an extra braiding factor. 
The result is
\beq
\begin{aligned}[]
[\cL(\gamma_{1,q})\cdot\cG]_{\alpha,m}
=
e^{-q \pi i b^2/2}
\frac{\sin(\pi b^2)}{\sin(2\pi b^2)}
\bigg(
&e^{\pi i qb(2\alpha-Q)}
\frac{\sin(\pi b(2\alpha-Q-m))}{\sin(\pi b(2\alpha-Q))}
\cG_{\alpha-b/2,m}
\\&+
e^{-\pi i qb(2\alpha-Q)}
\frac{\sin(\pi b(2\alpha-Q+m))}{\sin(\pi b(2\alpha-Q))}
\cG_{\alpha+b/2,m}
\bigg).
\end{aligned}
\label{dyon-torus}
\eeq
In section \ref{sec:algebra} we provide some consistency checks 
for the validity of this calculation by demonstrating that the dyonic 
operators satisfy the expected 't~Hooft commutation relations for 
$b=1$ and the expected operator product expansion for $b=0$.

\subsection{'t~Hooft Loop in the Theory with One $SU(2)_C$ and $N_F=4$}
\label{sec:sphere}

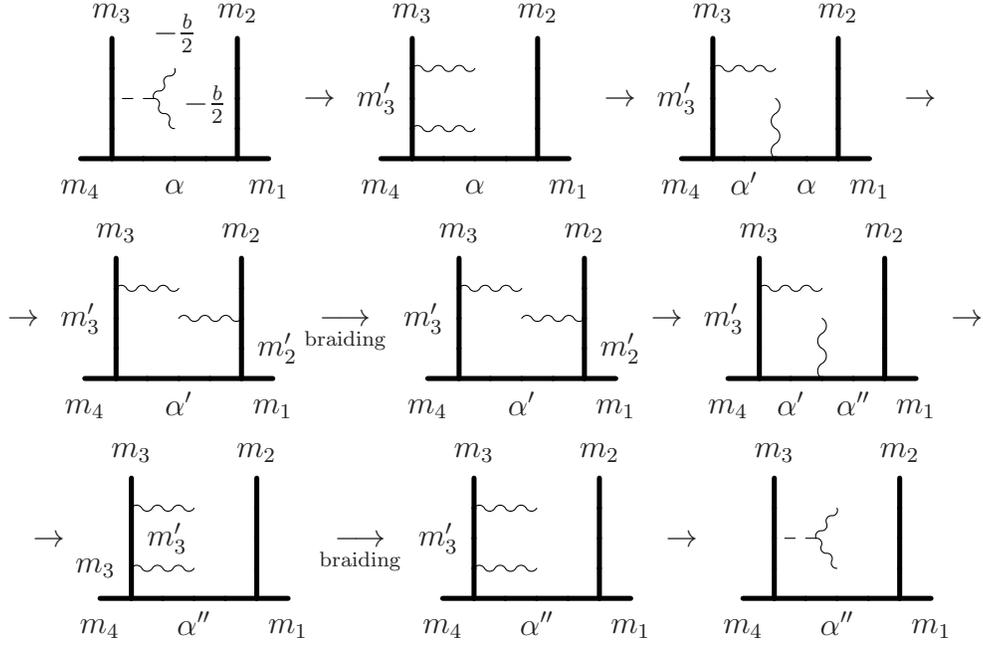
\begin{figure}[t]
\begin{center}
$$\begin{gathered}
\raisebox{0mm}[10mm][18mm]{\parbox{32mm}{\begin{center}
\begin{fmfgraph*}(25,20)
\fmfbottom{bl,br}
\fmftop{tl,tr}
\fmf{plain,width=2}{bl,b1,b2,b3,b4,b5,br}
\fmf{phantom}{tl,t1,t2,t3,t4,t5,tr}
\fmffreeze
\fmf{plain,width=2}{t1,c1,cc1,ccc1,b1}
\fmf{plain,width=2}{t5,c5,cc5,ccc5,b5}
\fmf{phantom}{t3,c3,cc3,ccc3,b3}
\fmffreeze
\fmf{boson,width=.5}{c3,cc2,ccc3}
\fmf{dashes,width=.5}{cc2,cc1}
\fmfv{label=$m_4$,label.angle=-90,label.dist=8}{bl}
\fmfv{label=$m_1$,label.angle=-90,label.dist=8}{br}
\fmfv{label=$\alpha$,label.angle=-90,label.dist=8}{b3}
\fmfv{label=$-\frac{b}{2}$,label.angle=90}{c3}
\fmfv{label=$-\frac{b}{2}$,label.angle=30,label.dist=4}{ccc3}
\fmfv{label=$m_3$,label.angle=90}{t1}
\fmfv{label=$m_2$,label.angle=90}{t5}
\end{fmfgraph*}\end{center}}}
\to\ 
\raisebox{0mm}[10mm][18mm]{\parbox{32mm}{\begin{center}
\begin{fmfgraph*}(25,20)
\fmfbottom{bl,br}
\fmftop{tl,tr}
\fmf{plain,width=2}{bl,b1,b2,b3,b4,b5,br}
\fmf{phantom}{tl,t1,t2,t3,t4,t5,tr}
\fmffreeze
\fmf{plain,width=2}{t1,c1,cc1,ccc1,b1}
\fmf{plain,width=2}{t5,c5,cc5,ccc5,b5}
\fmf{phantom}{t3,c3,cc3,ccc3,b3}
\fmffreeze
\fmf{boson,width=.5}{c3,c1}
\fmf{boson,width=.5}{ccc3,ccc1}
\fmfv{label=$m_4$,label.angle=-90,label.dist=8}{bl}
\fmfv{label=$m_1$,label.angle=-90,label.dist=8}{br}
\fmfv{label=$\alpha$,label.angle=-90,label.dist=8}{b3}
\fmfv{label=$m_3$,label.angle=90}{t1}
\fmfv{label=$m_3'$,label.angle=180}{cc1}
\fmfv{label=$m_2$,label.angle=90}{t5}
\end{fmfgraph*}\end{center}}}
\to\ 
\raisebox{0mm}[10mm][18mm]{\parbox{32mm}{\begin{center}
\begin{fmfgraph*}(25,20)
\fmfbottom{bl,br}
\fmftop{tl,tr}
\fmf{plain,width=2}{bl,b1,b2,b3,b4,b5,br}
\fmf{phantom}{tl,t1,t2,t3,t4,t5,tr}
\fmffreeze
\fmf{plain,width=2}{t1,c1,cc1,ccc1,b1}
\fmf{plain,width=2}{t5,c5,cc5,ccc5,b5}
\fmf{phantom}{t3,c3,cc3,ccc3,b3}
\fmffreeze
\fmf{boson,width=.5}{c3,c1}
\fmf{boson,width=.5}{cc3,b3}
\fmfv{label=$m_4$,label.angle=-90,label.dist=8}{bl}
\fmfv{label=$m_1$,label.angle=-90,label.dist=8}{br}
\fmfv{label=$\alpha'$,label.angle=-90,label.dist=5}{b2}
\fmfv{label=$\alpha$,label.angle=-90,label.dist=8}{b4}
\fmfv{label=$m_3$,label.angle=90}{t1}
\fmfv{label=$m_3'$,label.angle=180}{cc1}
\fmfv{label=$m_2$,label.angle=90}{t5}
\end{fmfgraph*}\end{center}}}
\to\\
\to\ 
\raisebox{0mm}[10mm][18mm]{\parbox{32mm}{\begin{center}
\begin{fmfgraph*}(25,20)
\fmfbottom{bl,br}
\fmftop{tl,tr}
\fmf{plain,width=2}{bl,b1,b2,b3,b4,b5,br}
\fmf{phantom}{tl,t1,t2,t3,t4,t5,tr}
\fmffreeze
\fmf{plain,width=2}{t1,c1,cc1,ccc1,b1}
\fmf{plain,width=2}{t5,c5,cc5,ccc5,b5}
\fmf{phantom}{t3,c3,cc3,ccc3,b3}
\fmffreeze
\fmf{boson,width=.5}{c3,c1}
\fmf{boson,width=.5}{cc3,cc5}
\fmfv{label=$m_4$,label.angle=-90,label.dist=8}{bl}
\fmfv{label=$m_1$,label.angle=-90,label.dist=8}{br}
\fmfv{label=$\alpha'$,label.angle=-90,label.dist=5}{b3}
\fmfv{label=$m_3$,label.angle=90}{t1}
\fmfv{label=$m_3'$,label.angle=180}{cc1}
\fmfv{label=$m_2$,label.angle=90}{t5}
\fmfv{label=$m_2'$,label.angle=0}{ccc5}
\end{fmfgraph*}\end{center}}}
\mathop{\longrightarrow}_\text{braiding}\ 
\raisebox{0mm}[10mm][18mm]{\parbox{32mm}{\begin{center}
\begin{fmfgraph*}(25,20)
\fmfbottom{bl,br}
\fmftop{tl,tr}
\fmf{plain,width=2}{bl,b1,b2,b3,b4,b5,br}
\fmf{phantom}{tl,t1,t2,t3,t4,t5,tr}
\fmffreeze
\fmf{plain,width=2}{t1,c1,cc1,ccc1,b1}
\fmf{plain,width=2}{t5,c5,cc5,ccc5,b5}
\fmf{phantom}{t3,c3,cc3,ccc3,b3}
\fmffreeze
\fmf{boson,width=.5}{c3,c1}
\fmf{boson,width=.5}{cc3,cc5}
\fmfv{label=$m_4$,label.angle=-90,label.dist=8}{bl}
\fmfv{label=$m_1$,label.angle=-90,label.dist=8}{br}
\fmfv{label=$\alpha'$,label.angle=-90,label.dist=5}{b3}
\fmfv{label=$m_3$,label.angle=90}{t1}
\fmfv{label=$m_3'$,label.angle=180}{cc1}
\fmfv{label=$m_2$,label.angle=90}{t5}
\fmfv{label=$m_2'$,label.angle=0}{ccc5}
\end{fmfgraph*}\end{center}}}
\to\ 
\raisebox{0mm}[10mm][18mm]{\parbox{32mm}{\begin{center}
\begin{fmfgraph*}(25,20)
\fmfbottom{bl,br}
\fmftop{tl,tr}
\fmf{plain,width=2}{bl,b1,b2,b3,b4,b5,br}
\fmf{phantom}{tl,t1,t2,t3,t4,t5,tr}
\fmffreeze
\fmf{plain,width=2}{t1,c1,cc1,ccc1,b1}
\fmf{plain,width=2}{t5,c5,cc5,ccc5,b5}
\fmf{phantom}{t3,c3,cc3,ccc3,b3}
\fmffreeze
\fmf{boson,width=.5}{c3,c1}
\fmf{boson,width=.5}{cc3,b3}
\fmfv{label=$m_4$,label.angle=-90,label.dist=8}{bl}
\fmfv{label=$m_1$,label.angle=-90,label.dist=8}{br}
\fmfv{label=$\alpha'$,label.angle=-90,label.dist=5}{b2}
\fmfv{label=$\alpha''$,label.angle=-90,label.dist=5}{b4}
\fmfv{label=$m_3$,label.angle=90}{t1}
\fmfv{label=$m_3'$,label.angle=180}{cc1}
\fmfv{label=$m_2$,label.angle=90}{t5}
\end{fmfgraph*}\end{center}}}
\to\\
\to
\raisebox{0mm}[10mm][18mm]{\parbox{32mm}{\begin{center}
\begin{fmfgraph*}(25,20)
\fmfbottom{bl,br}
\fmftop{tl,tr}
\fmf{plain,width=2}{bl,b1,b2,b3,b4,b5,br}
\fmf{phantom}{tl,t1,t2,t3,t4,t5,tr}
\fmffreeze
\fmf{plain,width=2}{t1,c1,cc1,ccc1,b1}
\fmf{plain,width=2}{t5,c5,cc5,ccc5,b5}
\fmf{phantom}{t3,c3,cc3,ccc3,b3}
\fmffreeze
\fmf{boson,width=.5}{c3,c1}
\fmf{boson,width=.5}{ccc3,ccc1}
\fmfv{label=$m_4$,label.angle=-90,label.dist=8}{bl}
\fmfv{label=$m_1$,label.angle=-90,label.dist=8}{br}
\fmfv{label=$\alpha''$,label.angle=-90,label.dist=5}{b3}
\fmfv{label=$m_3$,label.angle=90}{t1}
\fmfv{label=$m_3'$,label.angle=0}{cc1}
\fmfv{label=$m_3$,label.angle=180}{ccc1}
\fmfv{label=$m_2$,label.angle=90}{t5}
\end{fmfgraph*}\end{center}}}
\mathop{\longrightarrow}_\text{braiding}\ 
\raisebox{0mm}[10mm][18mm]{\parbox{32mm}{\begin{center}
\begin{fmfgraph*}(25,20)
\fmfbottom{bl,br}
\fmftop{tl,tr}
\fmf{plain,width=2}{bl,b1,b2,b3,b4,b5,br}
\fmf{phantom}{tl,t1,t2,t3,t4,t5,tr}
\fmffreeze
\fmf{plain,width=2}{t1,c1,cc1,ccc1,b1}
\fmf{plain,width=2}{t5,c5,cc5,ccc5,b5}
\fmf{phantom}{t3,c3,cc3,ccc3,b3}
\fmffreeze
\fmf{boson,width=.5}{c3,c1}
\fmf{boson,width=.5}{ccc3,ccc1}
\fmfv{label=$m_4$,label.angle=-90,label.dist=8}{bl}
\fmfv{label=$m_1$,label.angle=-90,label.dist=8}{br}
\fmfv{label=$\alpha''$,label.angle=-90,label.dist=5}{b3}
\fmfv{label=$m_3$,label.angle=90}{t1}
\fmfv{label=$m_3'$,label.angle=180}{cc1}
\fmfv{label=$m_2$,label.angle=90}{t5}
\end{fmfgraph*}\end{center}}}
\to\ 
\raisebox{0mm}[10mm][18mm]{\parbox{32mm}{\begin{center}
\begin{fmfgraph*}(25,20)
\fmfbottom{bl,br}
\fmftop{tl,tr}
\fmf{plain,width=2}{bl,b1,b2,b3,b4,b5,br}
\fmf{phantom}{tl,t1,t2,t3,t4,t5,tr}
\fmffreeze
\fmf{plain,width=2}{t1,c1,cc1,ccc1,b1}
\fmf{plain,width=2}{t5,c5,cc5,ccc5,b5}
\fmf{phantom}{t3,c3,cc3,ccc3,b3}
\fmffreeze
\fmf{boson,width=.5}{c3,cc2,ccc3}
\fmf{dashes,width=.5}{cc2,cc1}
\fmfv{label=$m_4$,label.angle=-90,label.dist=8}{bl}
\fmfv{label=$m_1$,label.angle=-90,label.dist=8}{br}
\fmfv{label=$\alpha''$,label.angle=-90,label.dist=5}{b3}
\fmfv{label=$m_3$,label.angle=90}{t1}
\fmfv{label=$m_2$,label.angle=90}{t5}
\end{fmfgraph*}\end{center}}}
\end{gathered}
$$
\vskip-5mm
\parbox{5in}{
\caption{Calculating the 't~Hooft loop on the four-punctured sphere 
as the monodromy associated to moving a $V_{1,2}$ field.
\label{fig:4puncture}}}
\end{center}
\end{figure}

The next theory we consider is ${\cal T}_{0,4}$, which is based on the four punctured
sphere. This theory admits a weakly coupled description in terms of an ${\cal N}=2$ 
$SU(2)$ gauge theory with four fundamental hypermultiplets. Our goal is to compute 
the expectation value of an 't~Hooft loop in this theory.

The monodromy corresponding to the 't~Hooft loop is captured by the 
series of moves illustrated in Figure~\ref{fig:4puncture}. First we insert a 
pair of degenerate fields in the edge labeled by $m_3$. One of the 
degenerate fields is then moved across the trivalent graph and then 
transported back. Note that when the degenerate field hits the 
vertex with external edges labeled by $m_1$ and $m_2$ braiding 
around the punctures must be performed.
Likewise, we must braid when the degenerate field is back into the 
original vertex labeled by $m_3$ and $m_4$.

Using the formulae for the fusion and braiding moves we find 
that the expression for the complete move is the following 
\beq
\begin{aligned}
T_{s,s'}&=\sum_{s_2,s_3=\pm}
G_{0m_3'}\left[\begin{smallmatrix} 
m_3 & -\frac{b}{2}\\ m_3 & -\frac{b}{2}\end{smallmatrix}\right]
G^{-1}_{m_3\alpha'}\left[\begin{smallmatrix} 
m_3'&-\frac{b}{2}\\ m_4 & \al\end{smallmatrix}\right]
G_{\al,m_2'}\left[\begin{smallmatrix} 
-\frac{b}{2} & m_2\\ \al' & m_1\end{smallmatrix}\right]
\left(B_{m_2'}^{-\frac{b}{2},m_2}\right)^2
\\&\hskip1cm{}\times
G^{-1}_{m_2'\alpha''}\left[\begin{smallmatrix} 
-\frac{b}{2} & m_2\\ \al' & m_1\end{smallmatrix}\right]
G_{\alpha'm_3}\left[\begin{smallmatrix} 
m_3'&-\frac{b}{2}\\ m_4 & \al''\end{smallmatrix}\right]
\left(B_{m_3'}^{-\frac{b}{2},m_3}\right)^{-2}
G^{-1}_{m_3'0}\left[\begin{smallmatrix} 
m_3&-\frac{b}{2}\\m_3&-\frac{b}{2}\end{smallmatrix}\right]\,,
\label{actionsphere}
\end{aligned}
\eeq
where 
\beq
\begin{aligned}
\alpha'=\alpha-\frac{sb}{2}\,,\qquad
\alpha''=\alpha-\frac{(s+s')b}{2}\,,\qquad
m_2'=m_2-\frac{s_2b}{2}\,,\qquad
m_3'=m_3-\frac{s_3b}{2}\,,
\end{aligned}
\eeq
and $s,s',s_2$ and $s_3$ are either $1$ or $-1$.

The loop operator corresponding to the curve traced by this move -- denoted 
by $\cL(\gamma_{2,0})$ -- acts on the conformal blocks on the four punctured 
sphere, which we label by $\cG_\alpha$ (we omit the labels $m_1,\ldots,m_4$ 
to avoid unnecessary clutter). The resulting monodromy 
is represented by the expression (\ref{actionsphere}) multiplying the conformal 
blocks, where the intermediate 
state $\alpha$ is replaced by $\alpha''=\alpha-kb/2$. 
We get the sum of three terms for $k=2,0,-2$
where compared to \eqn{actionsphere} we use $T_2=T_{++}$, 
$T_0=T_{+-}+T_{-+}$ and $T_{-2}=T_{--}$
\beq
[\cL(\gamma_{2,0})\cdot \cG]_\alpha
=T_2\,\cG_{\alpha-b}
+T_0\,\cG_{\alpha}
+T_{-2}\,\cG_{\alpha+b}\,.
\label{sphere-thooft}
\eeq
Explicitly the prefactors are 
\beq
\begin{aligned}
T_2&=
-4\frac{\sin\left(\pi b^2\right)
\sin\left(\pi b(\alpha+m_2-m_1-b)\right)
\sin\left(\pi b(\alpha-m_2+m_1-b)\right)}
{\sin\left(2\pi b^2\right)\sin\left(2\pi b(\alpha-b)\right)
\sin\left(\pi b(2\alpha-b)\right)}
\\&\hskip2cm\times
\sin\left(\pi b(\alpha+m_3-m_4-b)\right)
\sin\left(\pi b(\alpha-m_3+m_4-b)\right)
\end{aligned}
\eeq
\beq
\begin{aligned}
T_0&=
\frac{\left[\cos\left(\pi b(2\alpha-b)\right)\cos\left(\pi b(2m_2-b)\right)
-\cos\left(\pi b^2\right)\cos\left(\pi b(2m_1-b)\right)\right]}
{\cos^2\left(\pi b^2\right)\sin\left(2\pi b(\alpha-b)\right)
\sin\left(2\pi b\alpha\right)}
\\&\hskip1cm\times
\left[\cos\left(\pi b(2\alpha-b)\right)\cos\left(\pi b(2m_3-b)\right)
-\cos\left(\pi b^2\right)\cos\left(\pi b(2m_4-b)\right)\right]
\\&\hskip.5cm
+\frac{\cos\left(\pi b(2m_2-b)\right)\cos\left(\pi b(2m_3-b)\right)}
{\cos^2\left(\pi b^2\right)}
\end{aligned}
\label{T+-T-+}
\eeq
\beq
\begin{aligned}
T_{-2}&=
-4\frac{\sin\left(\pi b^2\right)
\sin\left(\pi b(\alpha+m_2+m_1-b)\right)
\sin\left(\pi b(\alpha-m_2-m_1-b)\right)}
{\sin\left(2\pi b^2\right)\sin\left(2\pi b\alpha\right)
\sin\left(\pi b(2\alpha-b)\right)}
\\&\hskip2cm\times
\sin\left(\pi b(\alpha+m_3+m_4-b)\right)
\sin\left(\pi b(\alpha-m_3-m_4-b)\right)
\end{aligned}
\eeq

In section \ref{sec:algebra} we provide some consistency checks 
for the validity of this calculation by demonstrating that the Wilson 
and 't~Hooft loop operators satisfy the expected 't~Hooft commutation 
relation for $b=1$ and the expected operator product expansion for $b=0$.

In the limit $b\to1$, where the Liouville correlator has an interpretation 
in term of the ${\cal N}=2$ $SU(2)$ gauge theory with $N_F=4$ on $S^4$ we get
\beq
\big\langle\CL(\gamma_{2,0})\big\rangle_{C_{g,n}}^{}
\,= \int d\alpha \sin^2(2\pi \alpha)\left[\overline\cG_{\alpha}\,T'_2\cG_{\alpha-1}
+\overline\cG_{\alpha}\,T'_0\cG_{\alpha}
+\overline \cG_{\alpha}\,T'_{-2}\cG_{\alpha+1}\right]\,,
\eeq
where 
\beq
\begin{aligned}
T'_2&=
\frac{2}{\sin^2\left(2\pi \alpha\right)}
\sin\left(\pi (\alpha+m_2-m_1)\right)
\sin\left(\pi (\alpha-m_2+m_1)\right)
\\&\hskip2cm \times
\sin\left(\pi (\alpha+m_3-m_4)\right)
\sin\left(\pi (\alpha-m_3+m_4)\right)
\end{aligned}
\eeq
\beq
\begin{aligned}
T'_0&=
\frac{\left[\cos\left( 2\pi\alpha\right)\cos\left(2\pi m_2 \right)
- \cos\left(2\pi m_1 )\right)\right]\left[\cos\left( 2\pi\alpha\right)\cos\left(2\pi m_3 \right)
- \cos\left(2\pi m_4 )\right)\right] }
{\sin^2\left(2\pi \alpha\right)}
\\&\hskip.5cm
+ \cos\left(2\pi m_2\right)\cos\left(2\pi m_3)\right)
\end{aligned}
\eeq
\beq
\begin{aligned}
T'_{-2}&=
-\frac{2}{\sin^2\left(2\pi \alpha\right)}
\sin\left(\pi (\alpha+m_2+m_1)\right)
\sin\left(\pi (\alpha-m_2-m_1)\right)
\\&\hskip2cm \times
\sin\left(\pi (\alpha+m_3+m_4)\right)
\sin\left(\pi (\alpha-m_3-m_4)\right)
\end{aligned}
\eeq
The 't~Hooft loop is given by a sum of three terms, one where the 
argument of the holomorphic contribution is shifted by $1$, the 
other by $-1$ and one is unshifted. In section \ref{sec:discuss}, 
we provide a possible gauge theory interpretation of this observation.

\subsection{More General Loop Operators}
\label{sec:ell}

We have in previous subsections demonstrated the prescription of 
calculating the Liouville loop operator $\cL(\gamma)$. This procedure 
can be applied to any collection of non-intersecting curves on the 
surface and by the conjectured relation between Liouville and 
the $\cN=2$ gauge theories (as well as the explicit check on Wilson 
loops) calculate general Wilson, 't~Hooft and dyonic loop operators 
in these gauge theories.

As reviewed in Section~\ref{sec:review}, the classification of non-intersecting 
curves on the Riemann surface exactly matches the possible charges 
carried by loop operators in the gauge theory \cite{Drukker:2009tz}. 
But we pointed out in Section~\ref{sec:pert} and in 
Section~\ref{sec:wilson}, there is more than one 
loop operator in Liouville theory associated to a given set of charges.
The three cases we discussed are associated to the 
Wilson loop which is multi-wound, the multi-trace operator and the 
loop in a higher dimensional irreducible representation. 

The distinction between the different loop operators is simply a change of 
basis, which can be seen both in the gauge theory and 
in Liouville theory. The multi-trace loops are in a product of the spin $1/2$ 
representation, which can be decomposed into irreducible 
representations. Similarly, the algebra of the degenerate fields 
is the quantum deformation of $sl(2,\bR)$.

In this regard our analysis is complete, we found a Liouville 
operator for gauge theory operators with arbitrary charges, 
which should furnish a complete basis on both sides of the 
correspondence.

The main advantage of the operators built out of the monodromy 
of the $V_{1,2}$ degenerate field along non-intersecting curves is 
that this definition is modular invariant. The classification of loop 
operators matches Dehn's classification of the curves and 
the explicit transformation rules under a change of pants 
decomposition was presented by Penner \cite{MR743669}. 
Another technical simplification is the fact that the fusion rules 
for $V_{1,2}$ are the simplest.

There could be other bases which have other advantages and 
we present now a proposal for one. It has the advantage that 
it is local on the graph $\Gamma_\sigma$ for a fixed pants 
decomposition $\sigma$ which simplifies the calculation in that duality 
frame. This comes clearly at the price of not having simple modular 
properties. But such a description is likely to match specific 
gauge theory descriptions in the corresponding weakly coupled 
limit of $\cT_{g,n}$.

We will consider loop operators made by splitting 
higher degenerate fields into collections of lower degenerate fields 
and fusing them back together along the Riemann surface. In particular 
given a pants decomposition $\sigma$, it decomposes the 
curve into arcs on the different pairs of pants, which is in fact a key step in 
Dehn's theorem. This operation groups different segments of 
the same curve as being parallel to each other on $\Gamma_\sigma$. 
It is then rather natural to replace a collection of parallel segments, 
with a single line for a higher degenerate field.%
\footnote{A similar idea, of allowing lines to fuse and split, 
was presented to us by D. Gaiotto, 
in the context of generalizing \cite{Drukker:2009tz} to theories 
with gauge group factors $SU(N)$ with $N>2$.}

Consider the case of the 't~Hooft loop in the theory with ${\cal N}=2$ $SU(2)$ gauge 
group with
$N_F=4$ discussed in Section~\ref{sec:sphere}. The loop is 
labeled by $\gamma_{2,0}$ and passes twice through the 
central edge in $\Gamma_\sigma$ (with the label $\alpha$ in 
Figure~\ref{fig:4puncture}). In fact one can consider a calculation 
where both of the degenerate fields are transported around the 
$m_3$, fused together, transported along the central edge, 
unfused into a pair of $V_{1,2}$ fields which then wrap the $m_2$ line.

Note that the fusion of the two degenerate fields will give the sum 
of the identity state and $V_{1,3}$. The contribution of the identity 
state is independent of $\alpha$, of $m_1$ and of $m_4$ and is 
the term on the last line of \eqn{T+-T-+}. The $V_{1,3}$ state gives 
all the other terms.

This decomposition of the loop operator into the contribution of the 
identity and $V_{1,3}$ along the intermediate edge should be mirrored 
by a decomposition of the analogous 't~Hooft operator in the gauge theory 
into a triplet and a signlet. This decomposition is rather subtle and 
as we saw in the Liouville calculation requires the choice of a pair 
of punctures with labels $m_2$ and $m_3$ (similarly this can be done with 
the pair $m_1$ and $m_4$). As mentioned before, a fully detailed 
construction of loop operators in these gauge theories does not 
exist and the subtleties in the reduction to irreducible representations 
have not been addressed. We do find some indication for this decomposition 
in the algebra of loop operators in Section~\ref{sec:algebra}.

\subsubsection{Building Loops from Smaller Constituents}

We would like to propose an algorithm that enables a relatively 
easy calculation of loop operators where all the arcs of the loop in 
a given pants decomposition are fused to higher degenerate fields 
(and roughly speaking should correspond to Wilson and 
't~Hooft loop operators in irreducible representations).

\begin{figure}[ht]
\begin{center}
\begin{tabular}{ccc}
\includegraphics[scale=.2]{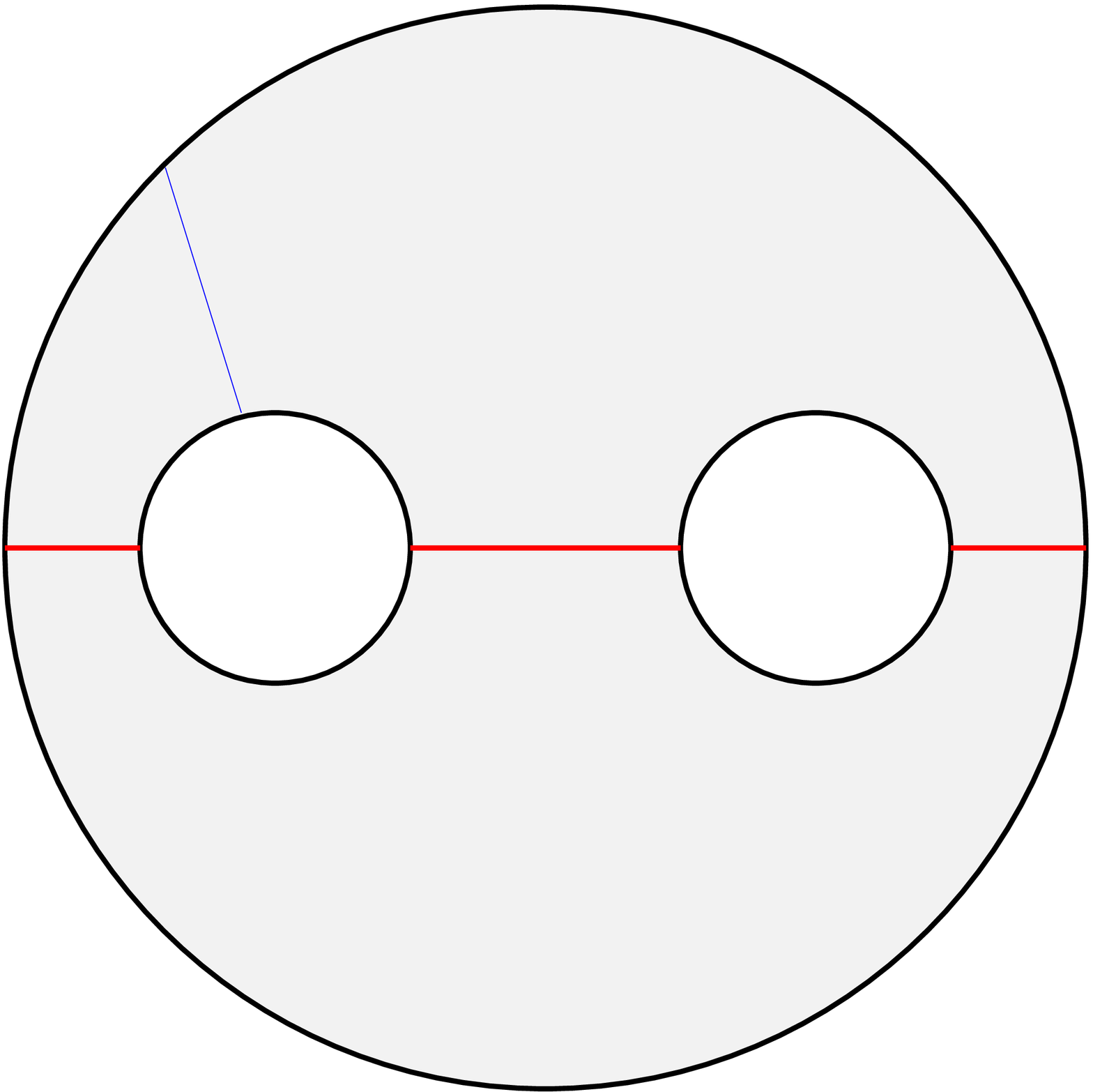}&
\includegraphics[scale=.2]{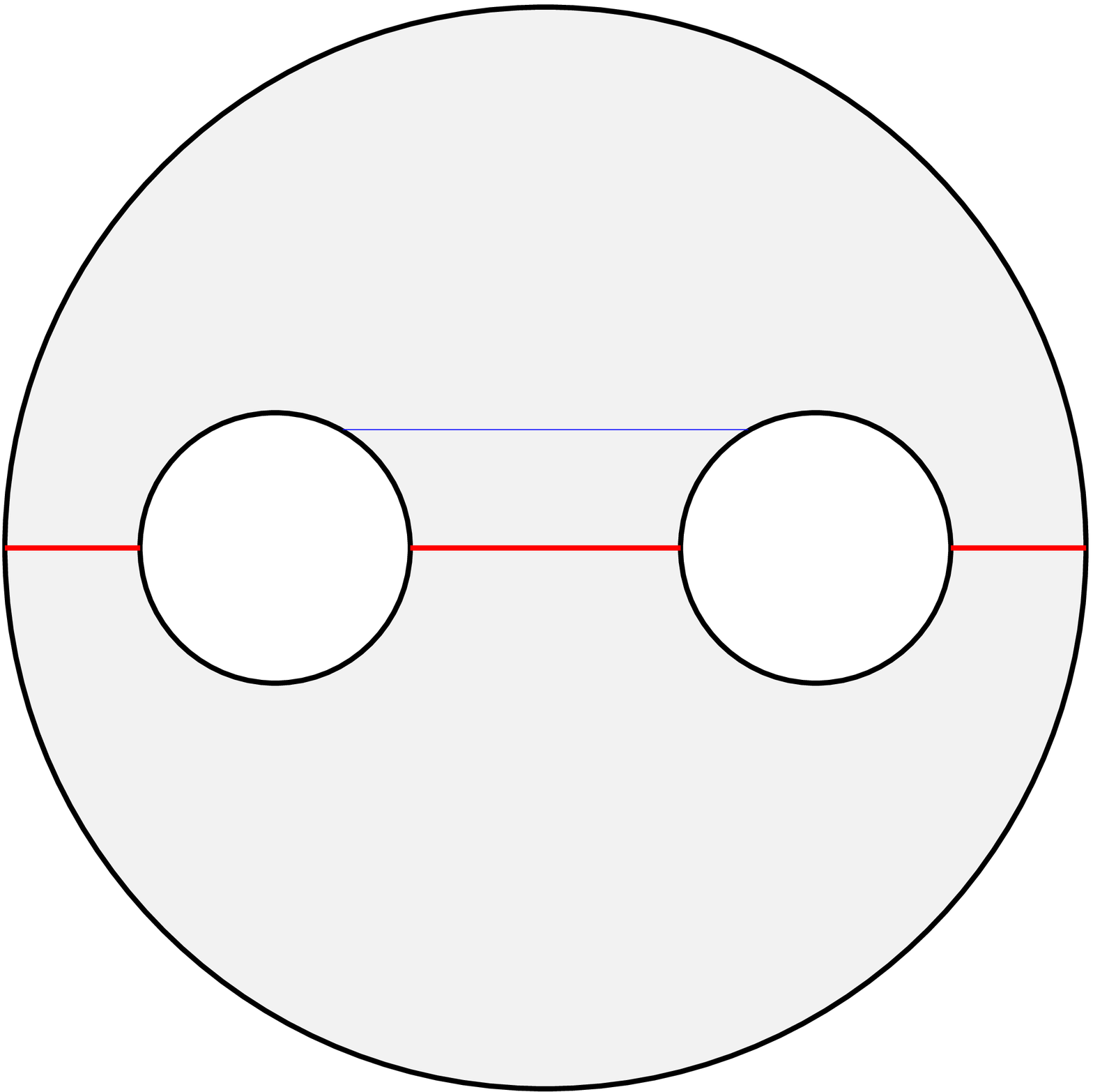}&
\includegraphics[scale=.2]{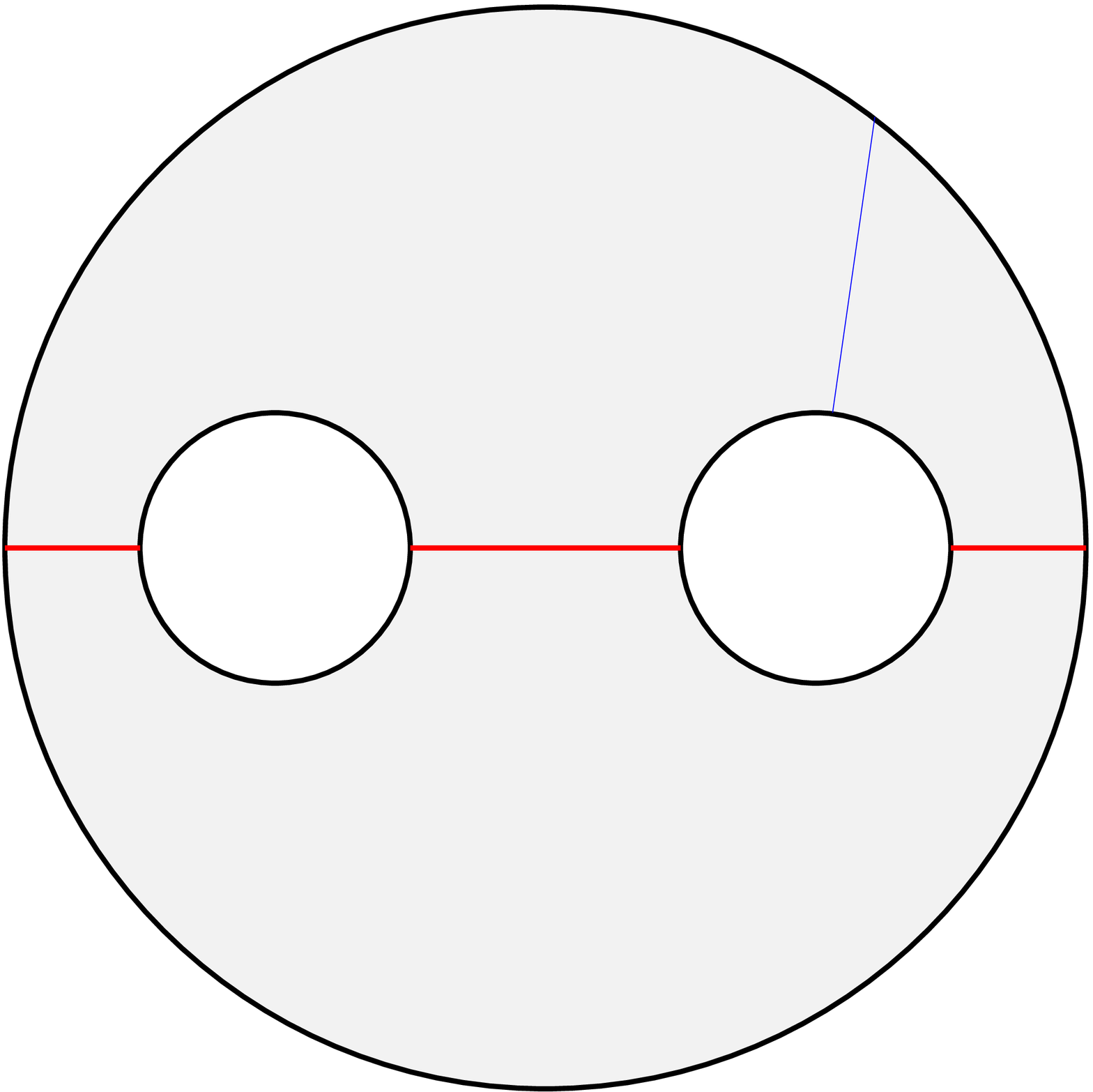}\\
$\ell_{12}$&$\ell_{23}$&$\ell_{13}$\\
\\
\includegraphics[scale=.2]{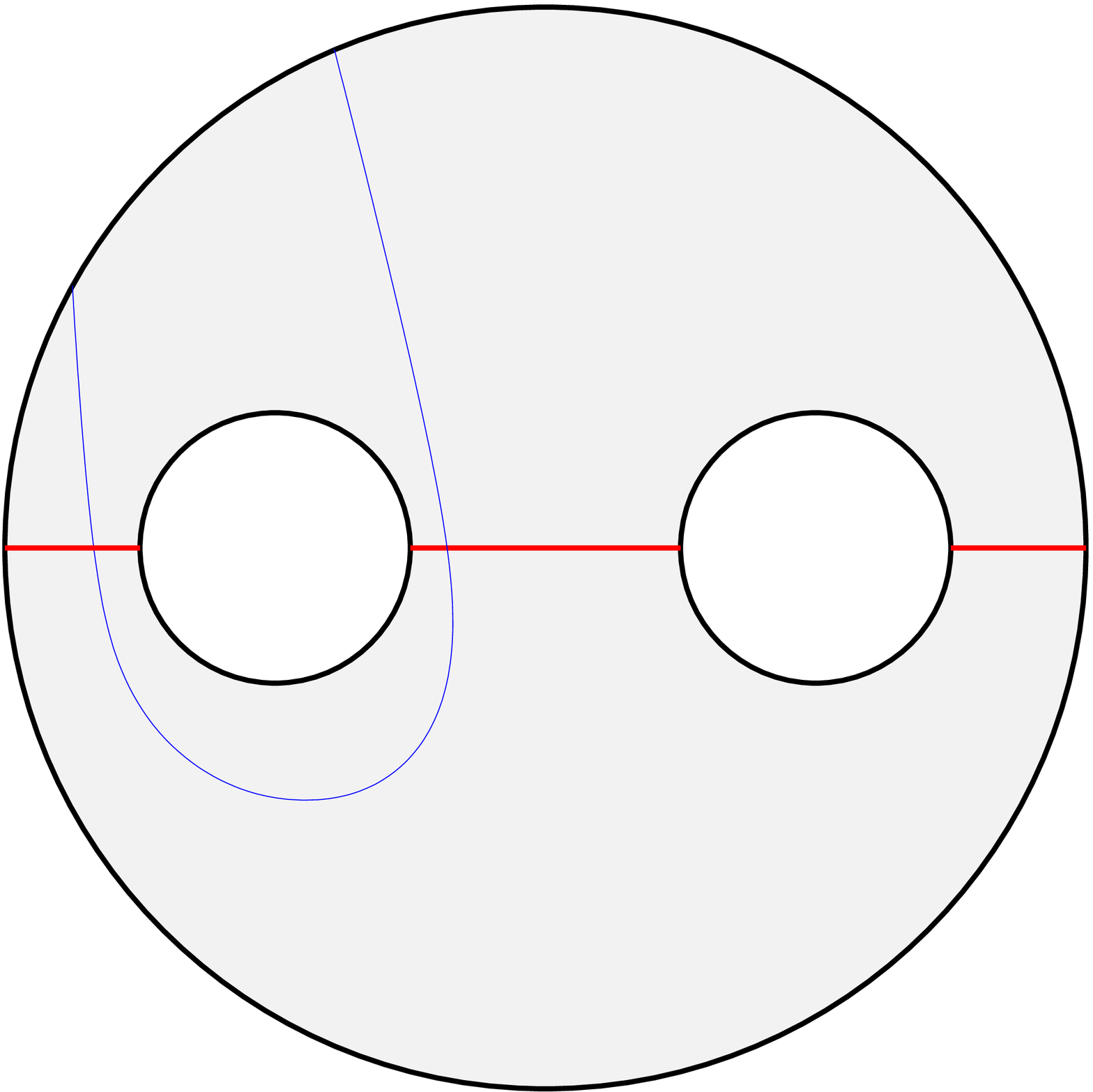}&
\includegraphics[scale=.2]{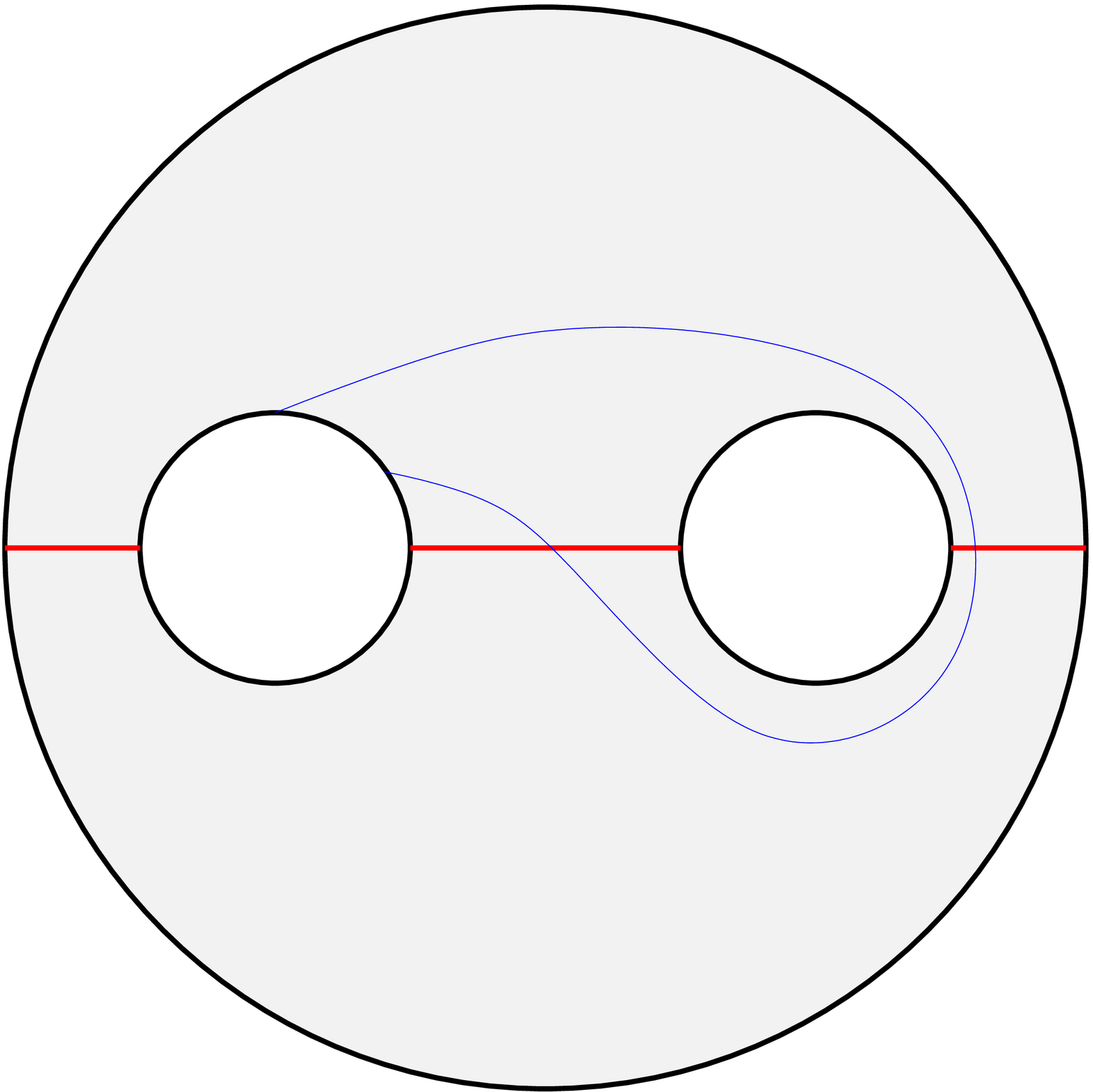}&
\includegraphics[scale=.2]{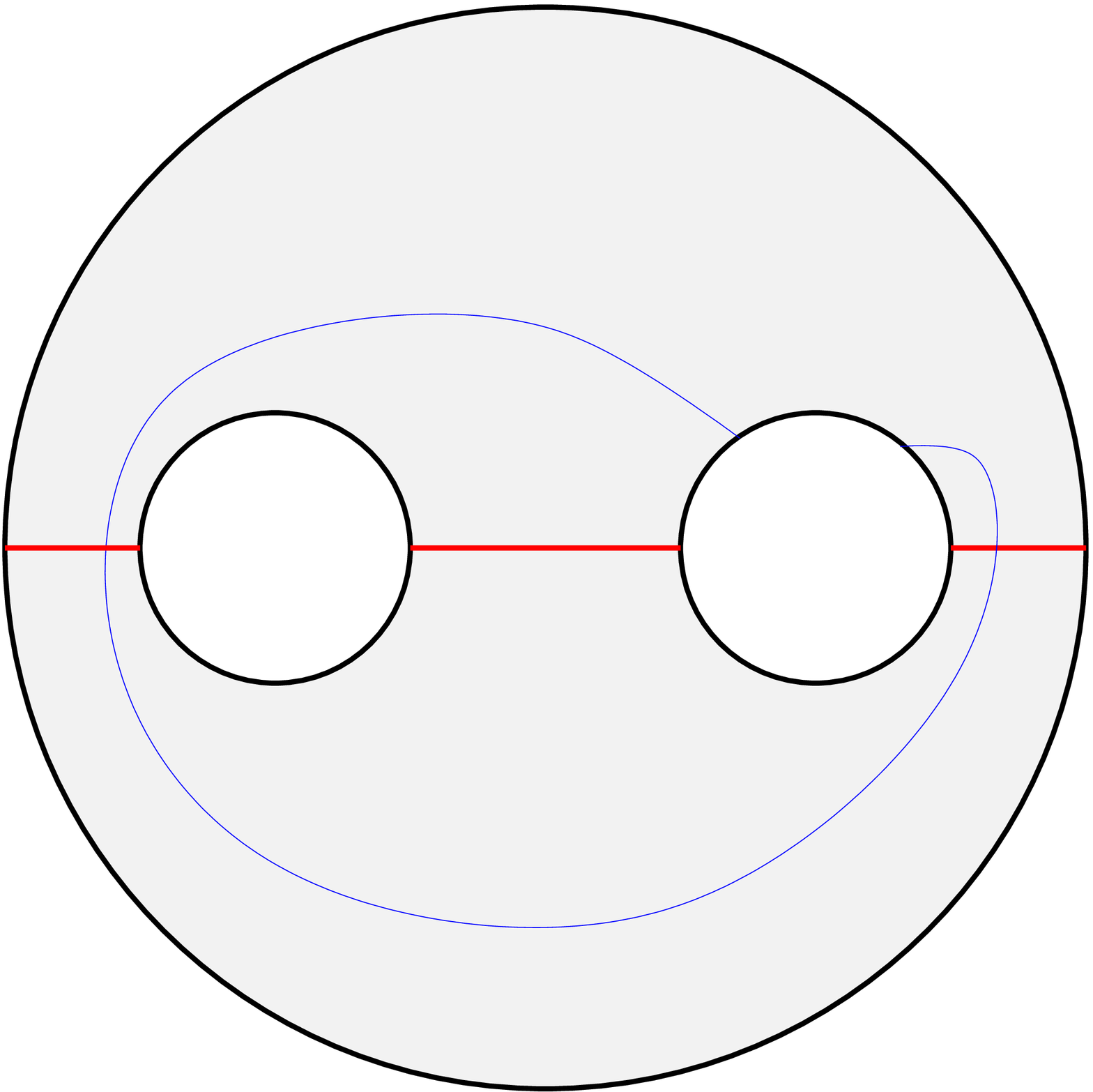}\\
$\ell_{11}$&$\ell_{22}$&$\ell_{33}$
\end{tabular}
\caption{Basic arcs on a pair-of-pants}\label{fig:arcs}
\end{center}
\end{figure}

We recall the statement of Dehn's theorem, which analyzes curves 
on surfaces with specific pants decomposition. On each pair of 
pants there will be a collection of arcs going between the three 
boundaries. 
By a homotopy one can guarantee that all the arcs on all the pairs of pants 
will start and end on the upper half of the boundary circles 
as shown in Figure~\ref{fig:arcs}, and follow one of these six basic paths.

Given the number of lines crossing the three boundaries of the pair of pants, 
$p_1$, $p_2$ and $p_3$, there is a unique collection 
of non-intersection arcs as in Figure~\ref{fig:arcs} with these number of 
endpoints.

If $p_i > p_j+p_k$, then we can use
\beq
\frac12(p_i-p_j-p_k)\ell_{ii} + p_j \ell_{ij}+p_k \ell_{ik}.
\eeq
On the other hand, if $p_j+p_k\ge p_i$ for all permutations of $\{1,2,3\}$,
then we can use
\beq
\frac12(p_1+p_2-p_3)\ell_{12} + \frac12(p_1+p_3-p_2)\ell_{13} +
\frac12(p_2+p_3-p_1)\ell_{23}.
\eeq

Instead of carrying the $V_{1,2}$ field along an arc and then on to another pair 
of pants, we want to treat all the arcs on this pair of pants in unison. 
For that we replace the $p_i$ lines crossing the boundary $i$ by a single 
line for the degenerate field $V_{1,p+1}$. We then consider the 
splitting and joining of these operators from one boundary to the other 
according to the multiplicities of the $\ell_{ij}$ curves.

More concretely we start by introducing pairs of degenerate fields 
$V_{1,p_i+1}$ at the center of the $i^\text{th}$ edge which are 
connected to the original graph through a handle carrying the identity 
state. If $q_i\neq0$, we split off a $V_{1,q_i+1}$ field, braid it around 
and refuse it back to a pair of $V_{1,p_i+1}$. This prescription is valid for 
$p_i\geq q_i$, otherwise we start with a pair of $V_{1,q_i+1}$, braid 
one around, split each to a pair $V_{1,p_i+1}$ and $V_{1,q_i-p_i+1}$ 
and combine the latter two and project the intermediate state to the identity.

We then have exactly the situation envisioned above, where at each 
boundary of a pair of pants there is an extra degenerate field with 
$V_{1,p_i+1}$. They are split and fused with each-other according to 
the degeneracies of the $\ell_{ij}$ lines and projected on to the identity.

This prescription can be used in the calculation of the loop 
operators discussed in the preceding subsections. In the case of the 
one-punctured torus, there is a single pair of pants on which 
$p_1=p_2=1$ and $p_3=0$. 
The relevant curve is $\ell_{12}$. We define first the operator 
$\hat w_{1,0}$ as a map
from the space of conformal blocks $\cC(C_{g,n})$ 
to the space of conformal blocks with an extra pair of 
degenerate fields $\hat\cC(C_{g,n+2})$, which is 
the inverse of the canonical identification and then
places them at the two sides of the glued pairs of pants. In 
our case the two degenerate fields end up on the edges 
labeled 1 and 2 of the three-leg graph representing the pair of pants.

We next consider the operator $\hat\ell_{12}$ acting on the enlarged space of 
conformal blocks on a single pair of pants, 
moves one degenerate field from boundary number 2 to boundary number 
1, combines the two and projects down back to the original space of 
conformal blocks.

Explicitly we can illustrate the action of $\hat\ell_{12}$ by the 
fusion and projection steps (in the case of the torus we also 
identify $\alpha_1=\alpha_2$, but for now we keep it generic)
\beq
\raisebox{0mm}[12mm][12mm]{\parbox{38mm}{\begin{center}
\begin{fmfgraph*}(30,20)
\fmfbottom{bl,br}
\fmftop{tl,tr}
\fmf{plain,width=2}{bl,b1,b2,b3,b4,br}
\fmf{phantom}{tl,t1,t2,t3,t4,tr}
\fmffreeze
\fmf{boson,width=.5}{t1,b1}
\fmf{plain,width=2}{t4,c4,cc4,b4}
\fmffreeze
\fmf{boson,width=.5}{c4,t2}
\fmfv{label=$\alpha_1'$,label.angle=-90,label.dist=5}{bl}
\fmfv{label=$\alpha_2'$,label.angle=0}{t4}
\fmfv{label=$\alpha_3$,label.angle=-90,label.dist=8}{br}
\fmfv{label=$\alpha_1$,label.angle=-90,label.dist=8}{b2}
\fmfv{label=$\alpha_2$,label.angle=0}{cc4}
\fmfv{label=$-\frac{b}{2}$,label.angle=180}{t1}
\fmfv{label=$-\frac{b}{2}$,label.angle=-110}{t2}
\end{fmfgraph*}\end{center}}}
\to
\raisebox{0mm}[12mm][12mm]{\parbox{41mm}{\begin{center}
\begin{fmfgraph*}(35,20)
\fmfbottom{bl,br}
\fmftop{tl,tr}
\fmf{plain,width=2}{bl,b1,b2,b3,b4,b5,br}
\fmf{phantom}{tl,t1,t2,t3,t4,t5,tr}
\fmffreeze
\fmf{boson,width=.5}{t1,b1}
\fmf{boson,width=.5}{t3,b3}
\fmf{plain,width=2}{t5,b5}
\fmfv{label=$\alpha_1'$,label.angle=-90,label.dist=5}{bl}
\fmfv{label=$\alpha_2'$,label.angle=0}{t5}
\fmfv{label=$\alpha_3$,label.angle=-90,label.dist=8}{br}
\fmfv{label=$\alpha_1$,label.angle=-90,label.dist=8}{b2}
\fmfv{label=$\alpha_1'$,label.angle=-90,label.dist=5}{b4}
\fmfv{label=$-\frac{b}{2}$,label.angle=180}{t1}
\fmfv{label=$-\frac{b}{2}$,label.angle=180}{t3}
\end{fmfgraph*}\end{center}}}
\to
\raisebox{0mm}[12mm][12mm]{\parbox{29mm}{\begin{center}
\begin{fmfgraph*}(25,20)
\fmfbottom{bl,br}
\fmftop{tl,tr}
\fmf{plain,width=2}{bl,b1,b2,b3,br}
\fmf{phantom}{tl,t1,t2,t3,tr}
\fmffreeze
\fmf{boson,width=.5}{tl,c1,t2}
\fmf{dashes, width=.5}{c1,b1}
\fmf{plain,width=2}{t3,b3}
\fmfv{label=$\alpha_1'$,label.angle=-90,label.dist=5}{bl}
\fmfv{label=$\alpha_2'$,label.angle=0}{t3}
\fmfv{label=$\alpha_3$,label.angle=-90,label.dist=8}{br}
\end{fmfgraph*}\end{center}}}
\to
\raisebox{0mm}[12mm][12mm]{\parbox{24mm}{\begin{center}
\begin{fmfgraph*}(20,20)
\fmfbottom{bl,br}
\fmftop{tl,tr}
\fmf{plain,width=2}{bl,b1,b2,b3,br}
\fmf{phantom}{tl,t1,t2,t3,tr}
\fmffreeze
\fmf{plain,width=2}{t2,b2}
\fmfv{label=$\alpha_1'$,label.angle=-90,label.dist=5}{bl}
\fmfv{label=$\alpha_2'$,label.angle=0}{t2}
\fmfv{label=$\alpha_3$,label.angle=-90,label.dist=8}{br}
\end{fmfgraph*}\end{center}}}
\eeq
Taking $\alpha_1'=\alpha_1-s_1b/2$ and 
$\alpha_2'=\alpha_2-s_2b/2$ with $s_{i}=\pm1$
we have for the renormalized conformal blocks $\cG$ \eqn{def-cG} 
\beq
[\hat\ell_{12}\cdot\hat\cG]_{\alpha_1,\alpha_2,\alpha_3}=
G_{\alpha_2,\al_1'}^{-1}\left[\begin{smallmatrix} 
-\frac{b}{2} & \alpha_2'\\ \al_1 & \al_3\end{smallmatrix}\right]
G_{\al_1,0}\left[\begin{smallmatrix} 
-\frac{b}{2} &-\frac{b}{2} \\
\al_1' & \al_1'\end{smallmatrix}\right]\cG_{\alpha_1',\alpha_2',\alpha_3}\,.
\eeq

Using the expressions \eqn{Gpm} and \eqn{fuse-proj} in Appendix B this is
\beq
\begin{aligned}[]
[\hat\ell_{12}\cdot\hat\cG]_{\alpha_1,\alpha_2,\alpha_3}
&=s_1s_2\,
\frac{\sin(\pi b(\alpha_3-\frac{Q}{2}+s_1s_2\frac{Q}{2}
-s_1(\alpha_2-\frac{Q}{2})
-s_2(\alpha_1-\frac{Q}{2}))}
{\sin(2\pi b^2)}
\,\cG_{\alpha_1',\alpha_2',\alpha_3}
\end{aligned}
\label{ell12}
\eeq
Similarly one can define analogous operators $\hat\ell_{23}$ and 
$\hat\ell_{31}$ which are given by permuting the indices, and these 
operations for other degenerate fields.

Using this operator we can immediately reproduce the 't~Hooft loop on the 
one-punctured torus calculated in Section~\ref{sec:torus}. We take 
$\alpha_1=\alpha_2=\alpha$ and $\alpha_3=m$. Then we have to act on the 
conformal block with $\hat w_{1,0}$, which will introduce $s=s_1=s_2$ 
and then act by $\hat\ell_{12}$
\beq
\cG_{\alpha,\alpha_3}
\quad\to\quad
[\hat w_{1,0}\cdot\cG]_{\alpha,m}=\sum_{s=\pm}\hat\cG^s_{\alpha',\alpha,m}
\quad\to\quad
[\hat\ell_{12}\cdot\hat w_{1,0}\cdot\cG]_{\alpha,m}=
\sum_{s=\pm}\cG_{\alpha',m}\,.
\eeq

Using \eqn{ell12} and including the action of $\hat w_{1,0}$ \eqn{w10} we get
\beq
\begin{aligned}
\big[\hat\ell_{12}\cdot\hat w_{1,0}\cdot\cG\big]_{\alpha,m}
&=\sum_{s=\pm}
\frac{\sin(\pi b(m-s(2\alpha-Q)))}
{\sin(2\pi b^2)}
\times
s\,\frac{\sin\left(\pi b^2\right)}
{\sin\left(\pi b(2\alpha-Q)\right)}
\,\cG_{\alpha-sb/2,m}
\\
&=
\sum_{s=\pm}
\frac{\sin(\pi b^2)}{\sin(2\pi b^2)}
\frac{\sin(\pi b(2\alpha-Q-sm)}{\sin(\pi b(2\alpha-Q))}
\,\cG^s_{\alpha-sb/2,m}\,.
\end{aligned}
\eeq
This indeed agrees precisely with \eqn{thooft-G}.

In the case of the four-punctured  sphere  there are two pairs of pants. The 
loop considered in Section~\ref{sec:sphere} has $p=2$ on one boundary of 
each pair of pants. In our present approach of doing the calculation with the 
highest possible degenerate fields we need to create a pair of 
$V_{1,3}$ fields with $\hat w_{2,0}$.

Then on each pair of pants we need to act by 
the operator $\hat\ell_{11}$ which takes the $V_{1,3}$ degenerate field 
attached to the first edge, replace it by a pair of $V_{1,2}$ fields, transports 
one of them around the second edge, fusing it back to the other
degenerate field and projecting on to the identity
\beq
\begin{aligned}
\raisebox{0mm}[10mm][15mm]{\parbox{29mm}{\begin{center}
\begin{fmfgraph*}(25,20)
\fmfbottom{bl,br}
\fmftop{tl,tr}
\fmf{plain,width=2}{bl,b1,b2,b3,b4,br}
\fmf{phantom}{tl,t1,t2,t3,t4,tr}
\fmffreeze
\fmf{boson,width=.5}{tl,c1,t2}
\fmf{boson,width=1}{c1,b1}
\fmf{plain,width=2}{t4,b4}
\fmfv{label=$\alpha_1''$,label.angle=-90,label.dist=5}{bl}
\fmfv{label=$\alpha_2$,label.angle=0}{t4}
\fmfv{label=$\alpha_3$,label.angle=-90,label.dist=8}{br}
\fmfv{label=$\alpha_1$,label.angle=-90,label.dist=8}{b2}
\fmfv{label=$-\frac{b}{2}$,label.angle=180}{tl}
\fmfv{label=$-\frac{b}{2}$,label.angle=0}{t2}
\fmfv{label=$-b$,label.angle=-135}{c1}
\end{fmfgraph*}\end{center}}}
&\to
\raisebox{0mm}[10mm][15mm]{\parbox{36mm}{\begin{center}
\begin{fmfgraph*}(30,20)
\fmfbottom{bl,br}
\fmftop{tl,tr}
\fmf{plain,width=2}{bl,b1,b2,b3,b4,b5,br}
\fmf{phantom}{tl,t1,t2,t3,t4,t5,tr}
\fmffreeze
\fmf{boson,width=.5}{t1,b1}
\fmf{boson,width=.5}{t3,b3}
\fmf{plain,width=2}{t5,b5}
\fmfv{label=$\alpha_1''$,label.angle=-90,label.dist=5}{bl}
\fmfv{label=$\alpha_2$,label.angle=0}{t5}
\fmfv{label=$\alpha_3$,label.angle=-90,label.dist=8}{br}
\fmfv{label=$\alpha_1'$,label.angle=-90,label.dist=5}{b2}
\fmfv{label=$\alpha_1$,label.angle=-90,label.dist=8}{b4}
\fmfv{label=$-\frac{b}{2}$,label.angle=180}{t1}
\fmfv{label=$-\frac{b}{2}$,label.angle=180}{t3}
\end{fmfgraph*}\end{center}}}
\to
\raisebox{0mm}[13mm][13mm]{\parbox{33mm}{\begin{center}
\begin{fmfgraph*}(25,20)
\fmfbottom{bl,br}
\fmftop{tl,tr}
\fmf{plain,width=2}{bl,b1,b2,b3,b4,br}
\fmf{phantom}{tl,t1,t2,t3,t4,tr}
\fmffreeze
\fmf{boson,width=.5}{t1,b1}
\fmf{plain,width=2}{t4,c4,cc4,b4}
\fmffreeze
\fmf{boson,width=.5}{c4,t2}
\fmfv{label=$\alpha_1''$,label.angle=-90,label.dist=5}{bl}
\fmfv{label=$\alpha_2$,label.angle=0}{t4}
\fmfv{label=$\alpha_3$,label.angle=-90,label.dist=8}{br}
\fmfv{label=$\alpha_1'$,label.angle=-90,label.dist=5}{b2}
\fmfv{label=$\alpha_2'$,label.angle=0}{cc4}
\fmfv{label=$-\frac{b}{2}$,label.angle=180}{t1}
\fmfv{label=$-\frac{b}{2}$,label.angle=-110}{t2}
\end{fmfgraph*}\end{center}}}
\to
\raisebox{0mm}[13mm][13mm]{\parbox{35mm}{\begin{center}
\begin{fmfgraph*}(25,20)
\fmfbottom{bl,br}
\fmftop{tl,tr}
\fmf{plain,width=2}{bl,b1,b2,b3,b4,br}
\fmf{phantom}{tl,t1,t2,t3,t4,tr}
\fmffreeze
\fmf{boson,width=.5}{t1,b1}
\fmf{plain,width=2}{t3,c3,cc3,b3}
\fmffreeze
\fmf{boson,width=.5}{c3,tr}
\fmfv{label=$\alpha_1''$,label.angle=-90,label.dist=5}{bl}
\fmfv{label=$\alpha_2$,label.angle=180}{t3}
\fmfv{label=$\alpha_3$,label.angle=-90,label.dist=8}{br}
\fmfv{label=$\alpha_1'$,label.angle=-90,label.dist=5}{b2}
\fmfv{label=$\alpha_2'$,label.angle=0}{cc3}
\fmfv{label=$-\frac{b}{2}$,label.angle=180}{t1}
\fmfv{label=$-\frac{b}{2}$,label.angle=-70}{tr}
\end{fmfgraph*}\end{center}}}
\\&\hskip-2cm
\to
\raisebox{0mm}[13mm][13mm]{\parbox{33mm}{\begin{center}
\begin{fmfgraph*}(25,20)
\fmfbottom{bl,br}
\fmftop{tl,tr}
\fmf{plain,width=2}{bl,b1,b2,b3,b4,br}
\fmf{phantom}{tl,t1,t2,t3,t4,tr}
\fmffreeze
\fmf{boson,width=.5}{t1,b1}
\fmf{plain,width=2}{t4,c4,cc4,b4}
\fmffreeze
\fmf{boson,width=.5}{c4,t2}
\fmfv{label=$\alpha_1''$,label.angle=-90,label.dist=5}{bl}
\fmfv{label=$\alpha_2$,label.angle=0}{t4}
\fmfv{label=$\alpha_3$,label.angle=-90,label.dist=8}{br}
\fmfv{label=$\alpha_1'$,label.angle=-90,label.dist=5}{b2}
\fmfv{label=$\alpha_2'$,label.angle=0}{cc4}
\fmfv{label=$-\frac{b}{2}$,label.angle=180}{t1}
\fmfv{label=$-\frac{b}{2}$,label.angle=-110}{t2}
\end{fmfgraph*}\end{center}}}
\to
\raisebox{0mm}[10mm][12mm]{\parbox{37mm}{\begin{center}
\begin{fmfgraph*}(30,20)
\fmfbottom{bl,br}
\fmftop{tl,tr}
\fmf{plain,width=2}{bl,b1,b2,b3,b4,b5,br}
\fmf{phantom}{tl,t1,t2,t3,t4,t5,tr}
\fmffreeze
\fmf{boson,width=.5}{t1,b1}
\fmf{boson,width=.5}{t3,b3}
\fmf{plain,width=2}{t5,b5}
\fmfv{label=$\alpha_1''$,label.angle=-90,label.dist=5}{bl}
\fmfv{label=$\alpha_2$,label.angle=0}{t5}
\fmfv{label=$\alpha_3$,label.angle=-90,label.dist=8}{br}
\fmfv{label=$\alpha_1'$,label.angle=-90,label.dist=5}{b2}
\fmfv{label=$\alpha_1''$,label.angle=-90,label.dist=5}{b4}
\fmfv{label=$-\frac{b}{2}$,label.angle=180}{t1}
\fmfv{label=$-\frac{b}{2}$,label.angle=180}{t3}
\end{fmfgraph*}\end{center}}}
\to
\raisebox{0mm}[10mm][12mm]{\parbox{24mm}{\begin{center}
\begin{fmfgraph*}(20,20)
\fmfbottom{bl,br}
\fmftop{tl,tr}
\fmf{plain,width=2}{bl,b1,b2,b3,br}
\fmf{phantom}{tl,t1,t2,t3,tr}
\fmffreeze
\fmf{boson,width=.5}{tl,c1,t2}
\fmf{dashes,width=.5}{c1,b1}
\fmf{plain,width=2}{t3,b3}
\fmfv{label=$\alpha_1''$,label.angle=-90,label.dist=5}{bl}
\fmfv{label=$\alpha_2$,label.angle=0}{t3}
\fmfv{label=$\alpha_3$,label.angle=-90,label.dist=8}{br}
\fmfv{label=$\alpha_1''$,label.angle=-90,label.dist=5}{b2}
\end{fmfgraph*}\end{center}}}
\to
\raisebox{0mm}[10mm][12mm]{\parbox{19mm}{\begin{center}
\begin{fmfgraph*}(15,20)
\fmfbottom{bl,br}
\fmftop{tl,tr}
\fmf{plain,width=2}{bl,b1,b2,b3,br}
\fmf{phantom}{tl,t1,t2,t3,tr}
\fmffreeze
\fmf{plain,width=2}{t2,b2}
\fmfv{label=$\alpha_1''$,label.angle=-90,label.dist=5}{bl}
\fmfv{label=$\alpha_2$,label.angle=0}{t2}
\fmfv{label=$\alpha_3$,label.angle=-90,label.dist=8}{br}
\end{fmfgraph*}\end{center}}}
\end{aligned}
\eeq
To write it down we take $\alpha_1'=\alpha_1-s_1b/2$, 
$\alpha_1''=\alpha_1'-s_1'b/2$ and $\alpha_2'=\alpha_2-s_2b/2$. 
Then we have 
\beq
\begin{aligned}[]
[\hat\ell_{11}\cdot\hat\cG]_{\alpha_1,\alpha_2,\alpha_3}&=
\sum_{s_1,s_2=\pm}
G^{-1}_{-b,\al_1'}\left[\begin{smallmatrix} 
-\frac{b}{2} & -\frac{b}{2}\\ \al_1'' & \al_1\end{smallmatrix}\right]
G_{\alpha_1\al_2'}\left[\begin{smallmatrix} 
-\frac{b}{2} & \alpha_2\\ \al_1' & \al_3\end{smallmatrix}\right]
\left(B_{\alpha_2'}^{-\frac{b}{2},\alpha_2}\right)^2
\\&\hskip2cm\times
G_{\al_2'\alpha_1''}^{-1}\left[\begin{smallmatrix} 
-\frac{b}{2} & \alpha_2\\ \al_1' & \al_3\end{smallmatrix}\right]
G_{\al_1'0}\left[\begin{smallmatrix} 
-\frac{b}{2} & -\frac{b}{2}\\ \al_1'' & \al_1''\end{smallmatrix}\right]
\cG_{\alpha_1'',\alpha_2,\alpha_3}
\\
&=\sum_{s_1,s_2=\pm}
G_{+,s_1}\left[\begin{smallmatrix} 
\al_1 & -\frac{b}{2}\\ \al_1'' &-\frac{b}{2}\end{smallmatrix}\right]
G_{s_2,-s_1}\left[\begin{smallmatrix} 
\al_2 & -\frac{b}{2}\\ \al_3&\alpha_1'\end{smallmatrix}\right]
e^{\pi b i(Q+s_2(2\alpha_2-Q))}
\\&\hskip2cm\times
G_{s_1's_2}\left[\begin{smallmatrix} 
\alpha_1'&-\frac{b}{2}\\ \al_3&\al_2 \end{smallmatrix}\right]
G_{-,-s_1'}\left[\begin{smallmatrix} 
-\frac{b}{2}& -\frac{b}{2}\\ \al_1''&\al_1'' \end{smallmatrix}\right]
\cG_{\alpha_1'',\alpha_2,\alpha_3}
\end{aligned}
\label{ell11}
\eeq

To find an explicit expression one uses \eqn{Gpm} and \eqn{fuse-proj} and 
performs the sum over $s_1$ and $s_2$. To simplify the final expression 
it is convenient 
to consider separately the cases with $\alpha_1''=\alpha_1-kb/2$ with 
$k=-2,0,2$. For $k=\pm2$ we need to take $s_1=s_1'=\pm1$ while 
for $k=0$ we need to sum over the two possibilities of $s_1=-s_1'$. 
The result is 
\begin{align}
[\hat\ell_{11}\cdot\hat\cG]&_{\alpha_1,\alpha_2,\alpha_3}
=-2ie^{i\pi b^2}\frac{
\sin\left(\pi b(\alpha_1+\alpha_2-\alpha_3-Q)\right)
\sin\left(\pi b(\alpha_1-\alpha_2+\alpha_3-Q)\right)}
{\sin\left(2\pi b^2\right)}
\cG_{\alpha_1-b,\alpha_2,\alpha_3}
\nonumber\\
&+2ie^{i\pi b^2}
\frac{\cos\left(\pi b(2\alpha_1-b)\right)\cos\left(\pi b(2\alpha_2-b)\right)
-\cos\left(\pi b^2\right)\cos\left(\pi b(2\alpha_3-b)\right)}
{\sin\left(2\pi b^2\right)}
\cG_{\alpha_1,\alpha_2,\alpha_3}
\nonumber\\
&+2ie^{i\pi b^2}\frac{
\sin\left(\pi b(\alpha_1+\alpha_2+\alpha_3-Q)\right)
\sin\left(\pi b(\alpha_1-\alpha_2-\alpha_3+Q)\right)}
{\sin\left(2\pi b^2\right)}
\cG_{\alpha_1+b,\alpha_2,\alpha_3}
\end{align}

To complete the calculation of the loop operator on the four punctured sphere 
we have to include 
the action of $\hat w_{2,0}$ \eqn{1tob}, then take a pair of these 
$\ell_{11}$ operators, one for each pair of pants, and sum over the three 
different values of $k$.

This gives the same as the loop calculated in Section~\ref{sec:sphere} 
without the term on the last line of \eqn{T+-T-+}. Indeed as stated before, 
this last term corresponds to the contribution from the identity state, or the 
trivial 't~Hooft loop and the expression we derive here should correspond 
to the loop operator in an irreducible representation.

The operators $\hat\ell_{22}$ and $\hat\ell_{33}$ can be defined in a similar 
fashion, as can the operators for higher degenerate fields.

\section{The Liouville-Teichm\"uller theory}
\label{liou-teich}

\ 
Liouville theory has a dual representation in terms of a quantum mechanical problem,
the quantum theory of the Teichm\"uller spaces. 
We first explain the basic quantization problem for the Teichm\"uller spaces
based on their natural symplectic structures.
We also explain how to describe the length of geodesics
on a Riemann surface as functions on Teichm\"uller space.
Then we recall the connection between Liouville theory and quantum
Teichm\"uller theory established in \cite{Teschner:2005bz,Teschner:2003at}.
Quantum Teichm\"uller theory possesses canonical observables,
namely {\it geodesic length operators}. We show that
they are precisely the Liouville loop operators we introduced in
Section~\ref{sec:bootstrap}, up to rescaling.

\subsection{Classical Liouville-Teichm\"uller Theory on Riemann Surfaces}

\subsubsection{Complex-Analytic Picture}

Teichm\"uller space is the space of deformations of the complex structure
on a topological surface $\Sigma_{g,n}$. For each
complex structure on a surface $\Sigma_{g,n}$ there
exists a unique metric of the form 
\begin{equation}\label{metric}
ds^2\,=\,e^{\vf}dzd\bz\,,
\end{equation}
which has constant negative curvature. A
metric of the form \rf{metric}
will have constant negative curvature $-8\pi \mu b^2$ iff it satisfies
the classical Liouville equation
\begin{equation}\label{EOM}
\pa\bar\pa \vf\,=\,2\pi \mu b^2\, e^{\vf}\,.
\end{equation}
The classical Liouville field $\vf$ is related to the quantum field $\phi$ used earlier by $\vf=2b\phi$.
We may therefore identify the Teichm\"uller space $\CT_{g,n}$ with the space
of solutions of the Liouville equation on the surface $\Sigma_{g,n}$.

We will consider the space of all solutions to the classical Liouville equation
on a surface $\Sigma_{g,n}$
as the phase space $\CT_{g,n}$ we are aiming to quantize.
The Liouville action functional $S^{\rm cl}\big[\vf]$,
\begin{equation}
S^{\rm cl}\big[\vf]\;=\; \frac{1}{8\pi}
\int_{C_{g,n}}d^2z \,\Big(\,\frac{1}{2}(\pa_a\vf)^2+8\pi\mu b^2e^{\vf}\,\Big)
+[{\rm boundary\;\, terms}]\,,
\end{equation}
with choice of boundary terms as given in \cite{MR882831,MR889594},
defines a natural symplectic form $\om$ on $\CT_{g,n}$, 
\begin{equation}\label{omdef}
\omega\;=\;2\pi i\,\pa\bar{\pa}S^{\rm cl},
\end{equation}
where $\pa$, $\bar{\pa}$ are the holomorphic and anti-holomorphic
components of the de Rham differential on $\CT_{g,n}$ respectively.
The symplectic form $\om$ coincides \cite{MR882831,MR889594}
with the Weil-Petersson symplectic
form , which is natural
from the point of view of Teichm\"uller theory, $\om=\om_{\rm\sst WP}$.
The problem to
quantize Liouville theory on a surface $\Sigma_{g,n}$ is therefore equivalent to
the problem of quantizing Teichm\"uller space with Poisson-bracket given by the
Weil-Petersson symplectic structure.

\subsubsection{Hyperbolic Picture}

Uniformization offers a complementary picture on the
Teichm\"uller spaces: Instead of the complex structure
let us now focus on the associated hyperbolic metric
as the relevant geometric structure on $C_{g,n}$.
Natural coordinates for the Teichm\"uller spaces can
then be defined in terms of the lengths of geodesics
on the surface with hyperbolic metric. These
coordinates offer two big advantages concerning the intended
quantization of the Teichm\"uller spaces: They are real,
so one may expect to find self-adjoint operators
as their quantum representatives, and the Poisson
bracket associated to the symplectic form \rf{omdef}
becomes very simple (linear) if the class of geodesics is suitably chosen.

\begin{figure}[ht]
\begin{center}\epsfxsize7cm
\epsfbox{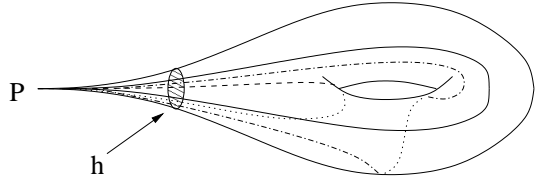}\hspace{1cm}
\end{center}
\caption{Triangulation of the once-punctured torus.}\label{triang}
\end{figure}

A particularly useful set of coordinates was introduced
by R. Penner in \cite{MR919235}. They can be defined for Riemann surfaces
that have at
least one puncture. One may assume having triangulated the surface
by geodesics that start and end at the punctures. As an example we have
drawn in Figure \ref{triang} a triangulation $\tau$ of
the once-punctured torus.
The length of these
geodesics will be infinite. In order to regularize this divergence
one may introduce one horocycle around each puncture and measure
only the length of the segment of a geodesic that lies between the
horocycles. Assigning to an edge $e$ its regularized length $l_e$
gives coordinates for the so-called decorated Teichm\"uller spaces.
These are fiber spaces over the
Teichm\"uller spaces which have fibers that parameterize the choices of the
``cut-offs'' as introduced by the horocycles.

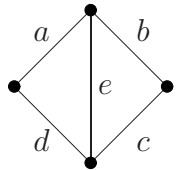
\begin{figure}[h]
\centering\setlength{\unitlength}{0.02in}
\begin{picture}(200,40)
\put(100,0){\line(-1,1){20}}\put(85,3){$d$}
\put(100,0){\line(1,1){20}}\put(112,3){$c$}
\put(80,20){\line(1,1){20}}\put(85,32){$a$}
\put(120,20){\line(-1,1){20}}\put(112,32){$b$}
\put(100,0){\line(0,1){40}}\put(102,18){$e$}
\put(100,0){\circle*{3}}
\put(80,20){\circle*{3}}
\put(100,40){\circle*{3}}
\put(120,20){\circle*{3}}
\end{picture}
\caption{The labeling of the edges}\label{rectlab}
\end{figure}

A closely related set of coordinates for
the Teichm\"uller spaces themselves was introduced by Fock in
\cite{Fock:dg-ga9702018}.
The coordinate $z_e$ associated to an edge $e$ of a triangulation
can be expressed in terms of the Penner-coordinates via
$z_e=l_a+l_c-l_b-l_d$, where $a$, $b$, $c$ and $d$ label the
other edges of the triangles that have $e$ in its boundary as indicated in
Figure \ref{rectlab}.

Instead of triangulations of the Riemann surfaces
it is often convenient to consider the corresponding {\it fat graphs}, which
are defined by putting a trivalent vertex into each triangle and by
connecting these vertices such that the edges of the triangulation are
in one-to-one correspondence to the edges of the fat-graph.

\subsubsection{Symplectic Structure in the Hyperbolic Picture}

As already mentioned, the Teichm\"uller spaces carry a natural symplectic form, called
Weil-Petersson symplectic form. We are therefore dealing with a
family of phase-spaces, one for each topological type of
the Riemann surfaces. One of the crucial virtues of the
Penner-coordinates is the fact that the Weil-Petersson symplectic
form has a particularly simple expression 
in these coordinates \cite{MR1163449}.
The corresponding Poisson-brackets are in fact {\it constant} for
Fock's variables $z_e$ \cite{Fock:dg-ga9702018},
\begin{equation}\label{poisson}
\{z_{e}^{},z_{e'}^{}\}=n_{e,e'}, \quad{\rm where}\quad
n_{e,e'}\in\{-2,-1,0,1,2\}.
\end{equation}
The value of $n_{e,e'}$ depends on how edges $e$ and $e'$ are imbedded
into a given fat graph. If $e$ and $e'$ don't have a common
vertex at their ends, or if one of $e$, $e'$ starts and
ends at the same vertex then $n_{e,e'}=0$. 
In the case that $e$ and $e'$ meet at two vertices 
one has $n_{e,e'}=2$ (resp. $n_{e,e'}=-2$)
if $e'$ is the first edge to the right%
\footnote{The orientation is induced by the imbedding of the fat-graph
into the surface.} 
(resp. left) of $e$ at both
vertices, and $n_{e,e'}=0$ otherwise.
In all the remaining cases $n_{e,e'}=1$ (resp. $n_{e,e'}=-1$)
if $e'$ is the first edge to the right (resp. left) of $e$ at the common
vertex.

If one considers a surface $\Sigma_{g,n}$ with genus $g$ and
$n$ boundary components one will find $n$
central elements in the Poisson-algebra defined
by (\ref{poisson}). These central elements $c_a$, $a=1,\ldots, n$
are constructed as $c_a=\sum_{e\in E_a} z_e$,
where $E_a$ is the set of edges in the triangulation that emanate from the
$a^{\rm th}$ boundary component. The value of $c_a$
gives the geodesic length of the $a^{\rm th}$ boundary component
\cite{Fock:dg-ga9702018}.

\subsubsection{Length Functions}

Having identified the phase space of Liouville theory on $C_{g,n}$
with the
Teichm\"uller space $\CT_{g,n}$ suggests to look for interesting observables.
The lengths of {\it closed} geodesics, considered as functions on $\CT_{g,n}$, are certainly interesting objects from
the geometric point of view. 

A nice feature of the Fock coordinates is that they
lead to a particularly simple way to reconstruct the length functions $l_\ga(P)$
corresponding to the point $P$ in Teich\-m\"uller space that is
parametrized by the variables $z_e(P)$.
Assume given a
path $\varpi_\ga$ on the fat graph homotopic to a simple closed
curve $\ga$ on $C_{g,n}$. Let the edges be labelled $e_i$, $i=1,\dots,r$
according to the order in which they appear on $\varpi_\ga$,
and define $\si_i$ to be $1$ if the path turns left
at the vertex
that connects edges $e_i$ and $e_{i+1}$, and to be equal to $-1$
otherwise. Consider the following matrix,
\begin{equation}\label{fuchsgen}
{\rm X}_{\ga}\;=\;{\rm V}^{\si_r}{\rm E}(z_{e_r})\dots {\rm V}^{\si_1}
{\rm E}(z_{e_1}),
\end{equation}
where the matrices ${\rm E}(z)$ and ${\rm V}$ are defined respectively by
\begin{equation}
{\rm E}(z)\;=\;\left(\begin{array}{cc} 0 & +e^{+\frac{z}{2}}\\
-e^{-\frac{z}{2}} & 0 \end{array}\right),\qquad
V\;=\;\left(\begin{array}{cc} 1 & 1 \\ -1 & 0 \end{array}\right).
\end{equation}
One
may then calculate the hyperbolic length of the closed geodesic
isotopic to $c$ via
\cite{Fock:dg-ga9702018}
\begin{equation}\label{glength}
L_{\ga}\,\equiv\,2\cosh\big(\fr{1}{2}l_{\ga}\big)\;=\;|{\rm tr}({\rm X}_{\ga})|.
\end{equation}

\subsection{K\"ahler Quantization}\label{kahlerquant}

A quick way to anticipate that the solution of the problem to quantize the Teichm\"uller spaces
is related to conformal field theory goes as follows, see 
\cite{Teschner:2003at} for a more detailed discussion.
Let us restrict to the case $g=0$ with $n$ conical singularities for notational simplicity. 
Complex analytic
coordinates for the moduli space $\CM_{0,n}$
of Riemann
surfaces with genus 0 and $n$ punctures are given by the
positions of the conical singularities $z_1,\dots,z_{n-3}$ if the
remaining three are assumed to be located at $0$, $1$ and
$\infty$, respectively 
\begin{equation}
\CM_{0,n}\;=\;\big\{ (z_1,\dots,z_{n-3});\ z_i\neq 0,1\;\,\text{and}\;\,
z_i\neq z_j\;\,\text{for}\;\,i\neq j\,\big\}
\end{equation}
The corresponding canonically conjugate
momenta w.r.t. $\om$ can be defined as
\begin{equation}
C_i=-\pa_{z_i}S^{\rm cl}\,.
\end{equation}
Indeed, since $S^{\rm cl}$ is the K\"ahler potential for $\omega$ we have the following
Poisson-brackets 
\begin{equation}
\{\,z_i,z_j\,\}\,=\,0\,=\,\{\,C_i,C_j\,\},\quad
\{\,z_i,C_j\,\}\,=\,\frac{1}{2\pi i}\de_{ij}\;.
\end{equation}
An observation from \cite{MR882831,MR1013154} which is important for us
is the fact that the momenta $C_i$ parameterize the 
classical energy-momentum tensor $T_{\vf}(z)\equiv-\fr{1}{4}(\vf_z)^2+\fr{1}{2}\vf_{zz}$ associated to a solution $\vf$ 
of \rf{EOM}. Indeed, it can be shown (see e.g. \cite{Takhtajan:2001uj})
that $T_\vf(z)$ can 
be expanded as 
\begin{equation}\label{Tdef}
T_{\vf}(z)\,=\,
\sum_{i=1}^{n-1}\left(\frac{\de_i}{(z-z_i)^2}+\frac{C_i}{z-z_i}
\right)\,.
\end{equation}
The asymptotic behavior of
$T_{\vf}(z)$ near $z=\infty$ may be represented as
\begin{equation}\label{Tas}
T_{\vf}(z)\;=\;\frac{\de_n}{z^2}+\frac{C_n}{z^3}+O(z^{-4})\,.
\end{equation}
The so-called accessory parameters $C_i$ are highly nontrivial functions on the moduli space
$\CM_{0,n}$ which are restricted by the relations
\begin{equation}
\sum_{i=1}^{n-1}C_i=0,\qquad
\sum_{i=1}^{n-1}(z_iC_i+\de_i)=\de_n,\qquad
\sum_{i=1}^{n-1}(z_i^2C_i+2\de_iz_i)=C_n.
\end{equation}

In analogy to the coherent state representation of quantum
mechanics it is then natural to consider a quantization scheme in
which states are represented by holomorphic multi-valued wave-functions
\begin{equation}\label{cohwavefct}
\Psi(z)\,=\,\langle\,z\,|\,\Psi\,\rangle\,,\qquad z=(z_1,\dots,z_{n-3})\,,
\end{equation} 
such that the operators $\sz_i$ corresponding to the classical observables $z_i$ are represented
as multiplication operators, $\sz_i\Psi(z)=z_i\Psi(z)$.
The state $|\,z\,\rangle$ introduced in \rf{cohwavefct} is thereby identified as an analog of a coherent state
(eigenstate of the "creation operators" ${\mathsf z}_i$) in quantum mechanics.

The operators $\SC_i$ associated to the momenta
$C_i$ conjugate to $z_i$ should be represented by the differential operators
$b^2\pa_{z_i}$ in such a representation. 
The energy-momentum
tensor $T_{\vf}$ would then be represented by the operator $b^2\ST(z)$, where
\begin{equation}\label{STdef}
\ST(z)\,=\,\sum_{i=1}^{n-1}\left(\frac{\De_{\al_i}}{(z-z_i)^2}+
\frac{1}{z-z_i}
\frac{\pa}{\pa z_i}\right)\,.
\end{equation}
We have introduced the quantum conformal dimensions $\De_{\al_i}$ which are related to the $\de_i$ by $\de_i=b^2\De_i+\CO(b^2)$.
This makes the
space of holomorphic wave-functions obtained in the
K\"ahler quantization of the Teichm\"uller
spaces into a module over the ring of holomorphic differential operators
on $\CT_{0,n}$.

This should be compared with the well-known statement
that the Virasoro conformal blocks can be represented by
means of holomorphic functions on $\CT_{0,n}$ denoted
by $\bra T(x)\,\Psi_{\al_n}(z_n)\dots \Psi_{\al_1}(z_1)\ket$
which form a module over the ring of holomorphic differential
operators on $\CT_{0,n}$ as is expressed by
the conformal Ward indentities
\begin{equation}\label{cfWard}
\big\bra \,T(x)\, \Psi_{\al_n}(z_n)\dots \Psi_{\al_1}(z_1)\,\big\ket=
\;\sum_{i=1}^{n-1}\left(\frac{\De_{\al_i}}{(x-z_i)^2}+
\frac{1}{x-z_i}
\frac{\pa}{\pa z_i}\right)
\big\bra\,\Psi_{\al_n}(z_n)\dots \Psi_{\al_1}(z_1)\,\big\ket.
\end{equation}
Comparison of \rf{STdef} and \rf{cfWard} 
suggests that wave-functions in the K\"ahler
quantization of $\CT_{0,n}$ can be identified with
Virasoro conformal blocks. The precise correspondence
will be reviewed in Subsection~\ref{sec:length-Kahler}
below.

\subsection{Quantization in a Real Polarization}

It is very difficult to describe the quantization of the length
operators directly in the K\"ahler quantization above. The
basic reason is that the relation between complex analytic coordinates
for $\CT_{g,n}$ and the geodesic length functions involves
uniformizing the surface $C_{g,n}$, which generically is
highly transcendental. It turns out to be possible, however, to bypass this difficulty
by first studying the quantum theory obtained by directly quantizing the Teichm\"uller
spaces using real coordinates, and then establishing the link of this quantization scheme
with the K\"ahler quantization.

\subsubsection{Algebra of Observables and Hilbert Space}
The simplicity of the Poisson brackets (\ref{poisson}) makes part of the
quantization quite simple. To each edge $e$ of a triangulation
of a Riemann surface $C_{g,n}$ associate a quantum operator
$\sz_e$. The algebra of observables $\CA(C_{g,n})$
will be the algebra with generators $\sz_e$, relations
\begin{equation}\label{comm}
[\sz_{e}^{},\sz_{e'}^{}]=2\pi i b^2 \{z_{e}^{},z_{e'}^{}\},
\end{equation}
and hermiticity assignment $\sz_e^{\dagger}=\sz_e^{}$.
The algebra $\CA(C_{g,n})$
has a center with generators
$\sfc_a$, $a=1,\ldots, n$ defined by $\sfc_a=\sum_{e\in E_a}\sz_e$,
where $E_a$ is the set of edges in the triangulation that emanates from the
$a^{\rm th}$ boundary component.
The representations of $\CA(C_{g,n})$ that we are going to consider
will therefore be such that the generators $\sfc_a$ are
represented as the operators of multiplication by real
positive numbers $l_a$. Geometrically one may interpret
$l_a$ as the geodesic length of the
$a^{\rm th}$ boundary component
\cite{Fock:dg-ga9702018}.
The vector $l=(l_1,\dots,l_n)$ of lengths of the boundary components
will figure as a label of the representation
$\pi(C_{g,n},\Lambda)$ of the algebra $\CA(C_{g,n})$.

To complete the definition of the
representation $\pi(C_{g,n},\Lambda)$ by operators
on a Hilbert space $\CH(C_{g,n})$ one just needs to find
linear combinations ${\mathsf x}_1,\dots,{\mathsf x}_{3g-3+n}$ and
$\spp_1,\dots,\spp_{3g-3+n}$ of the $\sz_{e}$ that satisfy
$[\spp_i,{\mathsf x}_j]=(2\pi i)^{-1}\delta_{ij}$.
The representation of $\CA(C_{g,n},\Lambda)$ on
$\CH(C_{g,n}):=L^2(R^{3g-3+n})$ is
defined by choosing
the usual Schr\"odinger representation for the ${\mathsf x}_i$, $\spp_i$.

It is very important to make sure that the resulting quantum theory
does not depend on the underlying triangulation in an essential
way. This can be done by constructing a family of unitary
operators $\SU_{\tau_2,\tau_1}$ that describe the change of
representation when passing from the quantum theory associated
to triangulation $\tau_1$ to the one associated to $\tau_2$
\cite{Chekhov:1999tn,MR1607296,Teschner:2005bz,MR2470108}.

\subsection{Geodesic Length Operators}

Length operators were first studied
in the pioneering works \cite{Chekhov:1999tn,Chekhov:2000tw}.
On first sight the problem looks
rather simple.
We may observe that the classical expression for $L_{\ga}\equiv
2\cosh\frac{1}{2}l_\ga$ as given by formula \ref{glength}
is a linear combination of monomials
in the variables $e^{\pm\frac{z_e}{2}}$ of a very particular form,
\begin{equation}\label{Lclass}
L_{\ga}\,
=\,
\sum_{\nu\in\CF}\,C_{\tau,\ga}(\nu)\,e^{(\nu,z)}\,,\qquad (\nu,z)\,=\,\sum_e\nu_ez_e
\end{equation}
where the summation is taken over a finite set $\CF$ of
vectors $\nu\in(\frac{1}{2}\BZ)^{3g-3+2n}$
with half-integer components $\nu_e$. The coefficients
$C_{\tau,\ga}(\nu)$ are positive
integers. In the quantum case one is interested in the definition
of length operators $\SL_{\tau,\ga}$ which should be representable
by expressions of the form,
\begin{equation}\label{Lquant}
\SL_{\tau,\ga}\,
=\,\sum_{\nu\in\CF}\,C_{\tau,\ga}({\nu})\,:\!e^{(\nu,\sz)}\!:_b^{}\,=\,
\sum_{\nu\in\CF}\,C_{\tau,\ga}^b({\nu})\,e^{(\nu,\sz)}\,,
\qquad(\nu,\sz)=\sum_e \nu_e\sz_e\,,
\end{equation}
where the notation $:\!\SO\!:_b^{}$ indicates use of a quantum ordering prescription.
In the second expression we have moved the effect of the ordering
into a quantum deformation $C_{\tau,\ga}^b(\nu)$ of the coefficients $C_{\tau,\ga}(\nu)$.

Note, on the other hand, that the following additional
properties seem to be indispensable if one wants to
interpret an operator of the general form \rf{Lquant}
as the quantum counterpart
of the functions $L_{\tau,\ga}=2\cosh\frac{1}{2}l_\ga$:
\begin{enumerate}
\item[(a)] {\bf Spectrum:}
$\SL_{\tau,\ga}$ is simple and takes values
in $[2,\infty)$. This is necessary and sufficient for the existence
of an operator $\sll_{\tau,\ga}$ - the {\it geodesic length operator} -
such that
$\SL_{\tau,\ga}=2\cosh\frac{1}{2}\sll_\ga$.
\item[(b)] {\bf Independence of triangulation:}
\[
\SU_{\tau_2\tau_1}^{-1}\cdot\SL_{\tau_1,\ga}^{}\cdot\SU_{\tau_2\tau_1}^{-1}
\,=\,\SL_{\tau_2,\ga}^{}\,,
\]
where $\SU_{\tau_2\tau_1}$ is the unitary operator
relating the representation associated to
triangulation $\tau_1$ to the one associated to $\tau_2$.
\end{enumerate}
Property (b) ensures that the collection of length
operators $\SL_{\tau,\ga}$ associated to the different
triangulations $\tau$ ultimately defines an operator $\SL_{\ga}$
that is {\it independent} of the triangulation.
It was observed in \cite{Chekhov:1999tn,Chekhov:2000tw}
that the deformation of the
coefficients $C_{\tau,\ga}^b({\nu})$ which is necessary for
having the properties (a) and (b) can be quite nontrivial in general,
in the sense that it can not be obtained from a simple
ordering prescription.

A general construction of length operators which fulfils the
requirements above was given in \cite{Teschner:2005bz}. This
construction coincides with the one in
\cite{Chekhov:1999tn,Chekhov:2000tw} whenever both can be applied.

\subsubsection{The Length Representation}

It can be shown that the length operators associated
to non-intersecting simple closed curves
commute with each other. This together with the
self-adjointness of the length operators
allows one to introduce bases of eigenfunctions
for the length operators.

One gets one such basis for each pants decomposition
of $C_{g,n}$. A key result for the connection
between quantum Liouville and quantum Teichm\"uller
theory is that for each marking $\si$ there exists
a basis for $\CH_{g,n}$ spanned by $|\,l\,\rangle_{\si}$,
$l=(l_1,\dots,l_{3g-3+n})$ which
obeys the factorization rules of conformal field theory
\cite{Teschner:2005bz}. This means in particular that
for any pair $\si_2$, $\si_1$ of markings one can always decompose the
unitary transformation $\SV_{\si_2\si_1}$ which relates
the representation corresponding to
marking $\si_1$ to the one corresponding to $\si_2$
as a product of operators which represent
the elementary fusion, braiding and
S-moves introduced in Section~\ref{sec:prelim}. The
unitary transformation $\SV_{\si_2\si_1}$ can be represented
as an integral operator of the form 
\begin{equation}
|\,l_2\,\rangle_{\si_2}\,=\,\int d\mu(l_1)\;V_{\si_2\si_1}(l_2,l_1) |\,l_1\,\rangle_{\si_1}\,.
\end{equation}
The explicit expressions for the kernel $V_{\si_2\si_1}(l_2,l_1)$ are known for the
cases where $\si_2$ and $\si_1$ differ by one of the elementary
moves.

\subsection{Relation between Length Representation and K\"ahler Quantization}
\label{sec:length-Kahler}

When discussing the more general case of $g\geq 0$ in the following we will use the gluing parameters 
$q_i$ as generalizations of the holomorphic coordinates $z_i$ used in 
\ref{kahlerquant}. 
The coherent states $|q\rangle$ are the eigenstates of the
operators $\sq_i$ corresponding to the holomorphic coordinates $q_i$, and the wave-functions
are holomorphic functions\footnote{More precisely sections of a projective line bundle on $\CT_{g,n}$.} 
$\Psi(q)=\langle \Psi|q\rangle$ on $\CT_{g,n}$.

The change of representation from length representation
to the holomorphic representation
is described by means of the matrix elements
\begin{equation}
\Psi_l^{\si}(q)\,=\,\langle\, l\,|\, q\,\rangle_\si^{}\,.
\end{equation}
The following characterization of these matrix elements was obtained in
\cite{Teschner:2003at}:
\begin{equation}\label{bl=wf}
\Psi_l^{\si}(q)\,=\,\CG_{\al,E}^{(\si)}(q)\,,
\end{equation}
where $\CG_{\al,E}^{(\si)}(q)$ is the Liouville conformal block%
\footnote{We indicate here that the dependence on the complex structure
$q$ which is suppressed in the rest of the text.}
\eqn{BPZblock}
associated to a marking $\si$ with external representations labeled
by $E=(m_1,\dots,m_n)$ and fixed intermediate
dimensions given by the parameters $\al=(\al_1,\dots,\al_{3g-3+n})$.
These parameters are related to the lengths $c_a$ of
the boundary components and to the lengths $l_i$
around the curves defining the pants
decomposition respectively as
\begin{equation}
m_a\,=\,\frac{Q}{2}+i\frac{c_a}{4\pi b}\,,\qquad
\al_i\,=\,\frac{Q}{2}+i\frac{l_i}{4\pi b}\,=\,\frac{Q}{2}+a_i\,,
\label{mal-cl}
\end{equation}
where $a=1,\dots,n$ and $i=1,\dots,3g-3+n$.
The main nontrivial result underlying \rf{bl=wf} is the
fact that both sides transform in the same way
under a change of marking $\si$. Comparing also the
asymptotic behavior at the boundaries of $\CT_{g,n}$
then allows one to conclude that the holomorphic functions
$\Psi_l^{\si}(q)$ and $\CG_{\alpha,E}^{(\si)}(q)$ 
must coincide \cite{Teschner:2003at}.

Let us then consider the resolution of the
identity, the so-called Bergman kernel, defined as
\begin{equation}
B(\bar q,{q}')\,=\,\int d\mu(l)\;\langle\, q\,|\, l\,\rangle_\si^{}\,{}^{}_\si\!\langle\,l\,|\,{q}'\,\rangle=
\int d\mu(l)\;\,({\Psi}_l^{\si}(q))^*\,\Psi_l^{\si}(q')\,.
\end{equation}
Comparing with \rf{correla-F} we see that
the correlation functions of Liouville theory on a Riemann surface $C_{g,n}$
can be represented as 
\begin{equation}
\Big<\prod_{a=1}^n V_{m_a}\Big>_{C_{g,n}}
\,=\,B(\bar q ,q)\,.
\end{equation}
The relation \rf{bl=wf} can be generalized to a direct
relation between the geodesic length operators
$\SL_\ga$ from quantum Teichm\"uller theory and the
Liouville loop operators $\CL(\ga)$,
\begin{equation}\label{bl=wf2}
(\SL_\ga\Psi_l^{\si})(q)\,=
\,\kappa(\cL(\gamma)\CG_{\alpha,E}^{(\si)})(q)\,,\qquad\kappa=2\cos\pi bQ\,.
\end{equation}
This relation is easily verified by considering the case
where $\ga$ is one of the curves defining the
pants decomposition associated to $\si$, say $\gamma_i$.
Both the operators $\SL_{\gamma_i}$ and $\kappa\CL(\ga_i)$ are
then diagonal in this basis and are represented by
multiplication with $2\cosh(l_{i}/2)$ and
$2\cosh\pi b(2\alpha_{i}-Q)$ respectively \eqn{w01},
which are indeed equal \eqn{mal-cl}.
The validity of \rf{bl=wf2} in general then follows
from the fact that both
sides transform the same way under changes of the marking.

We finally arrive at the following interpretation of the loop
operator expectation values within
quantum Teichm\"uller theory.
\begin{equation}
\kappa\,\langle\,\CL(\ga)\,\rangle_{C_{g,n}}\,=\,\langle\,q\,|\,\SL_\ga\,|\,q\,\rangle\,.
\label{loopislength}
\end{equation}
This means that the expectation value of the Liouville loop operator has the
geometric interpretation as the expectation value of the geodesic
length operator $\SL_\ga$ in the coherent state $|q\rangle$ which corresponds to a fixed
complex structure parameterized by $q$. It therefore solves the natural quantum counterpart
of the classical problem to calculate the geodesic length of $\ga$ in the metric associated 
by uniformization to the complex structure parameterized by $q$.

\section{Algebra of Loop Operators
in Liouville-Teichm\"uller Theory}
\label{sec:algebra}

Given a set of operators, it is very natural to study
the algebra formed by them.
For loop operators in gauge theory, this problem
goes back to 't~Hooft's original paper \cite{'tHooft:1977hy}
where magnetic loop operators were first introduced.
There it was found that Wilson and 't~Hooft loops in a spatial slice
at constant time, acting
on the Hilbert space of gauge theory, give 
rise to interesting commutation relations. The relations, 
which will be referred as the 't~Hooft commutation relations, 
encode information
about the electric and magnetic charges of the loop operators 
as well as how the loops are linked in the spatial slice. Another 
interesting algebraic property
to consider is the operator product expansion (OPE) of
loop operators. The OPE captures the decomposition of the 
product of two loop operators defined on two curves
in spacetime into a basis of loop operators in gauge theory in 
the limit where two operators are brought together.

In previous sections we have explicitly computed 
the action of Liouville loop operators supported on 
curves on the Riemann surface on the conformal blocks.
The action involves multiplication by functions of $\alpha_i$, $m_a$
as well as shifts of $\alpha_i$, where $\alpha_i$ and $m_a$ 
label the internal and external edges of the trivalent graph characterizing
the conformal block. Such multiplication and shifts can be thought of as
operators acting on the space of conformal blocks, and form an algebra
that can be computed explicitly in examples.
The description of this algebra in terms of quantum Teichm\"uller theory is
equivalent to that in Liouville theory,
as discussed in section \ref{liou-teich}.

In this section we study the algebra of loop operators
in Liouville theory and provide a physical interpretation of the algebra
in gauge theory.
First we explain the fundamental relation
satisfied by the Liouville loop operators.
The relation, known as the quantum skein relation,
completely determines the algebra of loop operators.

We then show that the commutation relations that follow
from the quantum skein relation at $b=1$ are 
precisely the ones found by 't~Hooft for the 
gauge theory loop operators that are Hopf-linked
in a three-dimensional constant time slice.

Furthermore, the quantum skein relations can also be viewed as a prediction for
the OPE of loop operators in gauge theory.
We compare this to the OPE of loop operators in 
the gauge theory deduced from S-duality, by generalizing an argument
given for a special case in \cite{Kapustin:2007wm}. 
We point out that the Liouville loop operators
that appear in the product of two loop operators agree, 
but that the numerical prefactors do not, and propose that 
the difference has to do with whether the loops are Hopf-linked or not.
We illustrate these general results,
by considering the concrete expressions 
derived earlier for the Liouville loop operators in 
$\cN=2^*$ super Yang-Mills theory
and the ${\cal N}=2$ $SU(2)$ gauge theory with $N_F=4$ .

\subsection{Quantum Skein Relation}
\label{sec:skein}

It is known from quantum Teichm\"uller theory
\cite{Chekhov:1999tn,Chekhov:2000tw}
that the geodesic length operators form a closed algebra 
with basic relation being the so-called {\it quantum skein relation},
see \cite{Chekhov:arXiv0710.2051}
for a nice review of the necessary results and references. 
As explained in Section~\ref{sec:length-Kahler},
the Liouville loop operator $\cL(\gamma)$ 
corresponding to a geodesic $\gamma$
is exactly the geodesic length operator $\SL_\gamma$,
up to rescaling by the constant $\kappa=2\cos (\pi b Q)$,
when we identify the Hilbert space of quantum Teichm\"uller theory
with the space of Liouville conformal blocks.
It follows that the Liouville loop operators
satisfy the quantum skein relation.%
 
\begin{figure}[ht]
\begin{center}
\begin{tabular}{cccccc}
\raisebox{-12mm}{
\includegraphics[scale=.34]{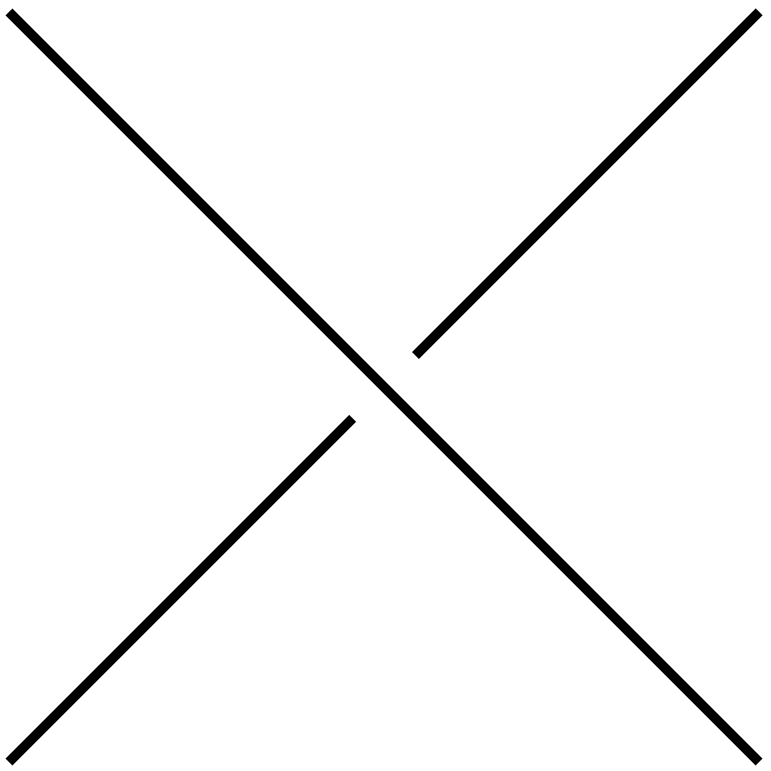}
}
&
$=
e^{-\pi i b^2/2}$
&
\raisebox{-12mm}{
\includegraphics[scale=.34]{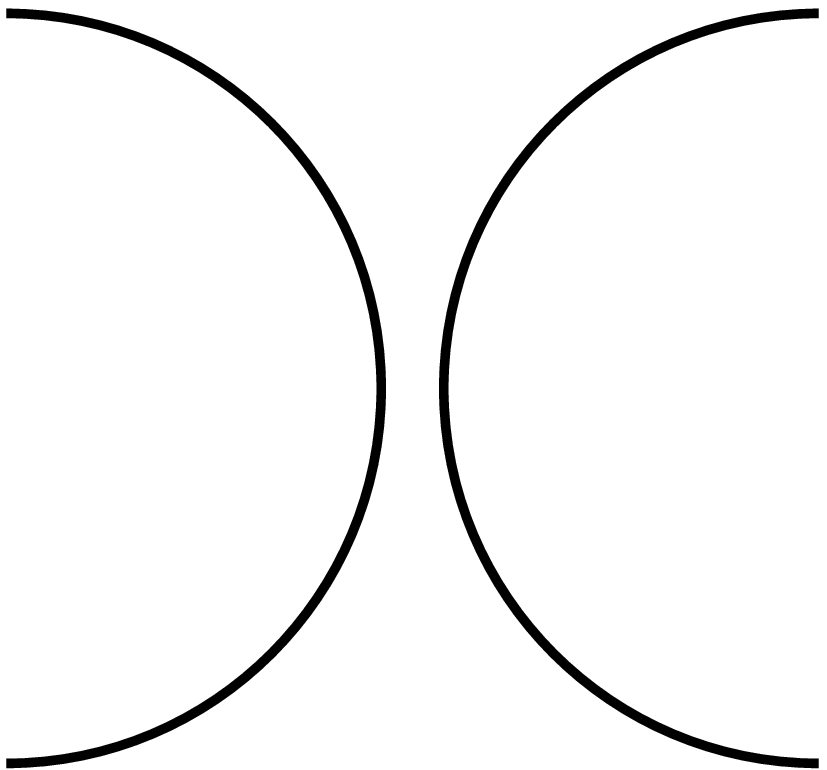}
}
&
$+$ $e^{\pi i b^2/2}$
&
\raisebox{-13mm}{
\includegraphics[scale=.34]{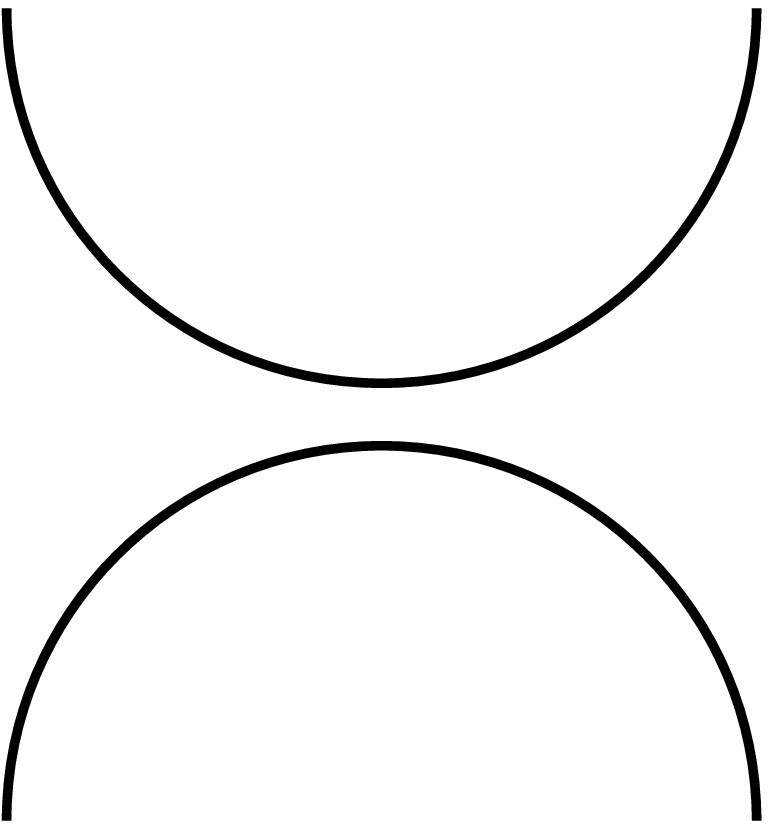}
}
\end{tabular}
\parbox{5in}{
\caption{A graphical expression of the quantum skein relation
among the length operators.
}\label{fig:skein}}
\end{center}
\end{figure}

\begin{figure}[ht]
\hspace{5mm}
\raisebox{-9mm}{
\psfrag{gamma}{\raisebox{3mm}{$\gamma$}}
\psfrag{gamma'}{$\gamma'$}
\includegraphics[scale=.25]{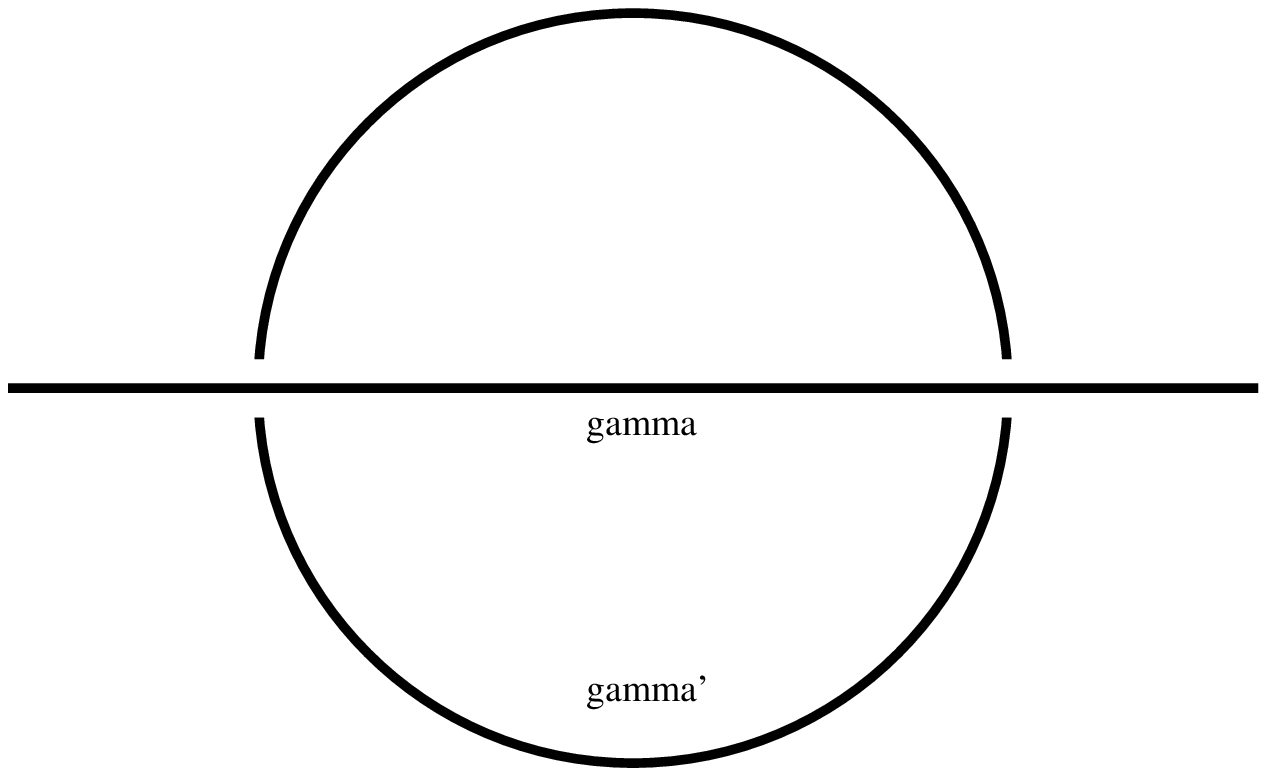}
}
$=$
\raisebox{-9mm}{
\includegraphics[scale=.25]{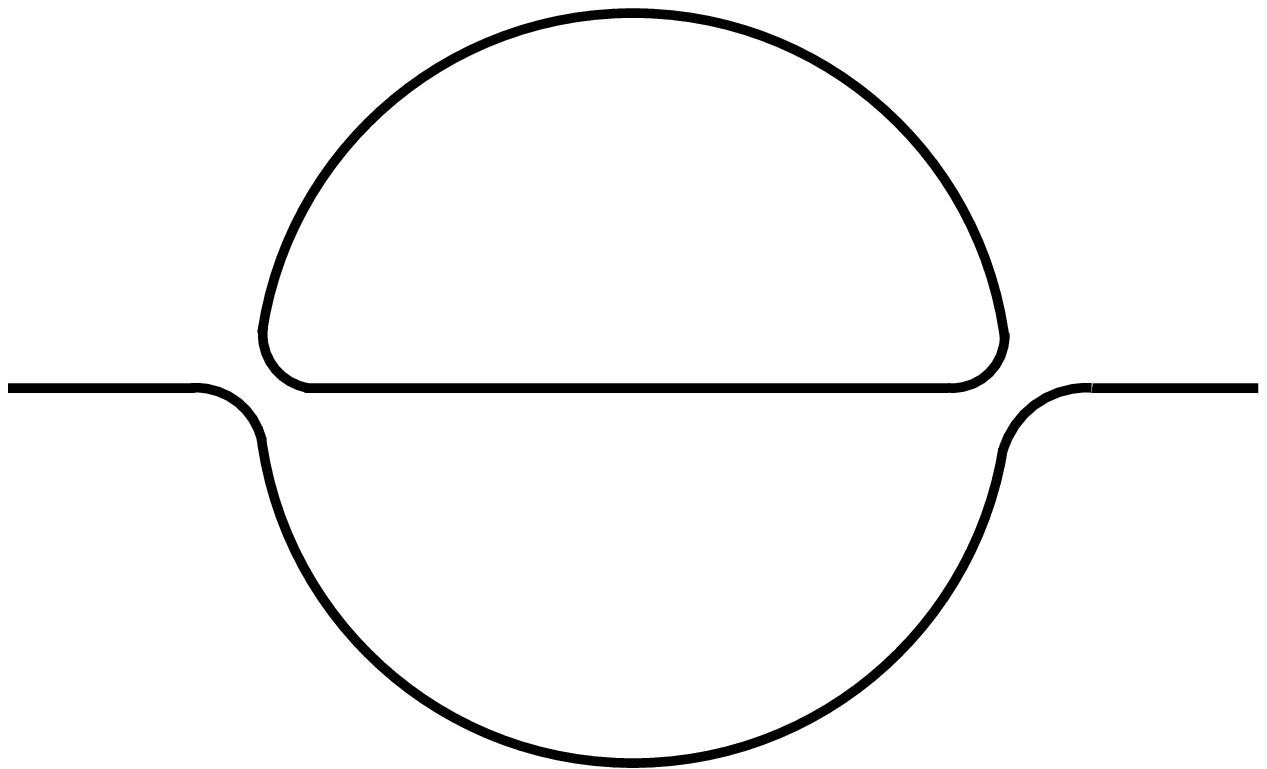}
}
$+e^{-\pi i b^2}$
\raisebox{-9mm}{
\includegraphics[scale=.25]{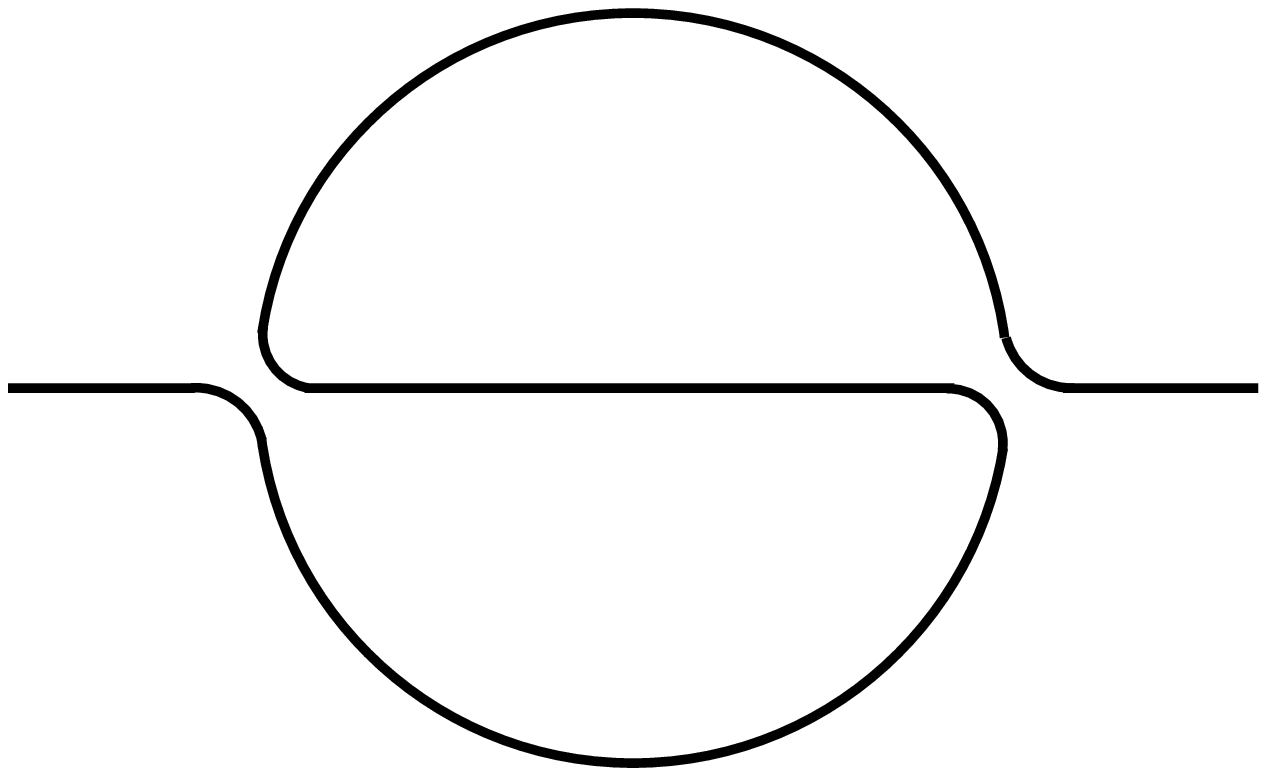}
}
\\
\vspace{2mm}
\hspace{60mm}
$+e^{\pi i b^2}$
\raisebox{-9mm}{
\includegraphics[scale=.25]{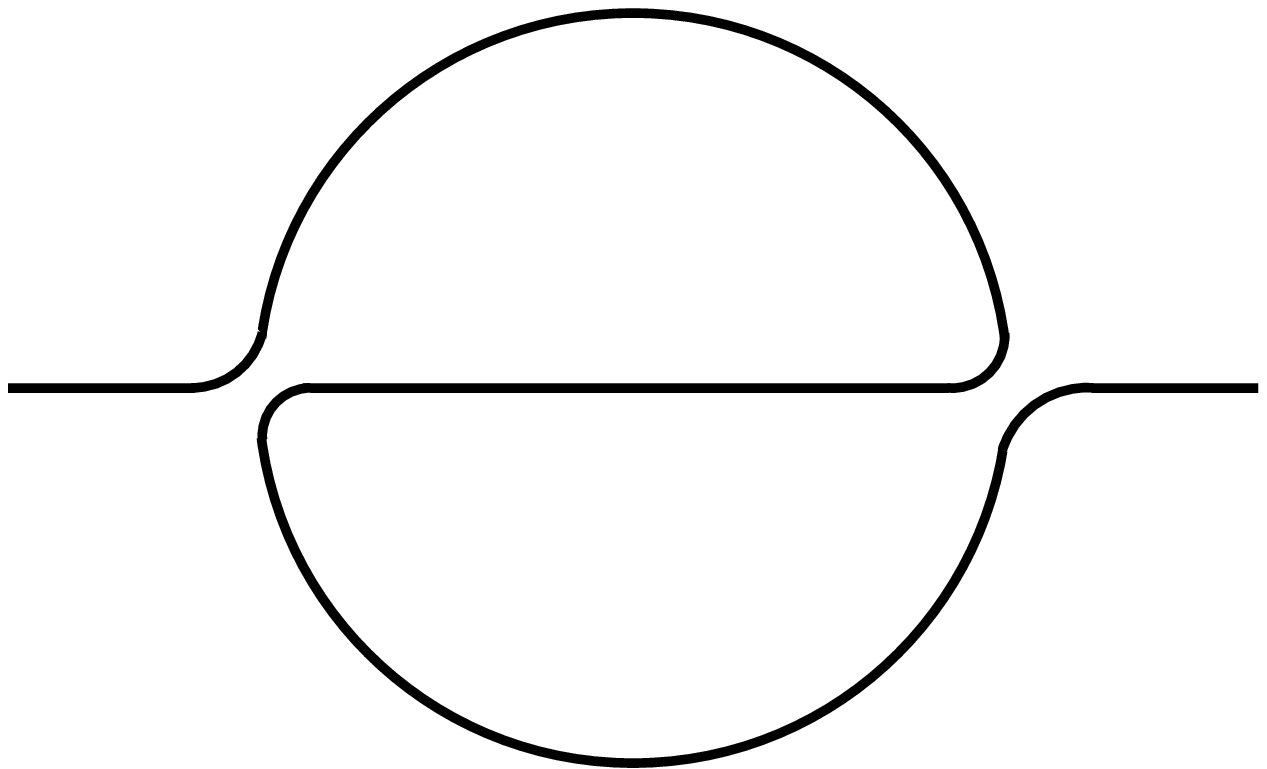}
}
$+$
\raisebox{-9mm}{
\includegraphics[scale=.25]{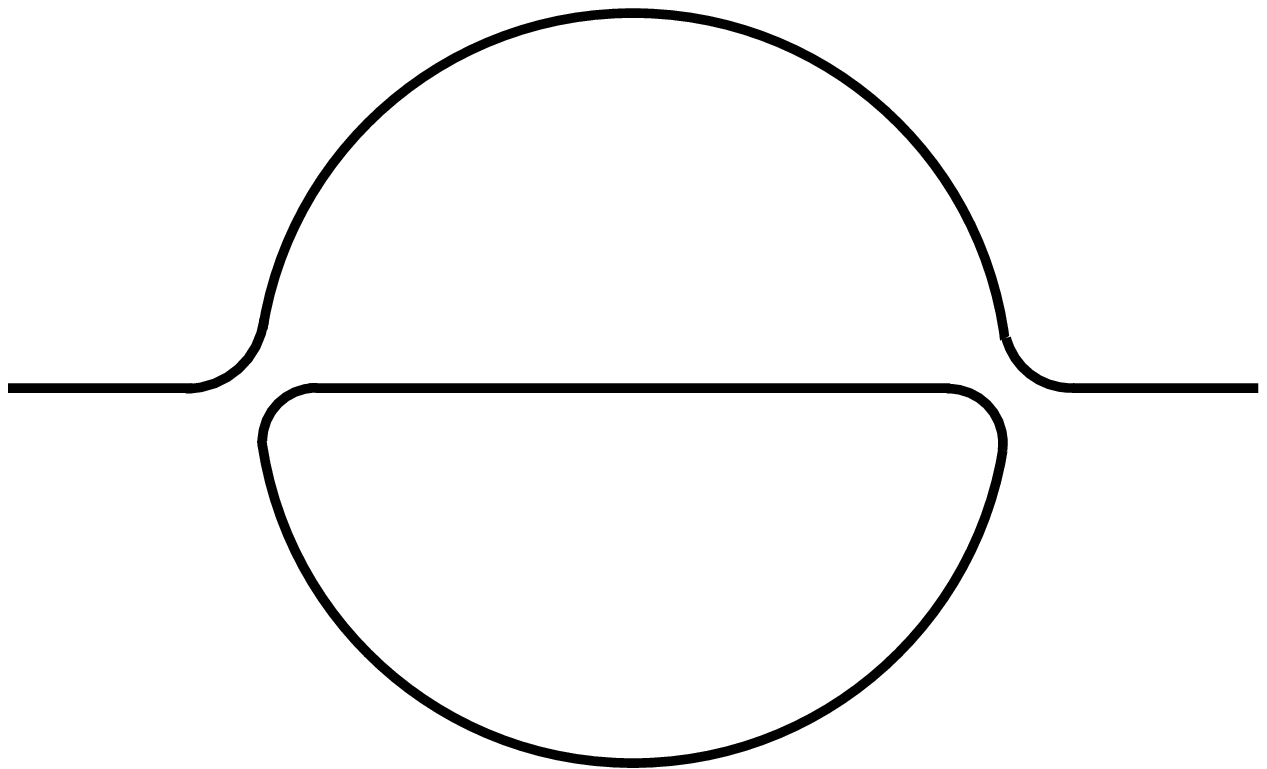}
}
\begin{center}
\parbox{5in}{
\caption{
The quantum skein relation for two curves
$\gamma$ and $\gamma'$ that intersect at two points.
$\SL_\gamma \SL_{\gamma'}$ on the  left hand side  is
expanded in length operators for non-intersecting curves.
Note that the ordering of $\gamma$ and $\gamma'$ matters
in this relation.
}\label{fig:skein2}}
\end{center}
\end{figure}

To explain the relation, let us first consider 
a pair of connected closed non-self-intersecting 
curves $\gamma$ and $\gamma'$
on the Riemann surface $C_{g,n}$
that intersect at one point.
There are two distinct ways to remove the intersection and reconnect
the two curves as shown in Figure~\ref{fig:skein},
where the line on top in the  left hand side  is identified with $\gamma$.
Let us denote by
$\gamma\star\gamma'$ and $\gamma\circ \gamma'$
the two resulting curves corresponding to the
first and second terms in the right hand side  of Figure~\ref{fig:skein}, respectively.
Then the four length operators $\SL_\gamma,\SL_{\gamma'},
\SL_{\gamma\star\gamma'}$ and $\SL_{\gamma\circ\gamma'}$ satisfy
the quantum skein relation
\beq
{\mathsf L}_\gamma\cdot
{\mathsf L}_{\gamma'}
=
e^{-\pi i b^2/2}{\mathsf L}_{\gamma\star \gamma'}
+
e^{\pi i b^2/2}{\mathsf L}_{\gamma\circ \gamma'}.
\label{eq:skein}
\eeq
It is also possible to derive the relation for the Liouville
loop operators directly
using the moves described in previous sections.%
\footnote{The composition of operators $\cL(\gamma)\cdot \cL(\gamma')$
can be realized as an operation that splits the identity
into two degenerate fields $V_{1,2}(z_1)$ and $V_{1,2}(z_2)$ 
at the intersection, moves $V_{1,2}(z_1)$ around along $\gamma$,
fuses them back to the identity, splits it again, moves
$V_{1,2}(z_1)$ around along $\gamma'$, and finally fuses
them back to the identity.

The operator $\cL(\gamma\star \gamma')$
on the other hand is an operation that splits the identity into
$V_{1,2}(z_1)$ and $V_{1,2}(z_2)$ 
at the intersection, moves $V_{1,2}(z_1)$ around along $\gamma$
back to the intersection point,
moves $V_{1,2}(z_1)$ around along $\gamma'$, and finally fuses 
them back to the identity.
Here we have assigned orientations to $\gamma$ and $\gamma'$
such that they consistently define an orientation of $\gamma\star \gamma'$.

Similarly the operator
$\cL_{\gamma\circ\gamma'}$ 
can be represented as an operation that splits the identity into
$V_{1,2}(z_1)$ and $V_{1,2}(z_2)$ 
at the intersection, moves $V_{1,2}(z_1)$ around along $\gamma$
back to the intersection point,
moves $V_{1,2}(z_2)$ around along $\gamma'$, and finally fuses
them  back to the identity.
Note that it is $z_2$, not $z_1$, that is moved around along $\gamma'$.

Writing out the fusion matrices corresponding to these moves,
one can derive the quantum skein relation.
}
When $\gamma$ and $\gamma'$ have more than one intersection,
one should apply the skein relation 
locally for all the intersections at the same time
as in Figure \ref{fig:skein2}.
A contractable curve is assigned the value 
$-e^{-\pi i b^2}-e^{\pi i b^2}=2\cos(\pi b Q)$.
See \cite{1751-8121-42-30-304007} for more examples.

Let us note that there is a natural basis of length/loop operators
in which we can apply the quantum skein relation.
The operators in this basis are in one-to-one correspondence
with the homotopy classes of non-self-intersecting closed curves
on the Riemann surface $C_{g,n}$ labeled by Dehn-Thurston parameters.
Let $\gamma$ be such a curve.
Then $\gamma$ in general has more than one connected components.
We denote by $K_I$ the number of components whose homotopy class is
that of a curve $\gamma^{(I)}$.
Then the operator that corresponds to $\gamma$
is given by
\beq
\SL_\gamma\equiv\prod_I \SL_{\gamma^{(I)}}^{K_I}.
\label{length-basis}
\eeq
Now let $\gamma$ and $\gamma'$ be
not-necessarily-connected simple closed curves.
The components of $\gamma$ generically 
intersect those of $\gamma'$ at several points.
By repeatedly using the quantum skein relation above,
one can express the product of any
operators of the form (\ref{length-basis})
as a sum of operators of the same form.

\subsection{Comparison with Gauge Theory}
\subsubsection{The Commutation Relations}

It follows that when $b=1$, we have
\beq
{\mathsf L}_{\gamma}\cdot {\mathsf L}_{\gamma'}
=(-1)^{\#(\gamma\cap\gamma')}{\mathsf L}_{\gamma'}\cdot {\mathsf L}_{\gamma},
\label{q-comm}
\eeq
where $ \#(\gamma\cap\gamma')$ is the number of intersections between
$\gamma$ and $\gamma'$.

We wish to provide
a gauge theory interpretation for the commutation relations (\ref{q-comm})
satisfied by the length operators.
As we now explain, this is precisely the commutation
relation satisfied by loop operators
that are linked in a constant time slice,
as originally studied by 't~Hooft \cite{'tHooft:1977hy}.

Let us briefly review the commutation relations found by 't~Hooft. 
It suffices for us
to consider a single gauge group $SU(2)$.
Let $L_{0,1}$ be the pure Wilson loop along curve 1 and $L_{1,0}$ 
the pure 't~Hooft loops along curve 2.
The two curves 1 and 2 are taken to be lie in the three-dimensional space
and we assume that they are Hopf-linked.%
\footnote{'t~Hooft's analysis is applicable in any gauge theory, also without
supersymmetry. To relate it to Liouville theory we need 
Hopf-linked loops in $\cN=2$ 
gauge theory which share some supersymmetry. Luckily this is the 
case for a pair of non-intersecting loops on $S^3$, as was shown for 
$\cN=4$ SYM in \cite{DGRT-more,DGRT-big}. The same argument 
carries over to our case.} 
We are interested in the canonical quantization of $SU(2)$ gauge theory
with matter that is neutral under the center $\bZ_2$ of the gauge group.
The product 
\beq
L_{0,1}L_{1,0}
\eeq
corresponds to inserting the Wilson loop $L_{0,1}$ at time $\epsilon>0$
and the 't~Hooft loop $L_{1,0}$ at time zero. Likewise, 
\beq
L_{1,0} L_{0,1}
\eeq
corresponds to inserting the Wilson loop $L_{0,1}$ at time $-\epsilon<0$
and the 't~Hooft loop $L_{1,0}$ at time zero. The 't~Hooft loop $L_{1,0}$ is defined by a background field configuration
that is singular along the curve 2 at time zero.
The singular configuration is such that at any time $t>0$,
there is a non-trivial background holonomy along a spatial curve that links
the curve 2.\footnote{In Euclidean spacetime, the spatial 't~Hooft
loop is obtained by inserting a magnetic monopole along the loop.
We can let the family of Dirac strings extend in the positive time direction.
Thus at any positive time, the Dirac string singularity is along the spatial 
loop, and going around it produces a phase (sign).\label{hopf}}
Thus the Wilson loop $L_{0,1}$ inserted at time $\epsilon>0$ picks
up the holonomy $-1$ relative to the Wilson loop inserted
at time $-\epsilon<0$, implying the commutation relation
\beq
L_{0,1}L_{1,0}=- L_{1,0}L_{0,1}
\eeq
for the loop operators acting on the Hilbert space of gauge theory.

The argument generalizes to more than one $SU(2)$ factor
in the gauge group and higher charges.
Let $\gamma$ and $\gamma'$ be connected simple closed geodesics
with Dehn-Thurston parameters $d$ and $d'$, respectively.
We can choose a pants decomposition where $\gamma$ is
a boundary of a pair of pants.
In other words, the gauge theory loop operator
$L_d$ is a pure Wilson loop
in, say the first $SU(2)$ gauge group.
In this case we have the commutation relation
\beq
L_d L_{d'}= (-1)^{p'_1} L_{d'} L_d,
\eeq
where $p'_1$ is the magnetic charge of $L_{d'}$
in the first gauge group.
Clearly $p'_1$ is the intersection number of $\gamma$ and $\gamma'$.

Thus we can write the commutation relation in a
frame-independent form as
\beq
L_d L_{d'}= (-1)^{\#(\gamma\cap\gamma')} L_{d'} L_d.
\eeq
We see that the loop operators in gauge theory,
supported on spacetime loops that are Hopf-linked 
obey precisely the same commutation relations -- the 't~Hooft commutation
relations --
as those for the corresponding geodesic length/Liouville loop operators
at $b=1$.
This observation
will lead to a conjecture and a definite prediction,
which we will formulate in Section~\ref{sec:discuss}.

\subsubsection{The Operator Product Expansion}
\label{sec:OPE}
As remarked at the end of Section~\ref{sec:skein},
there is a natural basis of geodesic length/Liouville loop
operators labeled by
not-necessarily-connected closed geodesics.
In this basis, the quantum skein relation 
allows one to decompose the product of 
operators 
into the sum of operators.
The correspondence of gauge theory and Liouville loop operators
proposed in Section~\ref{sec:bootstrap} then
leads us to view the quantum skein relation
as providing a rule for how to perform the corresponding decomposition,
namely the {\it operator product expansion} (OPE) 
of loop operators in gauge theory.

We observed above
that the commutation
relations at $b=1$ that follow from the quantum skein relation
are precisely those for gauge theory loop operators
on Hopf-linked spacetime loops.
At this value of $b$, the quantum skein relation
further requires that the gauge theory loop operators
satisfy the OPE
\beq
2L_d\, L_{d'}=-i L_{d\star d'}+i L_{d\circ d'},
\label{eq:OPE}
\eeq
where we have denoted by $d, d', d\star d'$ and $d\circ d'$
the sets of Dehn-Thurston parameters of the curves
$\gamma,\gamma',\gamma\star \gamma'$ and $\gamma\circ\gamma'$,
  respectively.
The OPE (\ref{eq:OPE}) then is
a prediction for the gauge theory loop operators,
to be discussed fully in Section~\ref{sec:discuss}.

In fact we can compute the OPE of loop operators
in the setup where the loops are not linked, by
generalizing an argument used in
 \cite{Kapustin:2007wm} for $\cN=4$ super Yang-Mills.
Since the loops are not linked,
the results do not necessarily have 
to agree with the Liouville loop operators. Interestingly, as we show below, 
the OPE of the non-linked loop operators exactly matches the 
{\it classical} skein relations, namely the relation (\ref{eq:skein}) at $b=0$.

In our $\cN=2$ gauge theories,
S-duality provides a specific identification
of the algebra of loop operators with the
group algebra of
\beq
(\Lambda_{m}\times \Lambda_w)/\cW,
\label{lattice}
\eeq
where 
$\Lambda_w\simeq \bZ^{3g-3+n}$
is
the weight lattice and $\cW\simeq (\bZ_2)^{3g-3+n}$ is the Weyl group,
both associated with the gauge group $G=SU(2)^{3g-3+n}$.
We have also defined the lattice $\Lambda_m$
of magnetic charges $(p_1,\ldots, p_{3g-3+n})\in \bZ^{3g-3+n}$
subject to  the Dirac quantization condition (\ref{dirac-cond}),
so that (\ref{lattice}) is the
space  of allowed Dehn-Thurston parameters
$d=(p_i,q_i).$%
\footnote{%
The Weyl group action replaces the conditions that $p_j\geq 0$,
and that if $p_j=0$ then $q_j\geq 0$.
}
Let us represent the group algebra of $\Lambda_{m}$
by  polynomials of $X_i, X_i^{-1}$,
and the group algebra of $\Lambda_w$ by
 polynomials of $Y_i, Y_i^{-1}$ where $i=1,\ldots, 3g-3+n$.
There is an obvious action of the Weyl group that
in particular exchanges $X_i$ with $X_i^{-1}$ and $Y_i$ with $Y_i^{-1}$.
The variable $Y_i$ carrying the electric charge is interpreted
as the Abelian Wilson loop with unit charge for the $i^{\rm th}$
$SU(2)$ group, in the Coulomb branch of the 
gauge theory where the $SU(2)$ is broken to a $U(1)$ subgroup.
Since the Wilson loop $L_{d_i}$ with the Dehn-Thurston parameters
all vanishing except for $q_i=1$ is in the fundamental
representation of the $i^{\rm th}$ $SU(2)$ gauge group, which has two weights
of opposite signs,
we associate to $L_{d_i}$ the polynomial $(Y_i+Y_i^{-1})/2$:
\beq
L_{d_i} \mapsto (Y_i+Y_i^{-1})/2.
\eeq
The factor of $1/2$ reflects the normalization of $L_{d_i}$ in
(\ref{eq:pestun}). For 't~Hooft loops, a similar interpretation is more subtle
and involves ``monopole bubbling'' \cite{Kapustin:2006pk}.
As explained in \cite{Kapustin:2007wm}, however, 
S-duality allows us to determine the polynomials
corresponding to an arbitrary non-linked loop operator $L_d$
associated to a connected simple closed geodesic,
and hence in fact a general closed non-self-intersecting
geodesic which always consists of
connected components.
This is because as in the case of $\cN=4$ super Yang-Mills with $SU(2)$ 
gauge group considered in
\cite{Kapustin:2006pk}, in the theories ${\cal T}_{g,n}$ any loop operator
can  be transformed to a pure Wilson loop by a 
duality transformation, which acts on the Dehn-Thurston parameters
\cite{Drukker:2009tz}.
Thus to an arbitrary loop operator corresponding to
a connected simple closed geodesic $\gamma_d$
with $d=(p_i,q_i)$, we associate a polynomial as
\beq
L_d\mapsto 
\frac{1}2
\left(
\prod_i Y_i^{p_i}X_i^{q_i}
+
\prod_i Y_i^{-p_i}X_i^{-q_i}
\right)\,.
\eeq
Given this identification of loop operators and the group algebra,
the OPE of non-linked loop operators is simply given
by the multiplication rules of the group algebra.
 
Let us now demonstrate that the algebra of
 non-linked  gauge theory loop operators 
is precisely that of the geodesic length/Liouville loop operators.
This is done by considering two connected simple closed
geodesics $\gamma$ and $\gamma'$ that share a single intersection point.
As before we choose a pants decomposition such that $\gamma$
is the closed geodesic corresponding to the first
gauge $SU(2)$ group.
Let $d=(0,\delta_{i1})$
and $d'=(p'_i,q'_i)$ be the vectors of Dehn-Thurston parameters
for $\gamma$ and $\gamma'$, respectively.
Then the OPE of the gauge theory loop operators $L_d$ and $L_{d'}$
is given by decomposing the product of the associated polynomials: 
\begin{align}
&
\left(
Y_1+Y^{-1}_1
\right)
\left(
\prod_i X_i^{p_i}Y_i^{q_i}
+\prod_i X_i^{-p_i}Y_i^{-q_i}
\right)
\\\nonumber
&\qquad=
\left(
\prod_i X_i^{p_i}Y_i^{q_i+\delta_{i1}}
+\prod_i X_i^{-p_i}Y_i^{-q_i-\delta_{i1}}
\right)
+
\left(
\prod_i X_i^{p_i}Y_i^{q_i-\delta_{i1}}
+\prod_i X_i^{-p_i}Y_i^{-q_i+\delta_{i1}}
\right).
\end{align}
Thus we get loop operators
associated with new Dehn-Thurston parameters
$(p_i,q_i+\delta_{i1})$
and 
$(p_i,q_i-\delta_{i1})$.
These are precisely the Dehn-Thurston parameters
for $\gamma\star \gamma'$ and $\gamma\circ \gamma'$.
Thus the OPE of the loop operators in this case is given by 
\beq
2 L_d L_{d'}= L_{d\star d'}+L_{d\circ d}.
\eeq
This is exactly what follows from the
classical skein relation, the skein relation for $b=0$.

We note that the operators which appear 
on the right hand side of (\ref{eq:OPE}), which is valid at $b=1$, 
are exactly the same as those appearing in the above equation 
at $b=0$, with modified coefficients. The fact that the same operators 
appear in the gauge theory calculation above supports our 
identification of loop operators in Liouville and the gauge theory 
and the conjectured form of the quantum OPE relations
(\ref{eq:OPE}).
It also
lends some hope that the relation 
\eqn{eq:OPE} at $b=1$ could  be shown to hold for linked 
loop operators in the gauge theory.

\subsection{Examples}

\subsubsection{$\cN=2^*$ Theory}
Let us study the algebra of loop operators in the
$\cN=2^*$ $ SU(2)$ gauge theory.
The Liouville theory realization was computed
for Wilson loops in Section~\ref{sec:wilson}, 
and for the minimal 't~Hooft loop as well as some dyonic 
loops in Section~\ref{sec:torus}.
We have the following Liouville loop operators
expressed as multiplication and difference operators:
\begin{align}
\cL(\gamma_{0,1})
&=
\frac{\cos(2\pi ba)}{\cos(\pi bQ)}\,,
\label{loop-torus11}\\
\cL(\gamma_{1,q})
&=
e^{-\frac{q}2 \pi i b^2}
\frac{\sin(\pi b^2)}{\sin(2\pi b^2)}
\bigg(
e^{2\pi i qb a}
\frac{\sin(2\pi b (a-m/2))}{\sin(2\pi b a)}
e^{-\frac{b}2 \partial_a}
\nn\\
&\hskip4cm
+e^{-2 \pi i qb a}
\frac{\sin(2\pi b(a+m/2))}{\sin(2\pi b a)}
e^{\frac{b}2 \partial_a}
\bigg).
\label{loop-torus1}
\end{align}
One can check that the quantum skein relation (\ref{eq:skein}) is
satisfied.%
\footnote{As discussed in Section~\ref{fig:torus},
we have chosen the precise phases of the Liouville loop operators 
so that they match those of the length operators.}
At $b=1$, the Liouville loop operators in 
(\ref{loop-torus11}) and (\ref{loop-torus1}) precisely satisfy the 't~Hooft 
commutation relation
\beq
\cL(\gamma_{1,q}) \cL(\gamma_{0,1})=
- \cL(\gamma_{0,1})\cL(\gamma_{1,q}).
\eeq

\subsubsection{${\cal N}=2$ $SU(2)_C$ with $N_F=4$}

Next, we consider ${\cal N}=2$ $SU(2)$ super Yang-Mills with four hypermultiplets
in the fundamental representation.
The Wilson loops $\cL(\gamma_{0,1})$ have the same expressions as
for $\cN=2^*$ (\ref{loop-torus11}).
The minimal 't~Hooft loop $\cL(\gamma_{2,0})$ (recall that the 
magnetic charge $p$ has to be even by Dirac quantization) computed in
(\ref{sphere-thooft})
is given by
\beq
\cL(\gamma_{2,0})
=T_2e^{-b\partial_\alpha}
+T_0
+T_{-2}
e^{b\partial_\alpha}
\eeq
with $T_2,T_0,T_{-2}$ given below (\ref{sphere-thooft}).
It is easy to check that the expected 't~Hooft commutation
is satisfied when $b=1$.

\section{Conclusions and Discussion}
\label{sec:discuss}

In this paper we have extended the correspondence \cite{Alday:2009aq}
between the $\cN=2$ theories ${\cal T}_{g,n}$
constructed by Gaiotto 
and Liouville theory on $C_{g,n}$ to include physical observables in the two theories.
We have advanced a precise correspondence between gauge theory loop operators -- for
Wilson, 't~Hooft and dyonic operators -- and operators in the Liouville conformal field theory.\footnote{This is based on the one-to-one map found in \cite{Drukker:2009tz} between the electric and magnetic charges of gauge theory loop operators and 
non-self-intersecting geodesics on the Riemann surface.} 
We have found a one-to-one correspondence between loop operators 
in Liouville theory, which we have
explicitly constructed, and gauge theory loop operators
$$
\text{loop operator in ${\cal T}_{g,n}$}
\quad\Longleftrightarrow\quad
\text{Liouville loop operator in $C_{g,n}$.}
$$
Our construction of Liouville loop operators, which are supported on 
geodesics in $C_{g,n}$, is modular invariant, and provides a natural 
basis for non-local observables in the Liouville conformal field theory. 
Using the relation between Liouville conformal field theory and quantum 
Teichm\"uller theory, we have also found a mapping between loop 
operators in gauge theory and geodesic length operators in quantum 
Teichm\"uller theory.

We have shown that our computation of Liouville loop operators exactly reproduces 
the expectation value of Wilson loop operators in the 
${\cal T}_{g,n}$ theories in \cite{Pestun:2007rz}. The modular
invariant construction of the Liouville loop operators provides 
us with a framework in which to calculate the expectation value 
of any dyonic operators by choosing a different pants decomposition 
of $C_{g,n}$, which 
maps the Wilson operator to a dyonic operator in the corresponding 
duality frame of ${\cal T}_{g,n}$. Our proposal, together with modular 
duality in Liouville theory 
\cite{Teschner:2001rv,Teschner:2003en,Teschner:2008qh} yields 
the expected action of the S-duality group $\Gamma({\cal T}_{g,n})$ 
on the loop operators in the ${\cal T}_{g,n}$ theory.

For specific examples of ${\cal T}_{g,n}$, we have explicitly computed 
the expectation value of 't~Hooft and dyonic operators. We have 
considered ${\cal N}=2^*$ and ${\cal N}=2$ $SU(2)$ gauge 
theory with $N_F=4$. Since no calculations of 't~Hooft operators 
have been performed beyond the ones for ${\cal N}=4$ super 
Yang-Mills in \cite{Gomis:2009ir,Gomis:2009xg}, our computations 
constitute predictions that would be interesting to confirm by a direct 
gauge theory analysis. A general algorithm for calculating 't~Hooft 
and dyonic operators in an arbitrary ${\cal T}_{g,n}$ theory has also 
been proposed, whereby the computation can be broken into a 
sequence of operations that depends on the details of the theory 
and on the choice of loop operator. 

We have also advanced a four dimensional field theory interpretation 
of the algebra of Liouville loop operators and geodesic length 
operators in quantum Teichm\"uller theory, which satisfy the 
quantum skein relation. The 't~Hooft commutation relations of gauge theory 
loop operators \cite{'tHooft:1977hy}, which describe the relations 
satisfied by loop operators that are Hopf-linked at a constant time 
slice in the four dimensional theory, have been identified with those
that follow from the
quantum skein relation of geodesic length operators at $b=1$. 
$$
\parbox{6cm}{'t~Hooft commutation relations of linked loop operators in ${\cal T}_{g,n}$}
\quad\Longleftrightarrow\quad
\parbox{6cm}{commutation relations 
from the quantum skein relation
at $b=1$.}
$$
Furthermore, we have shown that the OPE of  non-linked  gauge theory loop 
operators, predicted by the proposed action of S-duality \cite{Drukker:2009tz},
precisely matches the classical skein relation 
of geodesic length operators (at $b=0$). 
$$
\text{OPE of commuting loop operators in ${\cal T}_{g,n}$}
\quad\Longleftrightarrow\quad
\text{classical skein relation}. 
$$
We conjecture that gauge theory operators corresponding 
to the Liouville loop operators at $b=1$ are
Hopf-linked with each other on the equatorial $S^3$ in $S^4$. 
Furthermore, we conjecture that the OPE coefficients
of loop operators that correspond to Liouville loop operators 
at $b=1$ are given by the quantum skein relations at $b=1$.

An important open problem for the future is to extend Pestun's 
localization calculation \cite{Pestun:2007rz} for the Wilson 
loops of ${\cal N}=2$ gauge theories to general dyonic loop 
operators. Our Liouville calculation of the expectation value 
of loop operators provides some hints. As we have shown, 
the expression for the 't~Hooft operators is quite similar to 
that for Wilson loops. For 't~Hooft loops, there are extra terms 
in the matrix integral with shifted arguments which one may 
speculate correspond to the contribution of the various 
weights characterizing the magnetic charges of the loop 
operator. This interpretation is consistent with the calculation 
of 't~ Hooft operators in ${\cal N}=4$ super Yang-Mills 
\cite{Gomis:2009ir,Gomis:2009xg}, where it was found that 
there are multiple saddle point in the path integral. The 
physical interpretation of the extra saddle points has to do 
with monopole screening, whereby the singularity produced 
by the insertion of an 't~Hooft operator is screened by regular 
monopoles, resulting in saddle points with weaker singularities 
and which is in one-to-one correspondence with the weights labeling 
the representation of the 't~Hooft loop. Another important open 
problem is to provide a gauge theory interpretation of the 
prefactors multiplying the conformal blocks, which should 
correspond to the one loop determinant of fluctuations 
around the localization calculation for the 't~Hooft loop.

It is of importance to attain a deeper understanding of the 
physical and mathematical foundations that underlie
the gauge/Liouville correspondence. A possible point 
of contact is the parallel that exists between their Hilbert 
spaces and their inner products. The expression (\ref{pestun}) for the Wilson loop
expectation value that follows from localization
in gauge theory on $S^4$
has a natural interpretation as the expectation
value of the loop operator evaluated 
for the wave function defined by the Nekrasov instanton 
partition function.
Let $\theta$ be the latitude on $S^4$.
The instantons that contribute to the holomorphic part
are localized at the North Pole $\theta=0$, 
and the anti-instantons at the South Pole $\theta=\pi$.
The Wilson loop is located at an arbitrary latitude on the $S^4$, which without loss
of generality can be taken to be in 
the ``equator'' $S^3$ at $\theta=\pi/2$.
Thus it is natural to interpret the holomorphic (anti-holomorphic)
Nekrasov partition function
as the wave function of a state, obtained by performing the gauge path-integral
on the Northern (Southern) Hemisphere, 
in the BPS Hilbert space of the gauge theory
canonically quantized on $S^3$.

It is then very suggestive to compare the resulting
picture for the gauge theory loop expecation value with
the form the expectation value takes in quantum Teichm\"uller
theory, equation (\ref{loopislength}). It seems natural to propose
identifying the Hilbert space of quantum Teichm\"uller theory
with the BPS Hilbert space ${\cal H}_{\rm BPS}$
and the coherent state  $|q\rangle$
with the state in ${\cal H}_{\rm BPS}$ created by
performing the gauge theory path-integral over the lower
hemisphere. The geodesic length operators would thereby
directly be related to the projection of the gauge theory
loop operators onto the BPS Hilbert space.

Extensions of our results to higher rank gauge groups would 
also be worth pursuing. On the Riemann surface they involve 
conformal field theories with ${\cal W}$-symmetry 
\cite{Alday:2009aq,Wyllard:2009hg}. Increasing the rank of the 
gauge groups also admits an interesting generalization of 
quantum Teichm\"uller theory \cite{MR2233852}, which quantizes the 
moduli space of $SL(N)$ flat connections for $N>2$ 
\cite{MR1174252}. The corresponding gauge theories constructed 
out of the Riemann surface contain building blocks with no weakly 
coupled Lagrangian description \cite{Gaiotto-N=2}. Understanding 
the loop operators in the two dimensional conformal field theory as 
well as the generalization of the length operators in the quantization 
of the moduli space of $ SL(N)$ flat connections is bound to shed 
light in this interesting class of gauge theories.

Clearly it is of great interest to obtain a more direct interpretation of 
the correspondence between gauge theory and Liouville theory. 
Finding a more first principles interpretation
may also provide hints of how to extend the two dimensional 
interpretation of four dimensional ${\cal N}=2$ gauge theories away 
from the choice of localization parameters
satisfying $\epsilon_1/\epsilon_2=1$ 
relevant for the correspondence in 
Liouville. Progress in realizing the Nekrasov instanton 
partition function \cite{Nekrasov:2002qd,Nekrasov:2003rj} 
for arbitrary $\epsilon_1$ and $\epsilon_2$ have been made 
in topological string theory, where amplitudes sewn from a 
``refined" topological vertex \cite{Hollowood:2003cv,Iqbal:2007ii} 
were shown to reproduce Nekrasov's partition function. It 
would be interesting to incorporate the insights developed
in that context to extend these exciting correspondences.

\section*{Acknowledgements}
We would like to thank Harald Dorn, Davide Gaiotto, George Jorjadze,
Natalia Saulina and Herman Verlinde for helpful
discussions. N.D., J.G. and T.O. would like to thank the Benasque Center for Science
for its hospitality.
N.D is grateful to DESY Hamburg for its hospitality during the course of this 
work.
Research at the Perimeter Institute is
supported in part by the Government of Canada through NSERC and by
the Province of Ontario through MRI. J.G. also acknowledges further
support from an NSERC Discovery Grant and from and an ERA grant  by the Province of Ontario.
J.T. acknowledges support from the EC by the Marie Curie
    Excellence Grant MEXT-CT-2006-042695

\appendix

\section{Special Functions}
\label{spefunct}

\subsection{The Function $\Gamma_b(x)$}

The function $\Gamma_b(x)$ is a close relative of the double
Gamma function studied in \cite{Ba,Sh}. It 
can be defined by means of the integral representation
\begin{equation}
\log\Gamma_b(x)=\int\limits_0^{\infty}\frac{dt}{t}
\biggl(\frac{e^{-xt}-e^{-Qt/2}}{(1-e^{-bt})(1-e^{-t/b})}-
\frac{(Q-2x)^2}{8e^t}-\frac{Q-2x}{t}\biggl)\;\;.
\end{equation}
Important properties of $\Gamma_b(x)$ are
\begin{align}
{}& \text{(i) Functional equation:} \quad
\Gamma_b(x+b)=\sqrt{2\pi}b^{bx-\frac{1}{2}}\Gamma^{-1}(bx)\Gamma_b(x). \label{Ga_feq}\\
{}& \text{(ii) Analyticity:}\quad
\Gamma_b(x)\;\text{is meromorphic,}\nonumber\\ 
{}& \hspace{2.5cm}\text{poles:}\;\, 
x=-nb-mb^{-1}, n,m\in\BZ^{\geq 0}.\\
{}& \text{(iii) Self-duality:}\quad \Gamma_b(x)=\Gamma_{1/b}(x). \label{self-dual} 
\end{align}

\subsection{The Function $S_b(x)$}

The function $S_b(x)$ may be defined in terms of 
$\Gamma_b(x)$ as follows
\begin{equation}\label{sbdef}
S_b(x)=\Gamma_b(x)\,/\,\Gamma_b\big(Q-x)\;.
\end{equation}
An integral that represents $\log S_b(x)$ is
\begin{equation}
\log S_b(x)=\int\limits_0^{\infty}\frac{dt}{t}
\biggl(\frac{\sinh t(Q-2x)}{2\sinh bt\sinh b^{-1}t}-
\frac{Q-2x}{2t}\biggl)\;\;.
\end{equation}
The most important properties for our purposes are 
\begin{align}
{}& \text{(i) Functional equation:} \quad
S_b(x+b)=2\sin \pi b x\;
S_b(x). \label{sb_feq}\\
{}& \text{(ii) Analyticity:}\quad
S_b(x)\;\text{is meromorphic,}\nonumber\\ 
{}& \hspace{2.7cm}\text{poles:}\;\, 
x=-(nb+mb^{-1}), n,m\in\BZ^{\geq 0}.\\
{}& \hspace{2.7cm}\text{zeros:}\;\, 
x=Q+(nb+mb^{-1}), n,m\in\BZ^{\geq 0}.\nonumber \\
{}& \text{(iii) Self-duality:}\quad S_b(x)=S_{1/b}(x). \\
{}& \text{(iv) Inversion relation:}\quad S_b(x)S_b(Q-x)=1.\\
{}& \text{(v) Asymptotics:} \quad S_b(x)\sim e^{\mp\frac{\pi i}{2}x(x-Q)}\;\,{\rm for}\;{\rm Im}(x)\ra\pm\infty\\
{}& \text{(vi) Residue:} \quad {\rm res}_{x=c_b}S_b(x)
=(2\pi)^{-1}\label{sbRes}.
\end{align}

\subsection{$\up$ Function}
The $\up$ may be defined in terms of
$\Gamma_b$ as follows 
\begin{equation}
\up(x)^{-1} \equiv \Gamma_b(x)\Gamma_b(Q-x)\;.
\end{equation}
An integral representation convergent in the strip $0<{\rm Re}(x)<Q$ is 
\beq
\text{log}\up(x)=\int_{0}^{\infty}\frac{dt}{t}\left\lbrack\left(\frac{Q}{2}-x\right)^{2}e^{-t}-\frac{\text{sinh}^{2}(\frac{Q}{2}-x)\frac{t}{2}}{\text{sinh}\frac{bt}{2}\text{sinh}\frac{t}{2b}}\right\rbrack\;.
\eeq
Properties:
\begin{align}
{}& \text{(i) Functional equation:} \quad
\up(x+b)=\frac{\Ga_b(bx)}{\Ga_b(1-bx)}b^{1-2bx}\;
\up(x). \label{up_feq}\\
{}& \text{(ii) Analyticity:}\quad
\up(x)\;\text{is entire analytic,}\nonumber\\ 
{}& \hspace{2.7cm}\text{zeros:}\;\, 
x=-(nb+mb^{-1}), n,m\in\BZ^{\geq 0}.\\
{}& \hspace{2.7cm}\phantom{\text{zeros:}}\;\, 
x=Q+(nb+mb^{-1}), n,m\in\BZ^{\geq 0}.\nonumber \\
{}& \text{(iii) Self-duality:}\quad \up(x)=\Upsilon_{1/b}(x).
\end{align}

\section{Fusion Matrices}
\label{app:fusion}

In the following we will use the fusion coefficients involving some degenerate 
fields with $\alpha=-b/2$. We define
\beq
F_{s,s'}\left[\begin{smallmatrix}
\al_3 & -\frac{b}{2}\\
\al_4 & \phantom{r}\al_{1}%
\end{smallmatrix}
\right]\equiv
F_{\alpha_{1}-\frac{sb}{2},\alpha_3-\frac{s'b}{2}}
\left[\begin{smallmatrix}
\al_3 & -\frac{b}{2}\\
\al_4 & \phantom{r}\al_{1}%
\end{smallmatrix}
\right],
\eeq
where $s,s' =\pm$. 
\begin{equation}\label{fexp}\begin{aligned}
F_{++}&=\frac{\Gamma(b(2\al_1-b))\Gamma(b(b-2\al_3)+1)}{\Gamma(b(\al_1-\al_3-\al_4+b/2)+1)\Gamma(b(\al_1-\al_3+\al_4-b/2))} \\
F_{+-}&= \frac{\Gamma(b(2\al_1-b))\Gamma(b(2\al_3-b)-1)}{\Gamma(b(\al_1+\al_3+\al_4-3b/2)-1)\Gamma(b(\al_1+\al_3-\al_4-b/2))} \\
F_{-+}&=\frac{\Gamma(2-b(2\al_1-b))\Gamma(b(b-2\al_3)+1)}{\Gamma(2-b(\al_1+\al_3+\al_4-3b/2))\Gamma(1-b(\al_1+\al_3-\al_4-b/2))} \\
F_{--}&= \frac{\Gamma(2-b(2\al_1-b))\Gamma(b(2\al_3-b)-1)}{\Gamma(b(-\al_1+\al_3+\al_4-b/2))\Gamma(b(-\al_1+\al_3-\al_4+b/2)+1)} 
\end{aligned}
\end{equation}

Equivalently for the renormalized fusion coefficients we define
\beq
G_{s,s'}\left[\begin{smallmatrix}
\al_3 & -\frac{b}{2}\\
\al_4 & \phantom{r}\al_{1}%
\end{smallmatrix}
\right]\equiv
G_{\alpha_{1}-\frac{sb}{2},\alpha_3-\frac{s'b}{2}}
\left[\begin{smallmatrix}
\al_3 & -\frac{b}{2}\\
\al_4 & \phantom{r}\al_{1}%
\end{smallmatrix}
\right],
\eeq
These are very simple \cite{Ponsot:1999uf}
\begin{equation}
\begin{aligned}
G_{++}=& \frac{\sin\left(\pi b(\al_4+\al_3-\al_1-\frac{b}{2})\right)}
{\sin\left(\pi b(2\al_3-b)\right)}\\
G_{-+}=& \frac{\sin\left(\pi b(\al_3+\al_1-\al_4-\frac{b}{2})\right)}
{\sin\left(\pi b(2\al_3-b)\right)}
\end{aligned}\qquad
\begin{aligned}
G_{+-}=& \frac{\sin\left(\pi b(\al_4+\al_3+\al_1-\frac{3b}{2})\right)}
{\sin\left(\pi b(2\al_3-b)\right)}\\
G_{--}=& \frac{\sin\left(\pi b(\al_3-\al_1-\al_4+\frac{b}{2})\right)}
{\sin\left(\pi b(2\al_3-b)\right)}
\end{aligned}
\end{equation}
which can also be written as
\begin{equation}
\begin{aligned}
G_{s_1,-s_2}
&=s_1 \frac{\sin\left(\pi b(\alpha_4+s_1\alpha_3+s_2\alpha_1
-(1+s_1+s_2)\frac{b}{2})\right)}
{\sin\left(\pi b(2\al_3-b)\right)}\,.
\label{Gpm}
\end{aligned}
\end{equation}

Even more special cases involve a pair of degenerate fields. Creating 
them out of the identity is given by ($\alpha'=\alpha-sb/2$)
\begin{equation}
\raisebox{0mm}[10mm][12mm]{\parbox{14mm}{\begin{center}
\begin{fmfgraph*}(10,15)
\fmfbottom{bl,br}
\fmftop{tl,tr}
\fmf{plain,width=2}{bl,b1,br}
\fmf{phantom}{tl,t1,tr}
\fmffreeze
\fmf{boson,width=.5}{tl,c1,tr}
\fmf{dashes,width=.5}{c1,b1}
\fmfv{label=$\alpha$,label.angle=-90,label.dist=8}{bl}
\fmfv{label=$\alpha$,label.angle=-90,label.dist=8}{br}
\end{fmfgraph*}\end{center}}}
\to
\raisebox{0mm}[10mm][12mm]{\parbox{26mm}{\begin{center}
\begin{fmfgraph*}(20,15)
\fmfbottom{bl,br}
\fmftop{tl,tr}
\fmf{plain,width=2}{bl,b1,b2,b3,br}
\fmf{phantom}{tl,t1,t2,t3,tr}
\fmffreeze
\fmf{boson,width=.5}{t1,b1}
\fmf{boson,width=.5}{t3,b3}
\fmfv{label=$\alpha$,label.angle=-90,label.dist=8}{bl}
\fmfv{label=$\alpha$,label.angle=-90,label.dist=8}{br}
\fmfv{label=$\alpha'$,label.angle=-90,label.dist=5}{b2}
\fmfv{label=$-\frac{b}{2}$,label.angle=180}{t1}
\fmfv{label=$-\frac{b}{2}$,label.angle=0}{t3}
\end{fmfgraph*}\end{center}}}
\equiv\quad
G^{-1}_{0,\alpha'}
\left[\begin{smallmatrix}
-\frac{b}{2} & -\frac{b}{2}\\
\alpha &\alpha
\end{smallmatrix}\right]
=G_{-,s}
\left[\begin{smallmatrix}
\alpha & -\frac{b}{2}\\
\alpha &-\frac{b}{2}
\end{smallmatrix}\right]
=s\,\frac{\sin\left(\pi b^2\right)}
{\sin\left(\pi b(2\alpha-Q)\right)}\,,
\label{w10}
\end{equation}
Analogously fusing them back to the identity
\begin{equation}
\begin{aligned}
G_{\alpha',0}
\left[\begin{smallmatrix}
-\frac{b}{2} & -\frac{b}{2}\\
\alpha & \alpha
\end{smallmatrix}\right]
&=G_{s,-}
=s\,\frac{\sin\left(\pi b(2\alpha'-Q)\right)}
{\sin\left(2\pi b^2\right)}\,,
\\
G_{\alpha,0}
\left[\begin{smallmatrix}
-\frac{b}{2} & -\frac{b}{2}\\
\alpha' & \alpha'
\end{smallmatrix}\right]
&=G_{-s,-}
=-s\,\frac{\sin\left(\pi b(2\alpha-Q)\right)}
{\sin\left(2\pi b^2\right)}\,.
\end{aligned}
\label{fuse-proj}
\end{equation}

\subsection{Higher Degenerate Fields}
\label{app:higher}

Other quantities we will need involve the higher degenerate fields 
with $-qb/2$. The splitting of the identity into a pair of $-qb/2$ fields can 
be done recursively
\beq
\begin{aligned}
\label{pair-qb/2}
\raisebox{0mm}[13mm][13mm]{\parbox{40mm}{
\begin{fmfgraph*}(30,20)
\fmfbottom{bl,br}
\fmftop{tl,tr}
\fmf{plain,width=2}{bl,b1,b2,b3,b4,b5,br}
\fmf{phantom}{tl,t1,t2,t3,t4,t5,tr}
\fmffreeze
\fmf{dashes,width=.5,tension=3}{b2,c2}
\fmf{boson,width=1.5}{c2,t1}
\fmf{phantom}{c2,t3}
\fmffreeze
\fmf{boson,width=1.5}{c2,c4}
\fmf{boson,width=.5}{c4,t3}
\fmf{boson,width=1}{c4,t5}
\fmf{phantom}{c4,br}
\fmfv{label=$\alpha$,label.angle=-90,label.dist=8}{bl}
\fmfv{label=$\alpha$,label.angle=-90,label.dist=8}{br}
\fmfv{label=$-\frac{qb}{2}$,label.angle=180}{t1}
\fmfv{label=$-\frac{qb}{2}$,label.angle=-120}{c4}
\fmfv{label=$-\frac{b}{2}$,label.angle=180,label.dist=4}{t3}
\fmfv{label=$-\frac{(q-1)b}{2}$,label.angle=-90}{tr}
\end{fmfgraph*}}}
&\to
\raisebox{0mm}[13mm][13mm]{\parbox{47mm}{\hskip6mm
\begin{fmfgraph*}(30,20)
\fmfbottom{bl,br}
\fmftop{tl,tr}
\fmf{plain,width=2}{bl,b1,b2,b3,b4,b5,br}
\fmf{phantom}{tl,t1,t2,t3,t4,t5,tr}
\fmffreeze
\fmf{dashes,width=.5,tension=3}{b4,c4}
\fmf{boson,width=1}{c4,t5}
\fmf{phantom}{c4,t3}
\fmffreeze
\fmf{boson,width=1}{c4,c2}
\fmf{boson,width=.5}{c2,t3}
\fmf{boson,width=1.5}{c2,t1}
\fmf{phantom}{c2,bl}
\fmfv{label=$\alpha$,label.angle=-90,label.dist=8}{bl}
\fmfv{label=$\alpha$,label.angle=-90,label.dist=8}{br}
\fmfv{label=$-\frac{(q-1)b}{2}$,label.angle=-60}{t5}
\fmfv{label=$-\frac{(q-1)b}{2}$,label.angle=-90}{c2}
\fmfv{label=$-\frac{b}{2}$,label.angle=0,label.dist=4}{t3}
\fmfv{label=$-\frac{qb}{2}$,label.angle=180}{t1}
\end{fmfgraph*}}}
\to
\raisebox{0mm}[13mm][13mm]{\parbox{38mm}{\hskip10mm
\begin{fmfgraph*}(25,20)
\fmfbottom{bl,br}
\fmftop{tl,tr}
\fmf{plain,width=2}{bl,b1,b2,b3,b4,br}
\fmf{phantom}{tl,t1,t2,t3,t4,tr}
\fmffreeze
\fmf{boson,width=1.5}{tl,c1}
\fmf{boson,width=.5}{t2,c1}
\fmf{boson,width=1}{c1,b1}
\fmf{boson,width=1}{t4,b4}
\fmfv{label=$\alpha$,label.angle=-90,label.dist=8}{bl}
\fmfv{label=$\alpha$,label.angle=-90,label.dist=8}{br}
\fmfv{label=$\alpha^{(k')}$,label.angle=-90,label.dist=4}{b3}
\fmfv{label=$-\frac{(q-1)b}{2}$,label.angle=-135}{c1}
\fmfv{label=$-\frac{(q-1)b}{2}$,label.angle=-45}{t4}
\end{fmfgraph*}}}
\\&\to
\raisebox{0mm}[13mm][13mm]{\parbox{47mm}{\hskip6mm
\begin{fmfgraph*}(30,20)
\fmfbottom{bl,br}
\fmftop{tl,tr}
\fmf{plain,width=2}{bl,b1,b2,b3,b4,b5,br}
\fmf{phantom}{tl,t1,t2,t3,t4,t5,tr}
\fmffreeze
\fmf{boson,width=1.5}{t1,b1}
\fmf{boson,width=.5}{t3,b3}
\fmf{boson,width=1}{t5,b5}
\fmfv{label=$\alpha$,label.angle=-90,label.dist=8}{bl}
\fmfv{label=$\alpha$,label.angle=-90,label.dist=8}{br}
\fmfv{label=$\alpha^{(k)}$,label.angle=-90,label.dist=4}{b2}
\fmfv{label=$\alpha^{(k')}$,label.angle=-90,label.dist=4}{b4}
\fmfv{label=$-\frac{qb}{2}$,label.angle=180}{t1}
\fmfv{label=$-\frac{b}{2}$,label.angle=180}{t3}
\fmfv{label=$-\frac{(q-1)b}{2}$,label.angle=0}{t5}
\end{fmfgraph*}}}
\to
\raisebox{0mm}[13mm][13mm]{\parbox{38mm}{\begin{center}
\begin{fmfgraph*}(25,20)
\fmfbottom{bl,br}
\fmftop{tl,tr}
\fmf{plain,width=2}{bl,b1,b2,b3,b4,br}
\fmf{phantom}{tl,t1,t2,t3,t4,tr}
\fmffreeze
\fmf{boson,width=1.5}{t1,b1}
\fmf{boson,width=1}{tr,c4}
\fmf{boson,width=.5}{t3,c4}
\fmf{boson,width=1.5}{c4,b4}
\fmfv{label=$\alpha$,label.angle=-90,label.dist=8}{bl}
\fmfv{label=$\alpha$,label.angle=-90,label.dist=8}{br}
\fmfv{label=$\alpha^{(k)}$,label.angle=-90,label.dist=4}{b3}
\fmfv{label=$-\frac{qb}{2}$,label.angle=180}{t1}
\fmfv{label=$-\frac{qb}{2}$,label.angle=-45}{c4}
\end{fmfgraph*}\end{center}}}
\end{aligned}
\eeq
Using $\alpha^{(k)}=\alpha-kb/2$, $\alpha^{(k')}=\alpha-m'b/2$ and 
$k=k'+s_q$, we have for these fusion steps
\begin{align}
G_{0,\alpha-\frac{kb}{2}}
&\left[\begin{smallmatrix}
\alpha&-\frac{qb}{2} \\
\alpha&-\frac{qb}{2} 
\end{smallmatrix}\right]
=G_{-\frac{qb}{2},-\frac{(q-1)b}{2}}
\left[\begin{smallmatrix}
-\frac{qb}{2} &-\frac{b}{2} \\
0 &-\frac{(q-1)b}{2}
\end{smallmatrix}\right]
G^{-1}_{0,\alpha-\frac{k'b}{2}}
\left[\begin{smallmatrix}
-\frac{(q-1)b}{2} &-\frac{(q-1)b}{2} \\
\alpha&\alpha
\end{smallmatrix}\right]
\nonumber\\&\hskip2cm
\times
G^{-1}_{-\frac{(q-1)b}{2},\alpha-\frac{kb}{2}}
\left[\begin{smallmatrix}
-\frac{qb}{2} &-\frac{b}{2}\\
\alpha&\alpha-\frac{k'b}{2}
\end{smallmatrix}\right]
G_{\alpha-\frac{k'b}{2},-\frac{qb}{2}}
\left[\begin{smallmatrix}
-\frac{b}{2}&-\frac{(q-1)b}{2}\\
\alpha-\frac{kb}{2}&\alpha
\end{smallmatrix}\right]
\\\nonumber&
=G_{+-}
\left[\begin{smallmatrix}
-\frac{qb}{2} &-\frac{b}{2} \\
0 &-\frac{(q-1)b}{2}
\end{smallmatrix}\right]
G_{0,\alpha-\frac{k'b}{2}}
\left[\begin{smallmatrix}
\alpha&-\frac{(q-1)b}{2} \\
\alpha&-\frac{(q-1)b}{2} 
\end{smallmatrix}\right]
G_{-,s_q}
\left[\begin{smallmatrix}
\alpha-\frac{k'b}{2} &-\frac{b}{2}\\
\alpha&-\frac{qb}{2}
\end{smallmatrix}\right]
G_{-s_q,+}
\left[\begin{smallmatrix}
-\frac{(q-1)b}{2}&-\frac{b}{2}\\
\alpha&\alpha-\frac{kb}{2}
\end{smallmatrix}\right]
\end{align}
Using the explicit expressions for the fusion matrices we find
\beq
G_{0,\alpha-\frac{kb}{2}}
\left[\begin{smallmatrix}
\alpha&-\frac{qb}{2} \\
\alpha&-\frac{qb}{2} 
\end{smallmatrix}\right]
=-s_q\,\frac{\sin^2(\pi b^2(s_qq+k)/2)}
{\sin(\pi b^2q)\sin\left(\pi b(2\alpha-(k'+1)b)\right)}
G_{0,\alpha-\frac{k'b}{2}}
\left[\begin{smallmatrix}
\alpha&-\frac{(q-1)b}{2} \\
\alpha&-\frac{(q-1)b}{2} 
\end{smallmatrix}\right].
\eeq

Writing $k=s_1+\cdots +s_q$ we get for the full product
\beq
G_{0,\alpha-\frac{kb}{2}}
\left[\begin{smallmatrix}
\alpha&-\frac{qb}{2} \\
\alpha&-\frac{qb}{2} 
\end{smallmatrix}\right]
=(-1)^\frac{q+k}{2}\prod_{i=1}^q
\frac{\sin^2\left(\pi b^2(is_i+(s_1+\cdots +s_i))/2\right)}
{\sin(\pi b^2i)\sin\left(\pi b(2\alpha-(1+s_1+\cdots+ s_{i-1})b)\right)}\,,
\eeq
which has to be summed over all orderings of the $s_i$.

For example, for $q=2$ summing over the choices of $s_1$ and $s_2$ gives
\beq
\begin{aligned}
G_{0,\alpha-b}
\left[\begin{smallmatrix}\alpha& -b\\\alpha & -b\end{smallmatrix}\right]
&=\frac{\sin(\pi b^2)\sin(2\pi b^2)}
{\sin\left(\pi b(2\alpha-2b)\right)\sin\left(\pi b(2\alpha-b)\right)}\\
G_{0,\alpha}
\left[\begin{smallmatrix}\alpha& -b\\\alpha & -b\end{smallmatrix}\right]
&=-\frac{\sin^3(\pi b^2)}
{\sin\left(2\pi b^2\right)\sin\left(\pi b(2\alpha-b)\right)}
\left(\frac{1}{\sin\left(\pi b(2\alpha-2b)\right)}
+\frac{1}{\sin\left(2\pi b\alpha\right)}\right)\\
&=-\frac{\sin^2(\pi b^2)}
{\sin\left(\pi b(2\alpha-2b)\right)
\sin\left(2\pi b\alpha\right)}\\
G_{0,\alpha+b}
\left[\begin{smallmatrix} \alpha & -b\\\alpha & -b\end{smallmatrix}\right]
&=\frac{\sin(\pi b^2)\sin(2\pi b^2)}
{\sin\left(2\pi b\alpha\right)\sin\left(\pi b(2\alpha-b)\right)}
\end{aligned}
\label{1tob}
\eeq

Similarly we can go in the opposite direction
\beq
\begin{aligned}
\label{fusing-qb/2}
\raisebox{0mm}[13mm][13mm]{\parbox{38mm}{\begin{center}
\begin{fmfgraph*}(25,20)
\fmfbottom{bl,br}
\fmftop{tl,tr}
\fmf{plain,width=2}{bl,b1,b2,b3,b4,br}
\fmf{phantom}{tl,t1,t2,t3,t4,tr}
\fmffreeze
\fmf{boson,width=1.5}{t1,b1}
\fmf{boson,width=1}{tr,c4}
\fmf{boson,width=.5}{t3,c4}
\fmf{boson,width=1.5}{c4,b4}
\fmfv{label=$\alpha$,label.angle=-90,label.dist=8}{bl}
\fmfv{label=$\alpha$,label.angle=-90,label.dist=8}{br}
\fmfv{label=$\alpha^{(k)}$,label.angle=-90,label.dist=4}{b3}
\fmfv{label=$-\frac{qb}{2}$,label.angle=180}{t1}
\fmfv{label=$-\frac{qb}{2}$,label.angle=-45}{c4}
\end{fmfgraph*}\end{center}}}
&\to
\raisebox{0mm}[13mm][13mm]{\parbox{47mm}{\hskip6mm
\begin{fmfgraph*}(30,20)
\fmfbottom{bl,br}
\fmftop{tl,tr}
\fmf{plain,width=2}{bl,b1,b2,b3,b4,b5,br}
\fmf{phantom}{tl,t1,t2,t3,t4,t5,tr}
\fmffreeze
\fmf{boson,width=1.5}{t1,b1}
\fmf{boson,width=.5}{t3,b3}
\fmf{boson,width=1}{t5,b5}
\fmfv{label=$\alpha$,label.angle=-90,label.dist=8}{bl}
\fmfv{label=$\alpha$,label.angle=-90,label.dist=8}{br}
\fmfv{label=$\alpha^{(k)}$,label.angle=-90,label.dist=4}{b2}
\fmfv{label=$\alpha^{(k')}$,label.angle=-90,label.dist=4}{b4}
\fmfv{label=$-\frac{qb}{2}$,label.angle=180}{t1}
\fmfv{label=$-\frac{b}{2}$,label.angle=180}{t3}
\fmfv{label=$-\frac{(q-1)b}{2}$,label.angle=0}{t5}
\end{fmfgraph*}}}
\to
\raisebox{0mm}[13mm][13mm]{\parbox{38mm}{\hskip10mm
\begin{fmfgraph*}(25,20)
\fmfbottom{bl,br}
\fmftop{tl,tr}
\fmf{plain,width=2}{bl,b1,b2,b3,b4,br}
\fmf{phantom}{tl,t1,t2,t3,t4,tr}
\fmffreeze
\fmf{boson,width=1.5}{tl,c1}
\fmf{boson,width=.5}{t2,c1}
\fmf{boson,width=1}{c1,b1}
\fmf{boson,width=1}{t4,b4}
\fmfv{label=$\alpha$,label.angle=-90,label.dist=8}{bl}
\fmfv{label=$\alpha$,label.angle=-90,label.dist=8}{br}
\fmfv{label=$\alpha^{(k')}$,label.angle=-90,label.dist=4}{b3}
\fmfv{label=$-\frac{(q-1)b}{2}$,label.angle=-135}{c1}
\fmfv{label=$-\frac{(q-1)b}{2}$,label.angle=-45}{t4}
\end{fmfgraph*}}}
\\&\to
\raisebox{0mm}[13mm][13mm]{\parbox{47mm}{\hskip6mm
\begin{fmfgraph*}(30,20)
\fmfbottom{bl,br}
\fmftop{tl,tr}
\fmf{plain,width=2}{bl,b1,b2,b3,b4,b5,br}
\fmf{phantom}{tl,t1,t2,t3,t4,t5,tr}
\fmffreeze
\fmf{dashes,width=.5,tension=3}{b4,c4}
\fmf{boson,width=1}{c4,t5}
\fmf{phantom}{c4,t3}
\fmffreeze
\fmf{boson,width=1}{c4,c2}
\fmf{boson,width=.5}{c2,t3}
\fmf{boson,width=1.5}{c2,t1}
\fmf{phantom}{c2,bl}
\fmfv{label=$\alpha$,label.angle=-90,label.dist=8}{bl}
\fmfv{label=$\alpha$,label.angle=-90,label.dist=8}{br}
\fmfv{label=$-\frac{(q-1)b}{2}$,label.angle=-60}{t5}
\fmfv{label=$-\frac{(q-1)b}{2}$,label.angle=-90}{c2}
\fmfv{label=$-\frac{b}{2}$,label.angle=0,label.dist=4}{t3}
\fmfv{label=$-\frac{qb}{2}$,label.angle=180}{t1}
\end{fmfgraph*}}}
\to
\raisebox{0mm}[13mm][13mm]{\parbox{40mm}{
\begin{fmfgraph*}(30,20)
\fmfbottom{bl,br}
\fmftop{tl,tr}
\fmf{plain,width=2}{bl,b1,b2,b3,b4,b5,br}
\fmf{phantom}{tl,t1,t2,t3,t4,t5,tr}
\fmffreeze
\fmf{dashes,width=.5,tension=3}{b2,c2}
\fmf{boson,width=1.5}{c2,t1}
\fmf{phantom}{c2,t3}
\fmffreeze
\fmf{boson,width=1.5}{c2,c4}
\fmf{boson,width=.5}{c4,t3}
\fmf{boson,width=1}{c4,t5}
\fmf{phantom}{c4,br}
\fmfv{label=$\alpha$,label.angle=-90,label.dist=8}{bl}
\fmfv{label=$\alpha$,label.angle=-90,label.dist=8}{br}
\fmfv{label=$-\frac{qb}{2}$,label.angle=180}{t1}
\fmfv{label=$-\frac{qb}{2}$,label.angle=-120}{c4}
\fmfv{label=$-\frac{b}{2}$,label.angle=180,label.dist=4}{t3}
\fmfv{label=$-\frac{(q-1)b}{2}$,label.angle=-90}{tr}
\end{fmfgraph*}}}
\end{aligned}
\eeq
This is represented by the fusion matrices
\begin{align}
G_{\alpha-\frac{kb}{2},0}
&\left[\begin{smallmatrix}
-\frac{qb}{2} & -\frac{qb}{2}\\
\alpha &\alpha
\end{smallmatrix}\right]
=
G^{-1}_{-\frac{qb}{2},\alpha-\frac{k'b}{2}}
\left[\begin{smallmatrix}
-\frac{b}{2} & -\frac{(q-1)b}{2}\\
\alpha-\frac{kb}{2} &\alpha
\end{smallmatrix}\right]
G_{\alpha-\frac{kb}{2},-\frac{(q-1)b}{2}}
\left[\begin{smallmatrix}
-\frac{qb}{2} & -\frac{b}{2}\\
\alpha&\alpha-\frac{k'b}{2}
\end{smallmatrix}\right]
\nonumber\\&\hskip1in
\times
G_{\alpha-\frac{k'b}{2},0}
\left[\begin{smallmatrix}
-\frac{(q-1)b}{2} & -\frac{(q-1)b}{2}\\
\alpha &\alpha
\end{smallmatrix}\right]
G_{-\frac{(q-1)b}{2},-\frac{qb}{2}}
\left[\begin{smallmatrix}
-\frac{b}{2} & -\frac{(q-1)b}{2}\\
-\frac{qb}{2}&0
\end{smallmatrix}\right]
\\\nonumber&
\hskip-5mm
=G_{+,-s_q}
\left[\begin{smallmatrix}
\alpha-\frac{kb}{2} &-\frac{b}{2} \\
\alpha & -\frac{(q-1)b}{2}
\end{smallmatrix}\right]
G_{s_q,-}
\left[\begin{smallmatrix}
-\frac{qb}{2} & -\frac{b}{2}\\
\alpha&\alpha-\frac{k'b}{2}
\end{smallmatrix}\right]
G_{-+}
\left[\begin{smallmatrix}
-\frac{(q-1)b}{2}&-\frac{b}{2}\\
0&-\frac{qb}{2}
\end{smallmatrix}\right]
G_{\alpha-\frac{k'b}{2},0}
\left[\begin{smallmatrix}
-\frac{(q-1)b}{2} & -\frac{(q-1)b}{2}\\
\alpha &\alpha
\end{smallmatrix}\right]
\end{align}
Explicitly this is
\beq
G_{\alpha-\frac{kb}{2},0}
\left[\begin{smallmatrix}
-\frac{qb}{2} & -\frac{qb}{2}\\
\alpha &\alpha
\end{smallmatrix}\right]
=-s_q\,\frac{\sin^2\left(\pi b(2\alpha-(s_qq+k+2)\frac{b}{2})\right)}
{\sin(\pi b^2(q+1))\sin\left(\pi b(2\alpha-(k+1)b)\right)}
G_{\alpha-\frac{k'b}{2},0}
\left[\begin{smallmatrix}
-\frac{(q-1)b}{2} & -\frac{(q-1)b}{2}\\
\alpha &\alpha
\end{smallmatrix}\right].
\eeq

Writing $k=s_1+\cdots +s_q$ we get for the full product
\beq
(-1)^\frac{q-k}{2}\prod_{i=1}^q
\frac{\sin^2\left(\pi b(2\alpha-(2+is_i-(s_1+\cdots +s_i))\frac{b}{2})\right)}
{\sin(\pi b^2(i+1))\sin\left(\pi b(2\alpha-(1+s_1+\cdots+ s_i)b)\right)}\,,
\eeq
which has to be summed over all orderings of the $s_i$.

For example, for $q=2$ summing over the choices of $s_1$ and $s_2$ gives
\beq
\begin{aligned}
G_{\alpha-b,0}
\left[\begin{smallmatrix}
-b & -b\\\alpha & \alpha
\end{smallmatrix}\right]
&=\frac{\sin\left(\pi b(2\alpha-2b)\right)\sin\left(\pi b(2\alpha-3b)\right)}
{\sin\left(2\pi b^2\right)\sin\left(3\pi b^2\right)}\\
G_{\alpha,0}
\left[\begin{smallmatrix}
-b & -b\\\alpha & \alpha
\end{smallmatrix}\right]
&=-\frac{\sin\left(2\pi b\alpha\right)\sin\left(\pi b(2\alpha-2b)\right)
[\sin\left(2\pi b\alpha\right)+\sin\left(\pi b(2\alpha-2b)\right)]}
{\sin\left(2\pi b^2\right)\sin\left(3\pi b^2\right)\sin\left(\pi b(2\alpha-b)\right)}\\
&=-\frac{\sin\left(2\pi b\alpha\right)\sin\left(\pi b(2\alpha-2b)\right)}
{\sin\left(\pi b^2\right)\sin\left(3\pi b^2\right)}\\
G_{\alpha+b,0}
\left[\begin{smallmatrix}
-b & -b\\\alpha & \alpha
\end{smallmatrix}\right]
&=\frac{\sin\left(2\pi b\alpha\right)\sin\left(\pi b(2\alpha+b)\right)}
{\sin\left(2\pi b^2\right)\sin\left(3\pi b^2\right)}
\end{aligned}
\label{bto1}
\eeq

While we did not simplify these expressions in the general case, 
we did find by experimention the following useful formula for the 
product of two fusion steps with fixed $k$ (and checked it up to 
$q=5$).
\beq
G^{-1}_{0,\alpha-\frac{kb}{2}}
\left[\begin{smallmatrix}
-\frac{qb}{2} & -\frac{qb}{2}\\
\alpha &\alpha
\end{smallmatrix}\right]
G_{\alpha-\frac{kb}{2},0}
\left[\begin{smallmatrix}
-\frac{qb}{2} & -\frac{qb}{2}\\
\alpha &\alpha
\end{smallmatrix}\right]
=\frac{-1}{\sum_{k'} e^{\pi i k' bQ}}\,
\frac{\sin(\pi b(2\alpha-kb-Q))}{\sin(\pi b(2\alpha-Q))}\,.
\label{qfs}
\eeq
The sum in the denominator is over $k'=-q,-q+2,\cdots q$. 
Using this equation it is possible to derive the expression for the 
Wilson loop in the spin $j=q/2$ representation \eqn{w0q} in 
Section~\ref{sec:wilson}.

Each of the pair of $-qb/2$ degenerate fields can also be 
split recursively into $q$ of the basic degenerate fields
\beq
\label{split-qb/2}
\raisebox{0mm}[13mm][13mm]{\parbox{30mm}{\hskip11mm
\begin{fmfgraph*}(10,20)
\fmfbottom{bl,br}
\fmftop{tl,tr}
\fmf{plain,width=2}{bl,b1,br}
\fmf{phantom}{tl,t1,tr}
\fmffreeze
\fmf{boson,width=1}{tl,c1}
\fmf{boson,width=.5}{c1,tr}
\fmf{boson,width=1.5}{c1,b1}
\fmfv{label=$\alpha^{(q)}$,label.angle=-90,label.dist=4}{bl}
\fmfv{label=$\alpha$,label.angle=-90,label.dist=8}{br}
\fmfv{label=$-\frac{qb}{2}$,label.angle=-135}{c1}
\fmfv{label=$-\frac{(q-1)b}{2}$,label.angle=-180}{tl}
\fmfv{label=$-\frac{b}{2}$,label.angle=0}{tr}
\end{fmfgraph*}}}
\to
\raisebox{0mm}[13mm][13mm]{\parbox{37mm}{\hskip12mm
\begin{fmfgraph*}(20,20)
\fmfbottom{bl,br}
\fmftop{tl,tr}
\fmf{plain,width=2}{bl,b1,b2,b3,br}
\fmf{phantom}{tl,t1,t2,t3,tr}
\fmffreeze
\fmf{boson,width=1}{t1,b1}
\fmf{boson,width=.5}{t3,b3}
\fmfv{label=$\alpha^{(q)}$,label.angle=-90,label.dist=4}{bl}
\fmfv{label=$\alpha$,label.angle=-90,label.dist=8}{br}
\fmfv{label=$\alpha^{(1)}$,label.angle=-90,label.dist=4}{b2}
\fmfv{label=$-\frac{(q-1)b}{2}$,label.angle=180}{t1}
\fmfv{label=$-\frac{b}{2}$,label.angle=0}{t3}
\end{fmfgraph*}}}
\to
\cdots
\to
\raisebox{0mm}[13mm][13mm]{\parbox{50mm}{\begin{center}
\begin{fmfgraph*}(40,20)
\fmfbottom{bl,br}
\fmftop{tl,tr}
\fmf{plain,width=2}{bl,b1,b2,b3,b4,b5,b6,b7,b8,br}
\fmf{phantom}{tl,t1,t2,t3,t4,t5,t6,t7,t8,tr}
\fmffreeze
\fmf{boson,width=.5}{t1,b1}
\fmf{boson,width=.5}{t3,b3}
\fmf{boson,width=.5}{t6,b6}
\fmf{boson,width=.5}{t8,b8}
\fmfv{label=$\alpha^{(q)}$,label.angle=-90,label.dist=4}{bl}
\fmfv{label=$\alpha^{(q-1)}$,label.angle=-90,label.dist=4}{b2}
\fmfv{label=$\cdots$,label.angle=-70,label.dist=10}{b4}
\fmfv{label=$\alpha^{(1)}$,label.angle=-90,label.dist=4}{b7}
\fmfv{label=$\alpha$,label.angle=-90,label.dist=8}{br}
\fmfv{label=$-\frac{b}{2}$,label.angle=180}{t1}
\fmfv{label=$-\frac{b}{2}$,label.angle=180}{t3}
\fmfv{label=$\cdots$,label.angle=-70,label.dist=18}{t4}
\fmfv{label=$-\frac{b}{2}$,label.angle=0}{t6}
\fmfv{label=$-\frac{b}{2}$,label.angle=0}{t8}
\end{fmfgraph*}\end{center}}}
\eeq
Taking $\alpha^{(q)}=\alpha-kb/2$ with $k=\sum_{i=1}^q s_i$, we have 
for the first fusion step
\beq
\label{unfuseq}
G^{-1}_{-qb/2,\alpha^{(1)}}
\left[\begin{smallmatrix}
-\frac{(q-1)b}{2} & -\frac{b}{2}\\
\alpha^{(q)} &\alpha
\end{smallmatrix}\right]
=G_{+,s_1}
\left[\begin{smallmatrix}
\alpha & -\frac{b}{2}\\
\alpha^{(q)} & -\frac{(q-1)b}{2}
\end{smallmatrix}\right]
=\frac{\sin\left(\pi b^2(2\alpha-b+(qs-k)b/2)\right)}
{\sin\left(\pi b^2(2\alpha-b)\right)}\,.
\eeq

Splitting up the field with $-qb/2$ into $q$ fields with $-b/2$ is given by the 
product
\beq
\frac{1}{\sin^{q-1}\left(\pi b^2(2\alpha-b)\right)}
\prod_{i=2}^q
\sin\left(\pi b^2(2\alpha-b+(is_i-(s_1+\cdots +s_i))b/2)\right)\,.
\eeq
Clearly for $s_1=\cdots=s_q$ the product is just unity.

We can also go in the other direction, fusing $q$ basic degenerate fields and 
projecting on the symmetric combination, which is a single $qb/2$ field
\beq
\raisebox{0mm}[13mm][13mm]{\parbox{50mm}{\begin{center}
\begin{fmfgraph*}(40,20)
\fmfbottom{bl,br}
\fmftop{tl,tr}
\fmf{plain,width=2}{bl,b1,b2,b3,b4,b5,b6,b7,b8,br}
\fmf{phantom}{tl,t1,t2,t3,t4,t5,t6,t7,t8,tr}
\fmffreeze
\fmf{boson,width=.5}{t1,b1}
\fmf{boson,width=.5}{t3,b3}
\fmf{boson,width=.5}{t6,b6}
\fmf{boson,width=.5}{t8,b8}
\fmfv{label=$\alpha^{(q)}$,label.angle=-90,label.dist=4}{bl}
\fmfv{label=$\alpha^{(q-1)}$,label.angle=-90,label.dist=4}{b2}
\fmfv{label=$\cdots$,label.angle=-70,label.dist=10}{b4}
\fmfv{label=$\alpha^{(1)}$,label.angle=-90,label.dist=4}{b7}
\fmfv{label=$\alpha$,label.angle=-90,label.dist=8}{br}
\fmfv{label=$-\frac{b}{2}$,label.angle=180}{t1}
\fmfv{label=$-\frac{b}{2}$,label.angle=180}{t3}
\fmfv{label=$\cdots$,label.angle=-70,label.dist=18}{t4}
\fmfv{label=$-\frac{b}{2}$,label.angle=0}{t6}
\fmfv{label=$-\frac{b}{2}$,label.angle=0}{t8}
\end{fmfgraph*}\end{center}}}
\to
\cdots
\to
\raisebox{0mm}[13mm][13mm]{\parbox{37mm}{\hskip12mm
\begin{fmfgraph*}(20,20)
\fmfbottom{bl,br}
\fmftop{tl,tr}
\fmf{plain,width=2}{bl,b1,b2,b3,br}
\fmf{phantom}{tl,t1,t2,t3,tr}
\fmffreeze
\fmf{boson,width=1}{t1,b1}
\fmf{boson,width=.5}{t3,b3}
\fmfv{label=$\alpha^{(q)}$,label.angle=-90,label.dist=4}{bl}
\fmfv{label=$\alpha$,label.angle=-90,label.dist=8}{br}
\fmfv{label=$\alpha^{(1)}$,label.angle=-90,label.dist=4}{b2}
\fmfv{label=$-\frac{(q-1)b}{2}$,label.angle=180}{t1}
\fmfv{label=$-\frac{b}{2}$,label.angle=0}{t3}
\end{fmfgraph*}}}
\to
\raisebox{0mm}[13mm][13mm]{\parbox{30mm}{\hskip16mm
\begin{fmfgraph*}(10,20)
\fmfbottom{bl,br}
\fmftop{tl,tr}
\fmf{plain,width=2}{bl,b1,br}
\fmf{phantom}{tl,t1,tr}
\fmffreeze
\fmf{boson,width=1}{tl,c1}
\fmf{boson,width=.5}{c1,tr}
\fmf{boson,width=1.5}{c1,b1}
\fmfv{label=$\alpha^{(q)}$,label.angle=-90,label.dist=4}{bl}
\fmfv{label=$\alpha$,label.angle=-90,label.dist=8}{br}
\fmfv{label=$-\frac{qb}{2}$,label.angle=-135}{c1}
\fmfv{label=$-\frac{(q-1)b}{2}$,label.angle=-180}{tl}
\fmfv{label=$-\frac{b}{2}$,label.angle=0}{tr}
\end{fmfgraph*}}}
\eeq
Taking $\alpha^{(q)}=\alpha-kb/2$ with $k=\sum_{i=1}^q s_i$, we have 
for the last fusion step
\beq
G_{\alpha^{(1)},-qb/2}
\left[\begin{smallmatrix}
-\frac{(q-1)b}{2} & -\frac{b}{2}\\
\alpha^{(q)} & \alpha
\end{smallmatrix}\right]
=G_{s_1+}
=\frac{\sin\left(\pi b^2(q+s_1k)/2\right)}
{\sin\left(\pi b^2q\right)}\,.
\eeq

Doing this $q$ times we get the product
\beq
\prod_{i=2}^q
\frac{\sin\left(\pi b^2(i+s_i(s_1+\cdots +s_i))/2\right)}{\sin(\pi b^2i)}\,.
\eeq
Though this is not manifest, 
the product on the right hand side is symmetric under exchange of any 
of the $s_i\leftrightarrow s_j$. If we take $s_1,\cdots,s_{(q+k)/2}$ to be positive and the 
rest to be negative, then the numerator of the first $(q+k)/2$ terms will be 
$\sin(\pi b^2 i)$ and the subsequent ones $\sin(\pi b^2(2i-q-k)/2)$. The first 
$(q+k)/2$ will cancel against the denominator giving
\beq
\prod_{j=1}^{\frac{q-|k|}{2}}
\frac{\sin\left(\pi b^2j\right)}{\sin\left(\pi b^2(j+\frac{q+|k|}{2})\right)}\,.
\eeq
Clearly for $s_1=\cdots=s_q$ the product is just unity.

For example for $q=2$
\beq
\label{-b}
G_{\alpha^{(1)},-b}
\left[\begin{smallmatrix}
-\frac{b}{2} & -\frac{b}{2}\\
\alpha-\frac{(s_1+s_2)b}{2} & \alpha
\end{smallmatrix}\right]
=\frac{\sin\left(\pi b^2(3+s_1s_2)/2\right)}
{\sin\left(2\pi b^2\right)}
=\begin{cases}
1\,,&s_1=s_2\\
\frac{1}{2\cos(\pi b^2)}\,,& s_1=-s_2\,.
\end{cases}
\eeq
Combining three $-b/2$ fields to a single $-3b/2$ field gives
\beq
\frac{\sin\left(\pi b^2(3+s_1s_2)/2\right)\sin\left(\pi b^2(4+s_3(s_1+s_2))/2\right)}
{\sin(2\pi b^2)\sin(3\pi b^2)}
=\begin{cases}
1\,,&s_1=s_2=s_3\\
\frac{\sin(\pi b^2)}{\sin(3\pi b^2)}\,,& \text{otherwise.}
\end{cases}
\eeq

\end{fmffile}
\bibliography{refs}
\end{document}